\documentclass[10pt]{iopart}

\usepackage{braket,dsfont,color,amsfonts,amssymb,bm,amstext,mathrsfs}
\usepackage[bottom=1in,left=1in,right=1in,top=1in]{geometry}
\usepackage{graphicx}   
\usepackage{multirow}
\usepackage[normalem]{ulem}
\usepackage{qcircuit}
\usepackage{braket}

\usepackage{epsfig,amssymb}
\usepackage{changepage}
\usepackage{bbm}
\usepackage{bm}
\usepackage{amsthm}
\usepackage{xcolor}
\usepackage[caption=false]{subfig}
\usepackage{multirow}
\usepackage[normalem]{ulem}
\usepackage{qcircuit}
\usepackage{braket}

\usepackage{amsfonts,amssymb,dsfont}
\usepackage{graphicx}
\usepackage{soul}

\definecolor{rev}{rgb}{0,0,0}

\newcommand{\cumu}[1]{\big\langle\!\big\langle#1\big\rangle\!\big\rangle_B}
\newcommand{\expect}[1]{\big\langle#1\big\rangle_B}

\begin{document}


{\color{rev}
\title[]{Frequency estimation under non-Markovian spatially correlated quantum noise}}

\author{Francisco Riberi$^1$, Leigh M. Norris$^{2}$, F\'elix Beaudoin$^{3}$, and Lorenza Viola$^1$}

\address{$^1$Department of Physics and Astronomy, Dartmouth
College, 6127 Wilder Laboratory, \\ Hanover, NH 03755, USA}
\address{$^2$Johns Hopkins University Applied Physics Laboratory
11100 Johns Hopkins Road, \\ Laurel, MD, 20723, USA}
\address{$^{3}$Nanoacademic Technologies Inc., 666 rue Sherbrooke Ouest, Suite 802, Montr\'eal, 
\\ Qu\'ebec, Canada H3A 1E7}

\begin{abstract}
We study the estimation precision attainable by entanglement-enhanced Ramsey interferometry in the presence of spatiotemporally correlated non-classical noise. Our analysis relies on an exact expression of the reduced density matrix of the qubit probes under  general zero-mean Gaussian stationary dephasing, which is established through cumulant-expansion techniques and may be of independent interest in the context of non-Markovian open dynamics. By continuing and expanding our previous work [Beaudoin {\em et al.}, Phys.\,Rev.\,A\,{\bf 98}, 020102(R) (2018)], we analyze the effects of a {\em non-collective} coupling regime between the qubit probes and their environment, focusing on two limiting scenarios where the couplings may take only two or a continuum of possible values. In the paradigmatic case of spin-boson dephasing noise from a thermal environment, we find that it is {\color{rev} in principle} possible to suppress, {\em on average}, the effect of spatial correlations by {\em randomizing the location of the probes}, as long as enough configurations are sampled where noise correlations are negative. As a result, superclassical precision scaling is asymptotically restored for initial entangled states, including experimentally accessible one-axis spin-squeezed states.  
\end{abstract}


\submitto{\NJP}

\section{Introduction}
  
The field of quantum metrology is concerned with the study of measurement strategies that use quantum probes as sensors and exploit genuinely non-classical resources -- such as entanglement, squeezing, indistinguishability or  many-body interactions -- to enhance the precision to which a target physical quantity can be estimated \cite{SethMetrology,SmerziRMP,BraunRMP}.  {\em Entanglement-assisted metrology}, in particular, leverages the fact that multiple sensors may be available and able to be initialized in non-classical states and operated in parallel. In the simplest setting, measurements are carried out to estimate a single target parameter as precisely as possible, subject to specified operational constraints -- for instance, a fixed total time $T$ to carry out the experiment and a fixed number $N$ of sensors. It is well-known that, under these constraints, the best possible classical strategy yields a precision scaling of the form $N^{-1/2}$, the so-called \textit{standard quantum limit} (SQL), which is essentially set by the central limit theorem of classical statistics \cite{Smirnereview}. However, by exploiting non-classical correlations, the estimation precision may be improved beyond the SQL, under the same set of constraints. In an ideal noiseless setting, and assuming no interactions among the probes, the best quantum strategies can saturate the so-called \textit{Heisenberg limit} (HL), whereby the precision scales as $N^{-1}$ and thus achieves a remarkable $N^{-1/2}$ improvement over the SQL. Genuine $N$-particle entanglement is required to saturate the HL, as exhibited by the paradigmatic Greenberger-Horne-Zeilinger (GHZ) state (also referred to as NOON states in optical interferometry) \cite{Huelga,SethMetrology,SmerziRMP}; still, the SQL can be surpassed by using a number of other non-classical states, most notably, spin squeezed states as well as certain symmetric states \cite{Kita1993,DaveSqueezing,Nori2,Vuletic,Ouyang}.

Quantum-enhanced measurement strategies find use in such diverse areas as frequency estimation and magnetometry \cite{Bollinger2, Jones}, thermometry \cite{Stace}, high-precision timekeeping with atomic clocks \cite{Ye}, radar technologies \cite{Maccone}, and beyond. Notably, the use of squeezed light has proved instrumental for detecting the first unambiguous transient gravitational-wave signal \cite{Abbott}, and has enabled sub-SQL precision in low-intensity coherent Raman microscopy for biological applications \cite{Catxere}. Likewise, phase sensitivity with Heisenberg scaling has been reported in a Ramsey atomic interferometer where spin squeezing was enhanced with effective time-reversal techniques \cite{Monika,VladanSatin}. As experimental capabilities and measurement precision continue to improve, it becomes increasingly important to develop as general and complete as possible an understanding of the impact that noise sources have on the attainable quantum advantage.

Noisy quantum parameter estimation has been extensively studied over the past decade, for a variety of different environmental noise models. If noise is both spatially uncorrelated, so that it acts independently on each sensor, and temporally uncorrelated (``Markovian''), any metrological advantage afforded by entanglement is erased \cite{Huelga}. However, the presence of correlations in the noise can help, at least partially, to restore superclassical precision scaling. For noise that is still spatially local but temporally correlated (``non-Markovian''), the existence of a colored spectrum and initial Zeno dynamics may be exploited to achieve a superclassical {\em Zeno scaling limit} $\propto N^{-3/4}$ {\color{rev}\cite{{Matsuzaki},Chin2012,Macies2015}}. On the other hand, in a Markovian setting, perfect spatial correlations (``collective'' noise) may enable Heisenberg scaling via decoherence-free subspace encoding \cite{Dorner2012}; or, even if correlations are partial but of a  known structure, enhanced sensitivity may still be achievable via suitable entangled-state preparations \cite{Jeske2014} or active error-correction protocols that filter out the noise \cite{Layden}. In practice, spatial noise correlations tend to naturally emerge due to probe proximity \cite{Dorner2012,Monz,Lieven2020}. The occurrence of nontrivial temporal correlations is also generically expected, and has been directly verified across a variety of systems through ``quantum noise spectroscopy'' (QNS) experiments \cite{qns,Frey}. 

The important case where spatial and temporal correlations are {\em simultaneously} present has received comparatively little attention thus far. As in the spatially uncorrelated case, memory effects stemming from temporal noise correlations have been still found to be beneficial for reaching sub-SQL sensitivity in the presence of spatial correlations from a {\em classical} environment \cite{Szankowski}. In a high-precision regime, however, the classicality assumption should be treated with care, as quantum effects may be expected to ultimately play a role. Recent experiments on both single- \cite{Quintana2017,Yan2018} and two-qubit \cite{Uwe} superconducting devices have in fact directly probed \textit{non-classical} noise environments, that is, environments for which the \emph{non-commuting} nature of the underlying degrees of freedom manifests, in the frequency domain, through noise spectra that are asymmetric with respect to zero frequency \cite{ClerkRMP, Paz2017}. Crucially, coupling of the individual probes to such a correlated non-classical environment can generate uncontrolled entanglement among the probes. This leads to an additional source of uncertainty whose metrological impact has not been fully assessed yet. As a first step in this direction, Ref.\,\cite{FelixPRA} studied the problem of Ramsey frequency estimation in the presence of non-Markovian dephasing noise that is both spatially correlated and non-classical. Specifically, in the limiting case where the $N$ qubit sensors couple linearly and {\em collectively} to a bosonic environment, superclassical precision scaling failed to be reached by using entangled probes prepared in an experimentally accessible one-axis twisted state (OATS), with the bath-induced entanglement resulting in an exponential increase of the uncertainty away from the optimal measurement time.
 Interestingly, a strategy for leveraging the spatially correlated nature of the noise and restore a quantum advantage in this same bosonic setting was recently proposed in \cite{AncillaBased}, based on the use of an auxiliary qubit as a means for detecting and effectively undoing decoherence of the $N$ qubit probes. While such an approach is ideally suited for an initial GHZ state (and other states supported on a two-dimensional subspace), it does not extend in scalable form to other entangled states of practical relevance, such as spin squeezed states. 

Our main goal in this work is to further explore the combined effect and interplay of {\em spatiotemporal noise correlations and non-classicality} in Ramsey quantum frequency estimation protocols -- with emphasis on characterizing the extent to which relaxing the collectivity assumption of \cite{FelixPRA} affects the metrological outlook. We focus on the most adversarial (and, arguably, most common \cite{DegenRMP}) kind of noise, namely,  ``parallel'' {\em dephasing noise}, which couples through the {\em same} operators as the signal. While parallel noise that is not temporally correlated can be tackled by ancilla-free error-corrected sensing \cite{Layden2}, and spatiotemporally correlated noise that is not parallel can be countered by dynamical decoupling  (DD) \cite{multiDD,DurDD}, the occurrence of these features together prevents either of these techniques from being directly applicable as they stand. More concretely, we contrast the metrological performance of unentangled coherent spin states (CSSs) to those attainable by entangled probes initialized in either a GHZ state or an OATS. \cite{Kita1993,BraskPRX}. Allowing the qubits to couple to the bath in a way that breaks permutation symmetry can lead to an overall decrease of spatial correlations among the qubits when compared to the collective case which, as we show, may result in enhanced estimation precision. Interestingly, the benefit of varying probe distance in DC magnetometry using two entangled trapped ions has been analyzed in \cite{ruster}; in a similar vein, a scalable method for quantifying the strength of spatial correlations in environmental noise processes has been experimentally validated in \cite{Postler}.

An outline of the paper and a summary of its main results follow. Section \ref{sec:back} gives a brief introduction to the tools of estimation theory necessary to understand our work. In Sec.\,\ref{sec:GaussianDeph} we provide an {\em exact} representation for the reduced dynamics of a collection of probes subject to spatiotemporally correlated Gaussian stationary dephasing quantum noise, which may be of independent interest for open-quantum system studies. {\color{rev} While the approach in \cite{multiDD} relied explicitly on the bosonic nature of the bath to truncate the Magnus expansion of the time-dependent evolution operator, exact truncation is obtained here solely by exploiting Gaussianity, similar in spirit to the analogous exact result available for (invertible) observables in \cite{Paz2017}. Starting in Sec.\,\ref{sec:NQSB} for the rest of the paper, we analyze the paradigmatic linear spin-boson model as a concrete instance of the more general class of Gaussian quantum noise}. After summarizing the results found in \cite{FelixPRA} for the collective limit, we expand upon them by showing that the periodic frequency uncertainty minima found for a CSS initial state may be physically interpreted in terms of {\em disentanglement} among the qubit probes, in spite of nontrivial entanglement they still exhibit with the bath. Furthermore, we prove that the optimal choice of squeezing and rotation angles for an initial OATS in the absence of noise \cite{Kita1993} remains optimal under permutation-invariant noise (and, in fact, beyond). 

The next two sections include our main results regarding Ramsey frequency estimation for a dephasing spin-boson model with broken permutational symmetry. Specifically, in Sec.\,\ref{sec:eo} the simplest ``digital" departure from collectivity is analyzed, whereby the probes are spatially distributed in two ``lumps" of the same size, at a spatial distance $r$. We find that, while no scaling advantage can be gained with respect to the collective case, a constant factor improvement in the estimation precision is possible, and can be optimized at a finite value of the separation $r$ that minimizes the noise spatial correlations. In Sec.\,\ref{sec:random}, we introduce a {\em randomized estimation protocol}, by letting the qubit probes to be randomly Gaussian-distributed in space, and thus breaking permutation symmetry by effectively randomizing the position-dependent couplings. In a limit where the spatial dispersion of the Gaussian profile is wide enough relative to a characteristic bath length scale, we show that, as a net effect of the spatial averaging, the noise entering the qubit dynamics becomes spatially uncorrelated. As a result, asymptotic superclassical scaling is restored on average for both initial OATs and GHZ states: while the latter reaches the optimal $N^{-3/4}$ Zeno scaling, a precision scaling as $N^{-2/3}$ is found for an OATS in the limit of local non-Markovian noise -- a result not previously reported in the literature to the best of our knowledge. Concluding remarks are presented in Sec.\,\ref{conclusion}. 

{\color{rev} A number of additional results and the details of relevant derivations are included in the appendixes. In particular, in \ref{ZMGSDAp} we establish the above-mentioned explicit form of the reduced density operator for zero-mean Gaussian stationary dephasing, and showcase two different applications: first, we provide a characterization of sets of states that are {\em quantum-noise-insensitive}, depending on the symmetries of the noise; in particular, while states in the same entanglement class as the GHZ are well-known to be decoherence-free under fully correlated, collective dephasing \cite{QECBook}, we show that GHZ states are insensitive to the non-classical contributions of dephasing noise under arbitrary spatial correlations. Second, we establish a sufficient condition on the noise for the reduced dynamics to be {\em random unitary}, namely, representable in terms of classical stochastic fields \cite{RU1}. \ref{app:OATS} summarizes relevant derivations OATS dynamics under collective and even-odd spin-boson noise, whereas \ref{RCapp} provides full technical details of spin-boson dynamics under spatial randomization. Remarks about non-collective dephasing dynamics beyond the spin-boson setting are collected in \ref{app:phi1}.
}

\section{Background: Quantum frequency estimation by Ramsey interferometry} 
\label{sec:back}

While quantum metrology is a rich and diversified topic, Ramsey interferometry is both central to our investigations and representative of the key ideas involved in a quantum frequency estimation protocol. Three main steps are involved \cite{SmerziRMP,Smirnereview}: (i) {\em preparation} of the sensing system in an initial state, say, $\rho_0$; (ii) {\em encoding} the parameter of interest, say, an angular frequency $b$, into an  evolution period of duration $\tau$; (iii) {\em readout}, whereby an estimate $\hat{b}$ of $b$ is obtained through measurement of a suitable observable. This conceptually simple scheme is common to all interferometric sensors, from atomic clocks to gyroscopes, gravimeters, and gravitational wave detectors, to name a few. Crucially, any specification of an achievable estimation precision is only meaningful relative to what {\em resource constraints} are in place in a given setting. We take the probe system to comprise $N$ qubits, with associated Pauli matrices $\{ \sigma_n^{\alpha} \},$ $\alpha \in \{x,y,z\}$, $n=1, \ldots N$; accordingly, the system's Hilbert space ${\mathcal H}_{\rm S} \simeq ({\mathbb C}^2)^{\otimes N}$. While different settings have been envisioned \cite{BraunRMP}, we consider the standard {\em linear} interferometric regime in which only local (one-body) Hamiltonians are used in generating the parameter-encoding dynamics, and no interactions are permitted among the probes during the encoding period. In the absence of noise and letting $\hbar =1$, the Hamiltonian thus simply reads $H_{\rm S} = b \sum_{n=1}^N \sigma_n^z/2$. 

As a frequency-estimation cycle is typically repeated multiple times, the number of repetitions is related to the total time $T$ via $\nu \equiv T/\tau$. Let ${\mathbb P}(\mu | b)$ denote the probability that outcome $\mu$ is observed given that the parameter has the value $b$, so that the conditional probability of observing the sequence $\vec{\mu}\equiv \{ \mu_1, \ldots, \mu_\nu\}$ in $\nu$ independent measurements is ${\mathbb P}(\vec{\mu}|b)$. An {\em estimator} $\hat{b}(\vec{\mu})$ is a function that associates each set of measurement outcomes $\vec{\mu}$ with an estimate of the target parameter. Since the estimator is a function of random outcomes, it is itself a random variable and, as such, its statistical properties can be characterized in terms of cumulants. In particular, the mean and variance of the estimator are given by 
$$\langle \hat{b} (\vec{\mu}) \rangle_{\vec{\mu}} = \sum_{\vec{\mu}} {\mathbb P}(\vec{\mu}|b) \hat{b}(\vec{\mu}) , \quad 
\Delta \hat{b}^2 \equiv \langle \, (\hat{b}(\vec{\mu}) - \langle \hat{b} (\vec{\mu}) \rangle_{\vec{\mu}}   )^2 \,\rangle_{\vec{\mu}} = \sum_{\vec{\mu}} {\mathbb P}(\vec{\mu}|b)  ( \,\hat{b}(\vec{\mu}) - \langle \hat{b} (\vec{\mu}) \rangle_{\vec{\mu}} \,)^2,$$
where the expectation $\langle \cdot \rangle_{\vec{\mu}}$ is taken over {\em all} possible measurement outcomes and the quantum state. Irrespective of the explicit form of the estimator, two properties are especially desirable \cite{Kay,Helstrom}: 

\smallskip

\noindent
{\bf Definition.} {\em An estimator $\hat{b}$ is {\em unbiased} if its mean coincides with the true value, $\langle \hat{b} (\vec{\mu}) \rangle_{\vec{\mu}} = b$. Otherwise, the estimator is {\em biased}. An estimator $\hat{b}$ is {\em consistent} if it is convergent in probability to the true value for an infinitely large sample, that is, for every $\varepsilon >0$, the probability that the estimated value differs by more than $\varepsilon$ from the true value obeys 
$\lim_{\nu\rightarrow\infty} {\mathbb{P}} (| \hat{b} (\vec{\mu}) - b | > \varepsilon) =0$. Thus, we have $\lim_{\nu\rightarrow\infty}  \hat{b} (\vec{\mu})  =b$. }

\smallskip

\noindent 
Although consistency implies asymptotic unbiasedness, the converse is {\em not} true in general. The error of an estimator is naturally captured by the {\em mean squared error}, ${\rm MSE}(\hat{b}) \equiv \langle \,(\hat{b}(\vec{\mu}) - b )^2 \rangle_{\vec{\mu}}=\Delta \hat{b}^2+B^2$, where $B\equiv\langle{\hat{b}(\vec{\mu})}\rangle_{\vec{\mu}} - b$ is known as the {\em bias}. If $\hat{b}$ is unbiased, ${\rm MSE}(\hat{b})=\Delta\hat{b}^2$. In what follows, we will  neglect bias effects arising from finite measurement statistics {\color{rev}\cite{Rubio,Chabuda}} or possible noise-induced bias, and quantify the performance of $\hat{b}$ with the standard deviation or {\em precision}, $\Delta\hat{b}\equiv \sqrt{\Delta \hat{b}^2}$.

\subsection{Noiseless metrology bounds} 
One of the most important results in parameter-estimation theory stems from the fact that, under mild differentiability assumptions on the ``likelihood function'' $P(\vec{\mu}|b)$, a lower bound to the estimation precision may be established, in the form of a generalized uncertainty relation \cite{Kay}:

\smallskip

\noindent
{\bf Cram\'{e}r-Rao bound.} {\em The variance of any unbiased estimator $\hat{b}$ is bounded from below by the {\em Cram\'{e}r-Rao bound} (CRB), which is given in terms of the {\em classical Fisher information}, $F_{ \mathrm{cl}}$, as follows:}
\begin{eqnarray}
{\rm MSE}(\hat{b}) = \Delta \hat{b}^2 \geq \Delta \hat{b}^2_{\mathrm{CR}} = \frac{1}{ {\nu F_{ \mathrm{cl}} [{\mathbb P}(\vec{\mu}|b)] } }, \quad
F_{ \mathrm{cl}} [{\mathbb P}(\vec{\mu}|b)] \equiv 
\sum_{\vec{\mu}} \frac{1}{{\mathbb P}(\vec{\mu}|b)} \bigg( \frac{\partial {\mathbb P}(\vec{\mu}|b)}
{\partial b} \bigg)^2. 
\label{crb}
\end{eqnarray}

Any estimator achieving equality in the above CRB is termed {\em efficient}. An  upper bound to the Fisher information may be obtained by maximizing the classical expression over all possible generalized measurements that quantum mechanics allows, leading to $F_{ \mathrm{Q}} [\rho_b] \equiv \mathrm{max}_{\{{\cal E}\}} F_{ \mathrm{cl}} [ \rho_b]$, which is the {\em quantum Fisher Information} (QFI) \cite{CarlDistance}. Here, $\rho_b$ denotes the output state that serves as the pre-measurement state for readout and, formally, the maximization takes place over all possible POVM (positive operator-values measure) measurements, $\{ {\cal E }\}\equiv \{ E(\mu), \sum_\mu E(\mu)= {\mathbb I} \}$, in terms of appropriate POVM-element operators. Accordingly, a tighter lower bound than the one in Eq.\,(\ref{crb}) is provided by the {\em quantum CRB} \cite{Holevo,Helstrom}, 
\begin{equation}
\Delta \hat{b}^2_{\mathrm{CR}} \geq \Delta \hat{b}^2_{\mathrm{QCR}} = \frac{1}{{\nu F_{ \mathrm{Q}} [\rho_b] }}.
\label{qcr}
\end{equation} 
While hard to evaluate theoretically beyond specific cases, as well as to achieve by using experimentally accessible (local) measurements, the quantum CRB provides the ultimate precision limit allowed in the noiseless estimation scenario, with important advances toward both its determination and practical realization having been recently reported \cite{Datta,LiangBound}.

By using the fact that the QFI of {\em any} separable state $\rho_{\rm{sep}}$ of $N$ qubits is upper-bounded by $F_{ \mathrm{Q}} [\rho_{\rm{sep}}] \leq N$ \cite{SmerziRMP}, or by directly evaluating the CRB for the free precession dynamics generated by the above $H_S$ for any initially {\em separable} $\rho_0$,  it is straightforward to establish \cite{Smirnereview} that the achievable precision is given by the SQL, 
\begin{equation}
\Delta \hat{b}_{\mathrm{SQL}} = \frac{1}{\tau \sqrt{N \nu}}  = \frac{1}{\sqrt{N \tau T }}, 
\label{sql}
\end{equation} 
{\em independent} of the specific measurement and estimator used. It is well known that CSSs are optimal separable states for metrology, able to maximize the QFI and achieve the SQL \cite{DaveSqueezing,Bollinger2}. Let $\hat{J}\equiv (J_x,J_y, J_z),$ $J_\alpha \equiv \sum_n \sigma_n^\alpha/2$ denote collective spin angular momentum operators. CSSs can be seen as angular momentum {\em generalized coherent states} \cite{GCS}: they are constructed as product of $N$ qubits in pure states all pointing along the same direction $\vec{n}\equiv (\sin\theta \cos\phi, \sin\theta \sin\phi, \cos\theta)$, $\theta \in [0, \pi)$, $\phi\in [-\pi, \pi]$, leading to $\rho_{\mathrm{CSS}}\equiv |\psi\rangle\langle\psi|^{\otimes N}$, with $|\psi\rangle$ obeying the eigenvalue equation $(\hat{J}\cdot \hat{n}) |\psi\rangle = (N/2)  |\psi\rangle$. 

\subsection{Metrologically useful entanglement} 
Achieving superclassical precision within linear interferometry requires that the QFI  $F_{ \mathrm{Q}} [\rho_b] > N$, which turns out to be necessary and sufficient for entanglement in $\rho_b$ (hence in $\rho_0$) to translate into a sensitivity gain. Since a necessary (not sufficient) condition for attaining the maximum value $F_{ \mathrm{Q}} [\rho_{b,\rm{max}}] = N^2$ is that the state exhibits genuine $N$-partite entanglement \cite{SmerziRMP}, the achievable precision limit is now given by the HL, 
\begin{equation}
\Delta \hat{b}_{\mathrm{HL}} = \frac{1}{\tau N \sqrt{\nu}}  = \frac{1}{N \sqrt{\tau T}}.
\label{hl}
\end{equation} 
Thus, a scaling in precision that is faster by a factor of $N^{-1/2}$ with respect to the SQL is permitted, despite the {\em same} fixed resources (same  $N$ and $T$) being used. As mentioned, the HL is known to be saturated by GHZ states, $|{\rm GHZ}_N\rangle\equiv (|\!\uparrow\rangle^{\otimes N}\!+| \downarrow\rangle^{\otimes N})/\sqrt2$ \cite{Huelga,SethMetrology,SmerziRMP}. An optimal POVM measurement consists of determining the probability that all qubits are in the $|0\rangle$ state. Even in the noiseless scenario, however, GHZ states for large $N$ are not easily generated in practice. Spin-squeezed states possess metrologically useful entanglement in the sense that $N < F_{ \mathrm{Q}} [\rho_{b, \rm{sq}}] <N^2$, in addition to being easier to generate and measure experimentally. Generally, in a spin-squeezed state, the fluctuation of the collective-spin is reduced (``squeezed") along a particular direction at the cost of anti-squeezing the variance along an orthogonal direction \cite{Kita1993,Nori2}. Optimal spin-squeezed states can achieve Heisenberg scaling, i.e., $\Delta \hat{b}_{\rm{sq, opt}} =\mathcal{O} (N^{-1})$, albeit not the HL given above. 

In what follows, we shall consider spin-squeezed states obtained by nonlinear ``one-axis-twisting'' interactions \cite{Kita1993,BraskPRX}. Specifically, an OATS may be produced by first preparing a CSS along one direction, and then evolving under a Hamiltonian quadratic in one of the perpendicular spin components. For instance, let $\rho_{+\vec{\hat x}}= |+\rangle\langle +|^{\otimes N}$ be a CSS along the $+1$ eigenstate of $\sigma_n^x$; then 
\begin{equation}
\rho_{\rm{OATS}} (\beta,\theta) \equiv U_{\rm{sq}}(\beta,\theta)\rho_{+\vec{\hat x}}\,U_{\rm{sq}}^\dag (\beta,\theta), 
\qquad U_{\rm{sq}}(\beta,\theta) \equiv e^{-i\beta J_x}e^{-i\theta J_z^2/2}, 
\label{oat}
\end{equation}
for squeezing and rotation angles $\theta$ and $\beta$, is an OATS with minimum uncertainty along an angle in the $yz$-plane. When  $ \theta \lesssim 2 N^{-1/2}$ the resulting state is \textit{properly squeezed} \cite{SchulteEchoes}. Quantitatively, this is captured by the inequality $\xi_R <1$, where $\xi_R$ is the spin squeezing parameter introduced by Wineland \cite{DaveSqueezing}, which indicates that the state can be used to overcome the SQL in traditional Ramsey spectroscopy. Larger values of $\theta$ generate states that are ``oversqueezed", wrapping around the Bloch sphere. Interestingly, this class of states may still be highly entangled; when $\pi - 4 N^{-1/2} \leq \theta \leq \pi$ the ensuing states are close to rotated versions of the GHZ state, which saturates the HL in a noiseless setting. Starting from a properly squeezed OATS, the phase evolution that results from $H_S$ and encodes the target parameter $b$ can be detected through $\nu$ independent measurements of a transverse collective-spin observable, say, $J_y$, after an interrogation time $\tau$. The standard deviation in estimating $b$ may then be obtained via error propagation \cite{SmerziRMP}, 
\begin{equation}
\Delta \hat{b} (\tau) = \frac{\Delta J_y (\tau)}{ \sqrt{\nu} {|\partial\langle{J_y(\tau)\rangle}/\partial b|}},\qquad \Delta J_y (\tau) \equiv \sqrt{\langle{J_y^2(\tau) \rangle}- \langle{J_y(\tau)}\rangle^2}.
\label{eqnObs}
\end{equation}
Although the optimal precision scaling, $\Delta \hat{b}^{\text{OATS}}_{\rm{opt}} \propto N^{-5/6}$ \cite{Kita1993,Monika}, is not at the HL, OATSs are tractable analytically and experimentally accessible in various platforms.  Notably, OATSs have been generated in trapped ions and cold atoms in cavities \cite{Bohnet,Vuletic}, and have been used for quantum-enhanced magnetometry in a Bose-Einstein condensate \cite{Ober}. Proposals also exist for creating OATSs in diamond NV centers \cite{Lukin}. 

\section{Noisy frequency estimation: Spatiotemporally correlated Gaussian quantum dephasing} 
\label{sec:GaussianDeph}

We assume that each of the $N$ qubit sensors couples longitudinally to a dephasing quantum environment (or bath, B), through a bath operator $B_n$ acting on the Hilbert space ${\mathcal H}_{\rm B}$. In the interaction picture with respect to the free bath Hamiltonian, $H_{\rm B}$, we thus modify the previous noiseless-sensing Hamiltonian $H_S$ to a joint Hamiltonian on ${\cal H}_{\rm S} \otimes {\cal H}_{\rm B}$ of the form 
\begin{equation}
H_{\rm S} = \frac{1}{2} \sum_{n=1}^N  b\, \sigma_n^z \;\mapsto \; H_{\rm {SB}}(t)= \frac{1}{2} \sum_{n=1}^N \, 
\sigma_n^{z} \otimes [b + B_n(t)], 
\label{eqnSB}
\end{equation}
where, as before, $b$ is the angular frequency we wish to estimate, and $B_n(t) \equiv e^{i H_{\rm B} t}B_n e^{-i H_{\rm B} t}$. The special case of a {\em classical} bath may be formally recovered by letting the operators $B_n(t)$ be classical stochastic processes, in which case arbitrary commutators vanish. For a quantum bath, following \cite{multiDD}, it is useful to distinguish between a \emph{private} bath scenario, in which case each qubit couples to an individual environment and thus $[B_n(t), B_m(t')]=0$ for every qubit pair $n\ne m$ and for all $t, t'$, vs. a \emph{common} bath scenario, in which case $[B_n(t), B_m(t')]\ne 0$ for every $n,m$ for at least some $t,t'$. For either a classical or a quantum (common) bath, \emph{collective} dephasing corresponds to a permutation-invariant limit whereby $B_n(t) \equiv B(t)$, for all $t$, as extensively studied in the context of decoherence-free subspaces and noiseless subsystems \cite{QECBook}. A simple non-collective instance we will consider is an \emph{even-odd} setting, whereby $J$ out of $N\equiv 2J$ qubit probes couple to the bath via an operator $B_{\text{e}}(t)$, and the remaining $J$ via a second operator $B_{\text{o}}(t) \neq B_{\text{e}}(t)$. We will refer to the most general setting, where $B_n(t) \neq B_m(t)$ for all $n \neq m $, as  \emph{general non-collective} dephasing.

\subsection{Noise spatial and temporal correlations}

Throughout this work, we shall assume that the initial joint state is factorized, $\rho_{\mathrm {SB}} (0) \equiv \rho_0 \otimes\rho_{\mathrm B}$, and that the noise process described by the set $\{ B_n(t)\}$ is zero-mean, Gaussian and stationary -- which will be henceforth referred to as the \emph{zero-mean Gaussian stationary dephasing} (ZMGSD) setting. The zero-mean condition is met when expectation value of $B_n(t)$ is zero for all qubit indexes and at all times, that is, $\text{Tr}[B_n(t) \rho_{\mathrm{B}}]= \langle B_n(t) \rangle _{\mathrm{B}}= 0,$  for all $n,t $. Gaussianity requires that all bath operator cumulants with respect to the averaging operation $\langle \ldots\rangle_{\mathrm{B}}$ vanish for $n=3$ and higher \cite{multiDD,Paz2017}.  Thus, the noise statistical properties are fully characterized by the two-point correlation functions, 
\begin{equation*}
C_{nm}(t_1,t_2)\equiv \langle B_n(t_2) B_m(t_1) \rangle_{\mathrm B} = {\mathrm{Tr_B}}[ B_n(t_2) B_m(t_1) \rho_{\mathrm B}] , \quad 
t_2 \geq t_1 \geq 0. 
\end{equation*}
Finally, stationarity means that the noise statistics, hence the above correlations, are invariant under arbitrary time translations, hence we may let 
\begin{equation}
C_{nm}(t_1,t_2)= C_{nm} (t_2-t_1)\equiv C_{nm}(t)=\langle B_n(t) B_m(0) \rangle_{\mathrm B}.
\label{eqCorr}
\end{equation}
\noindent
Two important limiting cases of ZMGSD noise are worth highlighting: 
{\em spatially uncorrelated} (independent) noise corresponds to $C_{nm}(t) = \delta_{nm} f_n(t)$, for some function $f_n(t)= f_n(-t)^*$, whereas {\em temporally uncorrelated} (Markovian or ``white'') noise corresponds to $C_{nm}(t) = c_{nm} \delta(t)$, for some Hermitian matrix $c_{nm}=c_{mn}^*$ that encodes the spatial noise correlations. 

In the frequency domain, the Fourier transform of $C_{nm}(t)$ yields the {\em noise spectra} (or power spectral densities), $S_{nm}(\omega)$. If we let $S_{nm}(\omega) \equiv \frac{1}{2}[ S_{nm}^+(\omega) + S_{nm}^-(\omega)]$, then 
\begin{eqnarray}
S^{+}_{nm}(\omega) \equiv \int_{-\infty}^{\infty}\mathrm{d}s \,e^{-i\omega s} \,\langle \,\{B_n(s),B_m(0)\} \,\rangle_{\rm B} 
= S_{nm}(\omega) + S_{mn}(-\omega) \label{eqnSP1+},\\
S^{-}_{nm}(\omega) \equiv \int_{-\infty}^{\infty}\mathrm{d}s \,e^{-i\omega s} \,\langle \,[B_n(s),B_m(0)] \,\rangle_{\rm B} 
= S_{nm}(\omega) - S_{mn}(-\omega),
\label{eqnSP1-}
\end{eqnarray}
are often termed the ``classical'' ($+$) and ``quantum'' ($-$) spectra, respectively \cite{ClerkRMP,Paz2017,multiDD}. By definition, all quantum spectra vanish when the noise is classical; quantum spectra corresponding to $n\ne m$ vanish whenever the noise is private in the above sense.  Further to that, non-commutativity of noise operators manifests in different symmetry properties that classical vs. quantum spectra obey \cite{Paz2019}:
\begin{equation}
[S^+_{nm}(\omega)]^*= S^+_{mn}(\omega) =  S^+_{nm}(-\omega), \quad 
[S^-_{nm}(\omega)]^*= S^-_{mn}(\omega) = - S^+_{nm}(-\omega).
\label{sym}
\end{equation}
Thus, $S_{nm}(\omega) \ne S_{nm}(-\omega)$ in the presence of non-classical noise in general.

\subsection{Exact expressions for observable expectation values}
\label{sub:exact1}

As shown in Ref.\,\cite{FelixPRA}, the interplay between quantumness and noise correlations is responsible for qualitatively new uncertainty sources in the noisy sensing dynamics. Since $H_{\rm{SB}}(t)$ in Eq.\,(\ref{eqnSB}) generates pure-dephasing dynamics, we may evaluate $\langle{\sigma_n^{y}(t)}\rangle$ and $\langle{\sigma_{n}^{y}\sigma_m^{y}(t)}\rangle$ by invoking an exact result in terms of generalized cumulants of bath operators established in \cite{Paz2017}. Summing over all qubits and tracing out the bath, this leads to exact closed-form expressions for the time-dependent expectations of observables for arbitrary initial states $\rho_0$, and from there to obtaining the estimation procedure in Eq.\,(\ref{eqnObs}). In particular, if we measure along the $y$ axis, we may write
\begin{eqnarray}
\langle {J_y(t)}\rangle&=\!\sum_{n} 
e^{{-\chi_{nn}(t)/2}}   \,\Tr_{\rm S}\Big[ e^{-i\Phi_n(t)}\rho_0 {\sigma_n^{y}}/2\Big],	
\label{eqnmomentsJy1} \\
\langle{J_y^2(t)}\rangle &=
\frac{N}{4}+\!\!\sum_{n,m\neq n}\!\!\!   e^{-[\chi_{nn}(t)+\chi_{mm}(t)]/2} 
\,\Tr_{\rm S}
\Big[e^{-i \Phi_{nm}(t)}\rho_0 {\sigma_n^{y}\sigma_m^{y}}/4\Big]  , 
\label{eqnmomentsJy2} 
\end{eqnarray}
for ``effective propagators'' $\exp[-i\Phi_n(t)]$, $\exp[-i\Phi_{nm}(t)]$ {\color{rev} which, 
for completeness, we include here \cite{FelixPRA}:}
\begin{eqnarray}
\Phi_n(t)= b t \,\sigma_n^{z}+\!\!\sum_{\ell,\ell\neq n}\!\!\Psi_{n\ell}(t)\sigma_n^{z}\sigma_{\ell}^{z} 
\label{prop1}, \\
\Phi_{nm}(t)\!= b t \,(\sigma_n^{z}+\sigma_m^{z})-i  \chi_{nm}(t) \sigma_n^{z}\sigma_m^{z}+\!\!\!\!\sum_{\ell,\ell\neq nm}
 \!\!\left[\Psi_{n \ell}(t)\sigma_n^{z}\sigma_\ell^{z}+
\Psi_{m \ell}(t)\sigma_m^{z}\sigma_\ell^{z}\right].  
\label{prop2}
\end{eqnarray}
The above propagators depend on two sets of real quantities: the {\em decay parameters}, $\chi_{nm}(t)$, which describe loss of coherence in the $z$ basis, and the {\em phase parameters}, $\Psi_{nm}(t)$, which characterize bath-mediated entanglement and squeezing: 
\begin{eqnarray} 
\chi_{nm}(t) &\equiv & \frac1{2\pi}{\rm{Re}}\int_{0}^{\infty} \!\! d \omega\,F^+(\omega,t)\,S_{nm}^+(\omega), \quad 
\label{eqnChiPsi}
\Psi_{nm}(t) \equiv \frac1{2\pi}{\rm{Im}}\int_{0}^{\infty} \!\! d\omega\,F^-(\omega,t)\,S_{nm}^-(\omega),	
\end{eqnarray}
where $F^+(\omega,t)$ and $F^-(\omega,t)$ are, respectively, first- and second-order ``filter functions'' (FFs) which isolate the time dependence and allow, in general, for the inclusion of arbitrary open-loop dynamical control \cite{PazFF}. In our present setting, specializing to free evolution, we simply have 
\begin{eqnarray}
F^+(\omega, t)= \left |\int_0^t \! ds \, e^{i \omega s} \right |^2   
= \frac{2(1- \cos \omega t)}{\omega^2} , \label{FF1}\\
F^-(\omega,t)= \int_0^t \!ds \int_0 ^{s}\!ds' \,e^{i \omega (s-s')}= \frac{1-e^{i \omega t}+ i \omega t}{\omega^2}.
\label{FF2}
\end{eqnarray} 
For later considerations, it is useful to observe that the first-order FF has a definite parity with respect to both time and frequency, that is, $F^+(\omega,t)= F^+(\omega,-t)= F^+(-\omega,t)$, and so do the real and imaginary parts of the second-order FF, namely, $\text{Re}\, F^-(\omega,t)= \text{Re}\, F^-(\omega,-t)= \text{Re}\, F^-(-\omega,t)$, and $\text{Im}\, F^-(\omega,t)= -\text{Im}\, F^-(\omega,-t)= - \text{Im}\, F^-(-\omega,t)$. 

Crucially, a non-vanishing $\Psi_{nm}(t)$ from Eq.\,(\ref{eqnChiPsi}) implies a non-vanishing asymmetric spectrum, $S^-_{nm}(\omega)\neq 0$, and thus non-classical noise. Note, however, that the converse need {\em not} be true: it is possible to have non-classical temporally uncorrelated noise described by $C_{nm}(t)= c_{nm} \delta(t)$, which yields $\Psi_{nm}(t)=0$ as long as the correlation coefficients $c_{nm}= c_{mn}^{*}$ are \textit{purely real}, as one can check by direct computation.

\subsection{Exact representation of the reduced dynamics}
\label{sec:general}

While the expressions given in Eqs.\,(\ref{eqnmomentsJy1})-(\ref{eqnmomentsJy2}) suffice for evaluating the achievable estimation precision, additional physical insight may be gained by obtaining an exact expression of the reduced density matrix of the qubit probes in the presence of ZMGSD noise. This, in turn, will allow discussion of a number of general properties of the open-system model which may be of independent interest.
\smallskip

{\color{rev} Let $|\vec{\alpha} \rangle \equiv \bigotimes_{n=1}^N |\alpha_n \rangle$, where $|\alpha_n \rangle $ is an eigenstate of $\sigma_z^n$, $|\alpha_n \rangle \in { \{ |\! \uparrow \rangle , |\! \downarrow \rangle \} }$, with eigenvalues $\alpha_n = \pm 1$, define the $z$ basis.} The desired reduced density matrix elements are given by 
$$\langle \vec{\alpha}|\rho(t)|\vec{\beta}\rangle = \left \langle \langle \vec{\alpha}| U(t) \rho_0 U^{\dagger}(t) |\vec{\beta}\rangle \right \rangle_{\mathrm{B}}, \quad U(t)=\mathcal{T}_+ \exp \bigg\{\!\!-\frac{i}{2} \sum_{n=1}^N \sigma_n^z \; \int_0^t \, ds\, [b + B_n(s)]  \bigg\}.$$ 
By introducing the time-ordered bath evolution operators $U_{\vec{\alpha}}(t) \equiv U(t)|\vec{\alpha}\rangle$, we can then write:
$$ \langle \vec{\alpha}|\rho(t)|\vec{\beta}\rangle = \langle \vec{\alpha}|\rho_0|\vec{\beta}\rangle \left \langle U_{\vec{\alpha}}(t) U_{\vec{\beta}}^{\dagger}(t)  \right \rangle_{\mathrm{B}}, 
\quad  U_{\vec{\alpha}}(t) = \mathcal{T}_+ \exp \bigg\{\!\!-\frac{i}{2} \sum_{n=1}^N \alpha_n  \int_0^t \!\! ds\, [b + B_n(s)]  \bigg\} . 
\quad $$  
When the noise is ZMGSD, we can evaluate the trace $\left \langle U_{\vec{\alpha}}(t) U_{\vec{\beta}}^{\dagger}(t)  \right \rangle_{\mathrm{B}}$ exactly. Three main steps are involved. First, we express the time-ordered bath evolution operators inside the average in terms of their Taylor series,
\begin{eqnarray}
\left \langle U_{\vec{\alpha}}(t)\, U_{\vec{\beta}}^{\dagger}(t)  \right \rangle_{\mathrm{B}} &= \bigg \langle \sum_{n=0}^{\infty} \frac{(-i)^n}{n!} \int_0^{\infty}\,ds_1 \ldots \int_0^{t}\,ds_n\,
 \mathcal{T}_-\, H_{\vec{\alpha}}(s_1)\ldots H_{\vec{\alpha}}(s_n) \nonumber\\ 
 &\sum_{m=0}^{\infty} \frac{(i)^m}{m!} \int_0^{t}\,ds'_1\, \ldots \int_0^{\infty}\,ds'_m\,
 \mathcal{T}_+\, H_{\vec{\beta}}(s'_1)\ldots H_{\vec{\beta}}(s'_n) \bigg \rangle_{\mathrm{B}}, 
 \label{morecum}
\end{eqnarray}
with $H_{\vec{\alpha}}(s)\equiv \sum_{n=1}^N \alpha_n \;[b + B_n(s)]  $ and $(\mathcal{T}_-)$ denoting anti-time-ordering. We then write the ensuing moments, $\langle \mathcal{T}_-\, H_{\vec{\alpha}}(s_1)\ldots H_{\vec{\alpha}}(s_n) \mathcal{T}_+\, H_{\vec{\beta}}(s'_1)\ldots H_{\vec{\beta}}(s'_m) \rangle_{\mathrm{B}} $, as a linear combination of the cumulants. Finally, we exploit Gaussianity to truncate the resulting expression exactly to the second order {\color{rev} (see \ref{app:exact} for details)}. The end result is:
\begin{eqnarray}
&\left \langle U_{\vec{\alpha}}(t) U_{\vec{\beta}}^{\dagger}(t)  \right \rangle_B  = \exp \bigg\{ i \,\frac{b}{2} \sum_{n=1}^N (\alpha_n - \beta_n) - \frac{1}{4}\!\!\sum_{n,m =1}^N \bigg[
\beta_n \beta_m \int_0^t ds \int_0^s ds' \langle  B_n(s') B_m(s) \rangle_{\mathrm{B}} \nonumber \\ 
& + \alpha_n \alpha_m\! 
  \int_0^t ds \int_0^s ds' \langle B_n(s) B_m(s') \rangle_{\mathrm{B}} 
- \beta_n \alpha_m \!\int_0^t ds \int_0^t ds' \langle B_n(s) B_m(s') \rangle_{\mathrm{B}} \bigg]
\bigg\}.
\label{Leigh2}
\end{eqnarray}
We stress that Eq.\,(\ref{Leigh2}) is derived solely by leveraging {\em Gaussianity} of the noise operators -- at variance with the approach of \cite{multiDD}, where their bosonic statistics was used to truncate exactly the Magnus expansion of the propagators to the second order and, as mentioned in the Introduction, closer in spirit to the approach used in \cite{Paz2017} for obtaining expectation values of observables. 

After some rearranging of the exponents in Eq.\,(\ref{Leigh2}), we obtain an expression for the ZMGSD reduced density matrix element in terms of a decay function $\gamma(\vec{\alpha},\vec{\beta},t)$  and two phases $\varphi_0(\vec{\alpha},\vec{\beta},t)$ and $\varphi_1(\vec{\alpha},\vec{\beta},t)$: 
\begin{eqnarray}
\label{eq::MatrixEl}
\bra{\vec{\alpha}}\rho(t)\ket{\vec{\beta}}=e^{i b t\sum_{n=1}^N(\beta_n-\alpha_n)}\; 
e^{- \gamma(\vec{\alpha},\vec{\beta},t)+ i \varphi_0(\vec{\alpha},\vec{\beta},t)+ i \varphi_1(\vec{\alpha},\vec{\beta},t)}\bra{\vec{\alpha}}\rho_0\ket{\vec{\beta}}, 
\end{eqnarray}
where 
\begin{eqnarray}
\gamma(\vec{\alpha},\vec{\beta},t)&= \sum_{n,m=1}^N (\alpha_n - \beta_n) (\alpha_m - \beta_m) 
\kappa_{nm}(t), 
\label{gamma}\\
\varphi_0(\vec{\alpha},\vec{\beta},t)&=\sum_{n,m=1}^N (\beta_n \beta_m- \alpha_n \alpha_m) \xi_{nm}(t),
\label{phi0}\\
\varphi_1(\vec{\alpha},\vec{\beta},t)&=\sum_{n,m=1}^N (\beta_n \alpha_m-\alpha_n \beta_m ) \vartheta_{nm}(t). 
\label{phi1}
\end{eqnarray}
From Eqs.\,(\ref{gamma})-(\ref{phi1}), we see that the relevant dynamical quantities consist of a state-dependent contribution, which is a function of the set of $\sigma_n^z$ eigenvalues $\{\alpha_n, \beta_n, \;1\leq n \leq N \}$ corresponding to the specific matrix element $\langle \vec{\alpha}| \rho(t)| \vec{\beta}\rangle$, multiplied by time-dependent {\em dynamic coefficients}, $\kappa_{nm}(t), \xi_{nm}(t), \vartheta_{nm}(t)$, given by:
\begin{eqnarray}
\kappa_{nm}(t) = \frac{1}{16} \int_0^t\,ds\, \int_0^t \,ds'\, \langle \{B_n(s),B_m(s')\} \rangle_{\mathrm{B}}, \label{chi}\\
i\, \xi_{nm}(t)=\frac{1}{16} \int_0^t\,ds\, \int_0^s \,ds'\, \left(\, \langle [B_n(s),B_m(s')] \rangle_{\mathrm{B}} + \langle [B_m(s),B_n(s')] \rangle_{\mathrm{B}} \,\right), \label{xi}\\
i\, \vartheta_{nm}(t)= \frac{1}{16} \int_0^t\,ds\, \int_0^t \,ds'\, \langle [B_n(s),B_m(s')] \rangle_{\mathrm{B}} .
\label{vartheta}
\end{eqnarray}
Physically, the decay factor $\gamma(\vec{\alpha},\vec{\beta},t)$ is responsible for the decoherence of the off-diagonal terms. The phase $\varphi_0(\vec{\alpha},\vec{\beta},t)$ is related to bath-mediated interaction between qubits, whereas a non-vanishing $\varphi_1(\vec{\alpha},\vec{\beta},t)$ reflects the non-commutativity of operators $U_{\vec{\alpha}}(t)$ and $U_{\vec{\beta}}^{\dagger}(t)$. Again, while the above representation is similar to the one established in \cite{multiDD} for the case of a bosonic bath, no explicit use of the bath algebraic properties has been made here -- making this procedure applicable to arbitrary (stationary) Gaussian dephasing settings. Moving forward, when referring to the decay and phases of an arbitrary matrix element, we will omit $\vec{\alpha},\vec{\beta}$ in the arguments and simply write $\gamma(t), \varphi_0(t), \varphi_1(t)$.

Since the bath correlation $\langle B_n(\tau)B_m(0) \rangle_{\mathrm{B}} $ has units of frequency squared, the dynamic coefficients are, consistently, dimensionless. It is easy to show that they enjoy the following properties:
\begin{itemize}
\item Reality: $\kappa_{nm}(t)$, $\xi_{nm}(t), \vartheta_{nm}(t) \in \mathbb{R}$, for all $t.$ 
\item Parity under index permutation: $\kappa_{nm}(t)= \kappa_{mn}(t),\, \xi_{nm}(t)=\xi_{mn}(t),\, \vartheta_{nm}(t)=-\vartheta_{mn}(t)$.
\item Parity under time inversion: $\kappa_{nm}(t)= \kappa_{nm}(-t),\, \xi_{nm}(t)=-\xi_{nm}(-t),\, \vartheta_{nm}(t)=\vartheta_{nm}(-t)$.
\item Frequency representation: 
\begin{eqnarray}
\kappa_{nm}(t)= \frac{1}{32 \pi} \int_{- \infty}^{\infty} \! d\omega \, F^+(\omega,t) \,S_{nm}^+(\omega), 
\label{freqchi}\\
i \xi_{nm}(t)= \frac{1}{32\; \pi} \int_{- \infty}^{\infty} \! d\omega \, F^-(\omega,t)\, \left(S_{nm}^-(\omega) + S_{mn}^-(\omega) \right),
\label{freqxi}\\
i \vartheta_{nm}(t)= \frac{1}{32\; \pi} \int_{- \infty}^{\infty} \, d\omega \, F^+(\omega,t) \,S_{nm}^-(\omega).
\label{freqtheta}
\end{eqnarray}
 
\item Short-time behavior: We can leverage the short-time behavior of the first- and second-order FFs to derive the short-time scaling of the dynamic coefficients. Assume, as usual, that the spectra have vanishing support in the high-frequency limit, say, 
$S^{\pm}_{nm}(\omega) \approx 0,$ for $ |\omega| \gtrsim \omega_c$, with $\omega_c$ being a cutoff frequency.  By Taylor-expanding the expressions in Eqs.\,(\ref{FF1})-(\ref{FF2}) for $\omega_c t \ll 1$, we arrive at:
\begin{eqnarray}
\kappa_{nm}(t) \approx \frac{t^2}{32} \, \langle\{B_n(0),B_m(0)\} \rangle_{\mathrm{B}},  
\label{chist} \\
\xi_{nm}(t) \approx \frac{1}{6}\,\frac{t^3}{32}   \,\left( \langle [\dot{B}_n(0), B_m(0)]\rangle_{\mathrm{B}} + \langle [\dot{B}_m(0), B_n(0)]\rangle_{\mathrm{B}} \right),  
\label{xist} \\
\vartheta_{nm}(t) \approx \frac{t^2}{32} \, \langle [B_n(0), B_m(0)] \rangle_{\mathrm{B}} , 
\label{thetast}
\end{eqnarray}
where $\dot{B}_n(0) \equiv \frac{d}{d t} B_n(t)\big|_{t=0}$ denotes the time derivative.
\end{itemize}

\smallskip

By comparing the expressions of the decay and phase parameters entering the effective propagators for the variance of $J_y$, Eq.\,(\ref{eqnChiPsi}), with the expressions in Eqs.\,(\ref{freqchi})-(\ref{freqtheta}) of the more general dynamic coefficients appearing in the reduced dynamics, it is clear that we can relate them as follows:
\begin{eqnarray}
\chi_{nm}(t)=4\,\kappa_{nm}(t), \quad
\Psi_{nm}(t)= 4 \left[ \xi_{nm}(t) + \vartheta_{nm}(t) \right].
\label{rel}
\end{eqnarray}   
A number of observations are possible: 

\smallskip

{\bf (i)} If the noise is {\em collective}, whereby $B_n(t) \equiv B(t)$ for all $t$, one can easily show that $\vartheta_{nm}(t)\equiv 0$ and hence $\varphi_1(t)\equiv 0$ as well. In other words, in this scenario the bath operators $U_{\vec{\alpha}}(t)$ commute for all $|\vec{\alpha}\rangle$. This, in turn, allows us to describe the evolution classically, as we discuss below. From a metrological standpoint, a striking aspect of the collective noise regime is that, as we remarked in \cite{FelixPRA},  the quantum noise enters the reduced dynamics \emph{unitarily}, in the form of a squeezing operator: $\rho(t)=e^{-i\xi(t)J_z^2}[\,\rho(t)|_{\xi=0}\,] e^{i\xi(t)J_z^2},$ where $\rho(t)|_{\xi=0}$ is obtained from Eq.\,(\ref{eq::MatrixEl}), with $\xi_{nm}(t)=\vartheta_{nm}(t)=0$ and $\kappa_{nm}(t)= \kappa(t)\; \forall n,m$. Since the QFI is invariant under unitary transformations that do not depend on $b$~\cite{CarlDistance}, it follows that the ultimate achievable precision, $\Delta \hat{b}_{\mathrm{QCR}}$, is {\em unaffected by quantum noise}. This implies that there always exists an {\em optimal} measurement on the probe that can completely cancel the effect of $\xi(t)$ in principle, and produce an uncertainty $\Delta \hat{b}_\text{min}=\Delta \hat{b}_\text{min}|_{\xi=0}$ equivalent to the classical case.

\smallskip
{\bf (ii)} If the noise is {\em temporally uncorrelated}, with $C_{nm}(t) = c_{nm}(t) \delta(t)$ and $c_{nm}(t)=c_{mn}^*(t)$, we have $\xi_{nm}(t)\equiv 0$ hence $\varphi_0(t)\equiv 0$. However, $\vartheta_{nm}(t)$ can still be non-zero in principle, provided that the correlation matrix has {\em strictly complex} support, $\text{Im}\, c_{nm}(t) \neq 0$. If the coefficients $c_{nm}$ are purely real, then $\vartheta_{nm}(t)=0$ and $\Psi_{nm}(t)=0$, as noted in Sec.\,\ref{sub:exact1}.

\smallskip
{\bf (iii)} In order for the non-classical phase parameters to contribute to the reduced dynamics, $\Psi_{nm}(t)\ne 0$, either $\xi_{nm}(t)$ or $\vartheta_{nm}(t)$ (or both) must be non-vanishing for some pair of indexes $n,m$ and some time $t$.

\section{Case study: Linear spin-boson dephasing}
\label{sec:NQSB}

\subsection{Model Hamiltonian}

To consider a concrete, paradigmatic application of the above ZMGSD formalism, we will focus on a spin-boson dephasing setting, where the environment Hamiltonian is a collection of bosonic modes (say, phonons) linearly coupled to the probes. That is, 
the system-bath Hamiltonian is given in Eq.\,(\ref{eqnSB}), with
\begin{equation}
H_{\rm B}= \sum_{\vec{k}} \Omega_{\vec{k}} b^\dag_{\vec{k}} b_{\vec{k}}, \;\,\Omega_{\vec{k}} > 0, \quad 
B_n(t) = \sum_{\vec{k}} \Big( g_{\vec{k}} e^{i \vec{k}\cdot \vec{r}_n }\;e^{i \Omega_{\vec{k}} t } \; b^\dag_{\vec{k}} + g_{\vec{k}}^* e^{-i \vec{k}\cdot \vec{r}_n}\;e^{i \Omega_{\vec{k}} t } \; b_{\vec{k}} \Big).
\label{eqnSBH}
\end{equation}
Here, $\vec{k}$ is a wave-vector of modulus $|\vec{k}| \equiv k$ labeling frequency $\Omega_{\vec{k}}$, $\vec{r}_n$ is the position of the $n$th qubit probe, and the  creation (annihilation) operators obey the bosonic algebra: $[b^\dag_{\vec{k}}, b^\dag_{\vec{k'}}]=0$, $[b_{\vec{k}}, b_{\vec{k'}}]=0$, $[b^\dag_{\vec{k}}, b_{\vec{k'}}]= \delta_{\vec{k}, \vec{k'}}$. Note that $\vec{k}$ and $\vec{r}_n$ can be one-, two- or three-dimensional vectors, depending on the spatial dimensionality of the system. The interaction of the qubits with the bath is governed by the position-dependent couplings $g_{\vec{k}}^n =  g_{\vec{k}} e^{i \vec{k}\cdot \vec{r}_n }$, where $g_{\vec{k}} \in {\mathbb C}$ has units of frequency. In what follows, we shall further assume that the couplings are \emph{isotropic}, with $g_{\vec{k}}= g_{k}$, and that the temperature is low enough so that the relevant modes are acoustic phonons, obeying a {\em linear} dispersion relationship of the form $\Omega_{\vec{k}} \equiv v k= \Omega_{k}$, where $v$ is the speed of sound in the medium \cite{IschiPRB}. The collective limit is recovered if we let all qubits be at the same position: $\vec{r}_n= \vec{r}_0, \forall n$. Conversely, the model is non-collective as long as there are two qubits $n$ and $m$ with $\vec{r}_n \neq \vec{r}_m$.

In practice, the above discrete set of modes is replaced by a continuum of closely spaced modes, with a {\em spectral density} function given by
$$ J(\omega)= \sum_{k}|g_{k}|^2 [\delta(\omega - \Omega_{k})+\delta(\omega + \Omega_{k})]=J(-\omega) , $$
which we take to have the form $J(\omega) \equiv \alpha \omega_c (\omega/\omega_c)^s K(\omega,\omega_c)$. Here, $\alpha >0$ is a  dimensionless strength constant, the low-frequency behavior is governed by $\omega^s$, where $s>0$ is the \emph{Ohmicity parameter}, and the high-frequency contributions decay according to the {\em cutoff function} $K(\omega, \omega_c)$. In this work we will consider a supra-Ohmic regime, $s>1$, with a Gaussian cutoff function, $K(\omega, \omega_c)= e^{-\omega^2/\omega_c^2}$ or a ``softer'' exponential cutoff, $K(\omega, \omega_c)= e^{-\omega/\omega_c}$, both of which arise in solid-state phonon environments \cite{Irene,IschiPRB}.

\subsection{Spin-boson reduced dynamics}
\label{sbred}

Assuming that the environment is initially in thermal equilibrium at inverse temperature $\beta$, namely, $\rho_B= \exp(-\beta H_B)/\text{Tr}[\exp(-\beta H_B)]$, the above $B_n(t)$ generate ZMGSD noise. Therefore, the reduced density matrix elements can be expressed in terms of Eqs.\,(\ref{eq::MatrixEl})-(\ref{phi1}), and we can further specialize the general expressions in Eqs.\,(\ref{chi})-(\ref{vartheta}) to this setting. Using the fact that $\langle b_{\vec{k}}^\dag b_{\vec{k}} \rangle_{\mathrm{B}}= (\coth \left(\beta \Omega_{\vec{k}}/2 \right)-1)/2=\langle b_{\vec{k}} b_{\vec{k}}^\dag \rangle_{\mathrm{B}} -1$ in thermal equilibrium, we have:
\begin{eqnarray*}
& \langle\, \{B_{n}(s),B_m(s')\}\, \rangle_{\mathrm{B}}\!= \!\sum_{\vec{k}} |g_{k}|^2 \!\coth \!\left(\beta \Omega_{k}/2 \right) \!\left( e^{i \Omega_{k} (s-s')}e^{i \vec{k}\cdot \vec{r}_{nm} } 
\!+ e^{-i \Omega_{k} (s-s')}e^{-i \vec{k}\cdot \vec{r}_{nm} } \right), 
\label{anticomm}\\
&\langle\, [B_{n}(s),B_m(s')  ]\,\rangle_{\mathrm{B}} \!= \!\sum_{\vec{k}} |g_{k}|^2 \! \left( -e^{i \Omega_{k} (s-s')}e^{i \vec{k}\cdot 
\vec{r}_{nm} } \!+ e^{-i \Omega_{k} (s-s')}e^{-i \vec{k}\cdot \vec{r}_{nm}  } \right) ,
\label{comm}
\end{eqnarray*}
where $\vec{r}_{nm} \equiv (\vec{r}_n-\vec{r}_m)$ is the distance between qubits $n$ and $m$.
The classical and quantum spectra can then be readily obtained by substituting into Eqs.\,(\ref{eqnSP1+})-(\ref{eqnSP1-}):
\begin{eqnarray}
S_{nm}^+(\omega)=2 \pi \sum_{\vec{k}} |g_{k}|^2 \coth \left(\beta \Omega_{k}/2 \right) \left( \delta(\omega-\Omega_{k}) e^{i \vec{k}\cdot \vec{r}_{nm} } + \delta(\omega+\Omega_{k}) e^{-i \vec{k}\cdot \vec{r}_{nm} } \right), \label{Splus}\\
S^-_{nm}(\omega)=2 \pi \sum_{\vec{k}} |g_{k}|^2  \left(- \delta(\omega-\Omega_{k}) e^{i \vec{k}\cdot \vec{r}_{nm} } + \delta(\omega+\Omega_{k}) e^{-i \vec{k}\cdot \vec{r}_{nm} } \right).
  \label{Sminus}
\end{eqnarray}
In this way, all the quantities in Eqs.\,(\ref{Splus})-(\ref{Sminus}) depend exclusively on $k$, except for the  $e^{\pm i \vec{k} \cdot \vec{r}_{nm}}$ factors. It is then convenient to split the sum over $\vec{k}$ modes by grouping together all wave-vectors with the same modulus: $\sum_{\vec{k}}= \sum_{k}\; \sum_{\vec{k}: |\vec{k}|=k}$.  In the continuum limit, we can then account for the angular dependence of the $\vec{k} \cdot \vec{r}_{nm}$ inner product by averaging over all directions of modes $\vec{k}$ with the same modulus $k$. That is, 
\begin{eqnarray*}
\sum_{\vec{k}: |\vec{k}|=k} e^{ \pm i \vec{k}\cdot \vec{r}_{nm}} \mapsto \int \; d \widetilde{\Omega}\, 
e^{ \pm i k\; r_{nm} \cos \theta}\equiv  f_D(k\, r_{nm}) = f_D(\Omega_{k} t_{nm}) ,\\
\qquad  f_D(\Omega_{k} t_{nm})=\left\{ \begin{array}{lr}
       2 \cos(\Omega_{k}  t_{nm}), & D=1, \\
        2 \sin(\Omega_{k}  t_{nm})/(\Omega_{k}  t_{nm}) ,& D=2,\\
        4\pi \sin(\Omega_{k}  t_{nm})/(\Omega_{k}  t_{nm}), &  D=3 ,
        \end{array} \right. 
\end{eqnarray*}
where $\widetilde{\Omega}$ the $D$-dimensional solid angle and we have defined the position-dependent \emph{transit times} as $t_{nm}\equiv r_{nm}/v \geq 0$.  We can then rewrite the above classical and quantum spectra as:
\begin{eqnarray*}
S^+_{nm}(\omega)&=2\pi J(\omega) f_D(\omega t_{nm})\coth(\beta |\omega|/2), 
\qquad S^-_{nm}(\omega)&= 2\pi J(\omega) f_D(\omega t_{nm}) \,\text{sgn}(\omega) . 
\end{eqnarray*}

Replacing in Eqs.\,(\ref{freqchi})-(\ref{freqtheta}) and exploiting the parity properties of quantum and classical spectra, Eq.\,(\ref{sym}), the spin-boson dynamic coefficients can  be written as:
\begin{eqnarray}
 \kappa_{nm}(t)=\frac{1}{4} \int_0^{\infty} d\omega\, J(\omega)\; \frac{1- \cos\left( \omega t \right)}{\omega^2} f_D \left( \omega t_{nm} \right) \coth(\beta \omega/2), 
 \label{spchi}\\
  \xi_{nm}(t) =\frac{1}{4} \int_0^{\infty} d\omega\, J(\omega)\;  \frac{\omega t- \sin\left( \omega t \right)}{\omega^2} f_D \left( \omega t_{nm} \right),
  \label{spxi}\\
  \vartheta_{nm}(t)=0 . 
  \label{spvartheta}
\end{eqnarray}  
We remark that the dependence upon the transit time in the above expressions differs (even in $D=1$) from the one that is obtained by {\em neglecting the angular dependence} of the $\vec{k} \cdot \vec{r}_{nm}$ inner product. While the latter is unimportant in the collective limit (where $t_{nm}\equiv 0$) and is indeed often neglected in simplified treatments in the literature \cite{Palma,multiDD}, it plays an important role in the more general non-collective settings we aim to study. In particular, in conjunction with the isotropy of the coupling constants, $g_{\vec{k}}= g_{k}$, it enforces the fact that $\vartheta_{nm}(t)\equiv 0$. Consequently, $\varphi_1(t) \equiv 0,$ for all $|\vec{\alpha}\rangle, |\vec{\beta}\rangle$, which in turn makes {\color{rev} the dephasing spin-boson dynamics {\em random unitary}, as we discuss in \ref{sub:RU}.}

The above frequency integrals can be analytically evaluated by considering a low-temperature regime where $\coth(\beta \omega/2) \approx 1$. In the short-time limit $\omega_c t \ll 1$, the presence of $K(\omega,\omega_c)$ allows us to write Eqs.\,(\ref{chist})-(\ref{thetast}) in term of a dimensionless parameter $\omega_c t$ and factors that depend on {\em dimensionless transit times} $x_{nm}\equiv\omega_c t_{nm}$, proportional to the qubit separation $r_{nm}$. We obtain 
\begin{eqnarray*}
\kappa_{nm}(t) \approx \kappa^2 (x_{nm}) (\omega_c t)^2, \qquad \xi_{nm}(t) \approx \xi^3(x_{nm}) (\omega_c t)^3 .
\label{sbchist}  
\end{eqnarray*} 
For an exponential cutoff function as used in \cite{FelixPRA}, for instance, direct calculation yields
\begin{eqnarray}
\kappa^2(x_{nm})=\alpha \,\frac{\Gamma(s+1)}{4}\, \frac{\cos[(s+1) \arctan (x_{nm})]}{(1+x_{nm}^2)^{(s+1)/2}},  \\  
\xi^3(x_{nm})=\alpha \,\frac{\Gamma(s+2)}{24} \frac{\cos[(s+2) \arctan(x_{nm})]}{(1+x_{nm}^2)^{s/2+1}} ,
\label{stconst}
\end{eqnarray}
where $\Gamma (x)$ is the Euler's Gamma function. Note that for, the spin-boson Hamiltonian, both the bath correlations and the dynamic coefficients \textit{depend on the qubits positions} through the position-dependent transit time, even if not explicitly stated in the argument of said functions.

\subsection{Ramsey metrology under collective spin-boson dephasing revisited} 

In Ref.\,\cite{FelixPRA}, the collective noise limit of the noisy dynamics generated by Eq.\,(\ref{eqnSBH}) was investigated. In this regime, the decay and phase coefficients are qubit-independent: 
$$  \chi_{nm}(t) = \chi(t)=\,4\,\kappa(t),\qquad  \Psi_{nm}(t) = \Psi(t)=\,4\,\xi(t), \;\forall n,m, \forall t.$$ 
A non-vanishing phase parameter $\Psi(t)\neq 0$ is then indicative of dephasing noise that is non-classical and both spatially and temporally correlated. 

The relevant frequency uncertainty $\Delta \hat{b}(t)$ can be estimated from Eq.\,(\ref{eqnObs}). Quantitative results have been obtained by contrasting the behavior of CSS vs. OATS uncertainties for $N$ qubits through a Ramsey protocol of {\em fixed total time} $T \equiv \nu \tau$. Below, we recall the salient points, by also highlighting two new results: A discussion of the physical nature of the CSS periodic uncertainty minima, and a proof that under the current measurement scheme, the squeezing and rotation angles  of Eq.\,(\ref{ideal}), minimizing the uncertainty in a noiseless setting are still optimal at short times in the presence of the dephasing noise here considered.

\subsubsection{CSS and uncertainty minima.}
\begin{figure}[t!]
    \centering
   \includegraphics[width=7.5cm]{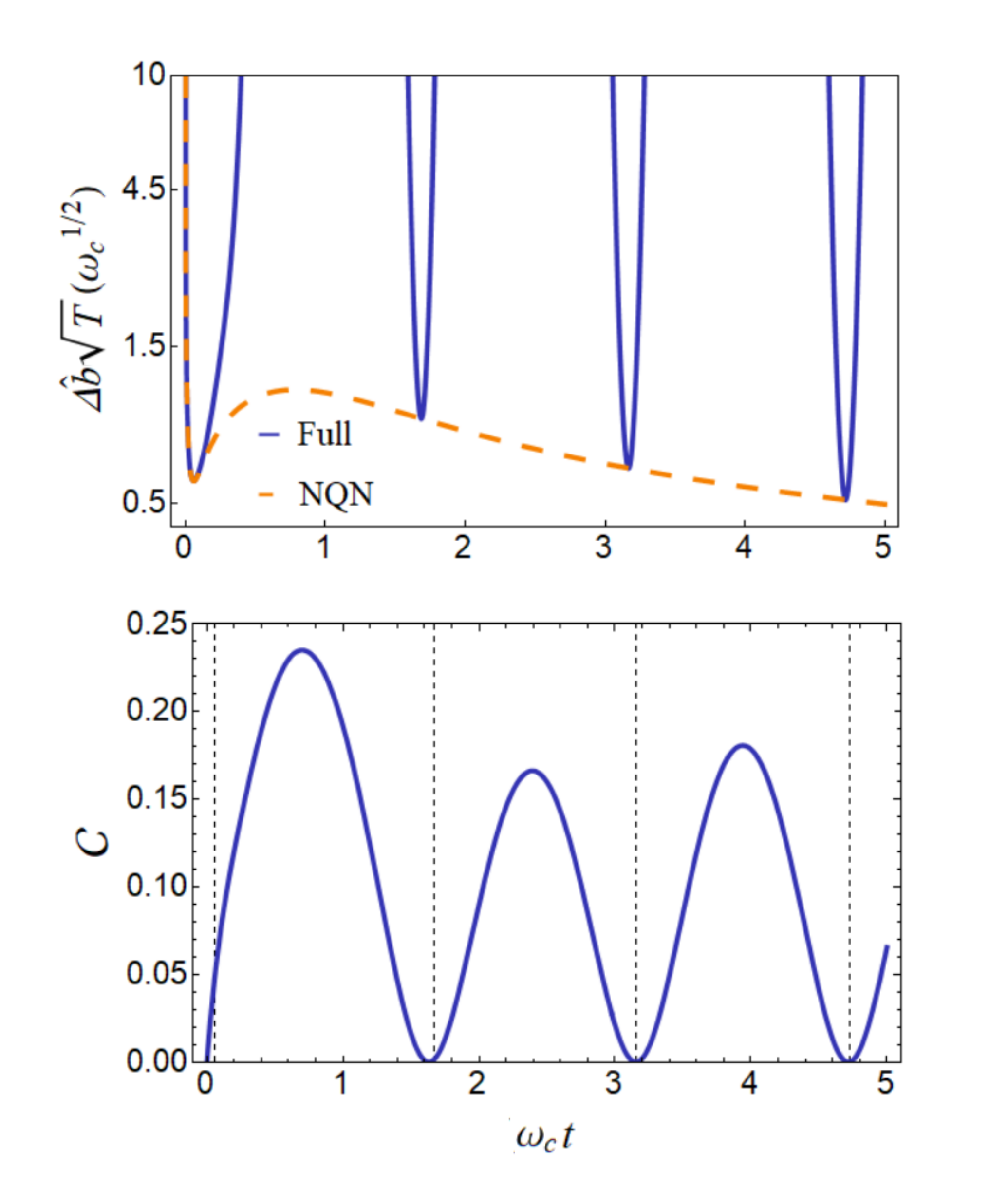}
   \vspace*{-4mm}
    \caption{{\bf Initial 	CSS with $\bm{N=100}$ qubits under collective spin-boson dephasing.} Top: Estimation precision $\Delta \hat{b} \sqrt{T}$ (in 	units of $\omega_c^{1/2}$) as a function of dimensionless time (compare Fig.\,1(b) in Ref.\,\cite{FelixPRA}). The dashed (orange) line corresponds to $\Psi(t)\equiv 0$. Bottom: Time-dependence of the two-qubit concurrence ${\mathcal C}(t)$. The vertical grid lines mark the instants at which $\Delta \hat{b}(t)$ has a minimum. An exponential cutoff 
     is assumed for the noise spectral 	density, and $s=3,\alpha=1$.}
    \label{conc}
\end{figure}

For an initial CSS, the estimation uncertainty under collective noise was found to obey the following exact expression \cite{FelixPRA}: 
\begin{equation}
\Delta \hat{b}(t)^2_{\text{coll}}= \frac{(N+1)e^{\chi(t)}-(N-1)e^{-\chi(t)}\cos(2\Psi(t))^{N-2}}{2 N T t \cos(\Psi(t))^{2N-2}} .
\label{dbcCSS}
\end{equation}
Due to quantum noise, in particular the $\cos(\Psi(t))^{2N-2}$ term in the denominator, the uncertainty grows \emph{exponentially} with qubit number, away from the optimal measurement time $\tau_{\text{opt}}$. In the short-time limit, we can approximate $\chi (t) \approx \chi_0^2\,(\omega_c t)^2$ and $\Psi(t) \approx \Psi_0^3 \,(\omega_c t)^3$, with $\chi_0^2 \equiv \alpha\, \Gamma(1+s)$ and $\Psi_0^3 \equiv \alpha \,\Gamma(s+2)/6$ being numeric prefactors for a spectral density with exponential cutoff. Optimizing $\Delta \hat{b} (t)_{\text{coll}}$ for $N \gg 1$, we obtain  
\begin{equation}
 \omega_c \tau_{\text{opt,coll}}^{\text{CSS}} =\chi_0^{-1}\; N^{-1/2}, \quad 
\Delta \hat{b}_{\text{opt,coll}}^{\text{CSS}}\approx  \left( 2 \, \chi_0 \right)^{1/2}\; (\omega_c  /T)^{1/2} N^{-1/4} .
\label{felix-css}
\end{equation}
In Fig.\,\ref{conc}(a), we plot $\Delta \hat{b}(t)_{\text{coll}}$, along with the {\em no-quantum noise} (NQN) limit, $\Delta \hat{b}_0(t)$, which is obtained from Eq.\,(\ref{dbcCSS}) by letting $\Psi(t)\equiv0$. Note that we have $\Delta \hat{b}_0(t) \leq  \Delta \hat{b}(t)$ at all times, so in the collective case the presence of quantum noise can provably only be detrimental for frequency estimation. 

As we can see from the top plot, $\Delta\hat{b}(t)_{\text{coll}}$ exhibits periodic dips; then proceeds to grow unboundedly. The first dip is the result of the short-time dynamics behavior and thus qualitatively different from the others, that simply arise when the $\cos(\Psi(t))^{2N-2}$ in the denominator of Eq.\,(\ref{dbcCSS}) approaches $1$. 

Physical insight into the meaning of these dips may be gained by evaluating the concurrence, $\mathcal{C}$, between (any) two qubits, by tracing out the remaining $N-2$ ones from the reduced density matrix. In Fig.\,\ref{conc}(b), the concurrence is plotted as a function of time. We see that $\mathcal{C}(t)$ becomes zero periodically, at times that approximate increasingly better the minima of the dips as the evolution proceeds. The fact that the minima do not occur at precisely the same times is attributable to the decay coefficient $\chi(t)$, which affects $\mathcal{C}(t)$ and $\Delta \hat{b}^2(t)_{\text{coll}}$ differently. In the regime $ \omega_c t \ll1$, $\chi(t)$ becomes a constant for $s>1$, and thus, asymptotically, $\mathcal{C}=0$ precisely at the uncertainty minima. This means that while the qubits are still non-trivially entangled with the bath, as reflected by the fact that the $N$-qubit state is no longer pure, any two-qubit reduced density matrix becomes separable, and may be written as a mixture of unentangled states at the minima. Accordingly, the precision attains its best possible value when the bath-mediated entanglement among the probes, associated to $\Psi(t)$, is itself at a minimum.

\subsubsection{OATS and optimal squeezing-rotation angles.} 
\label{sub:optimal}
The second initial state analyzed in \cite{FelixPRA} was an OATS, $\rho_{\rm{OATS}}(\beta, \theta)$, as defined in Eq.\,(\ref{oat}), with squeezing and rotation angles chosen to minimize the initial variance along $y$. Let us now briefly describe how these angles are obtained. Initially, regardless of the presence of noise, the variance 
 along the $y$ axis can be written as
\begin{equation}
\Delta J_y^2 = \frac{N}{4}\bigg\{  \bigg[1+ \frac{1}{4} \left( N-1\right) A  \bigg]+ \frac{1}{4} \left( N-1\right) \sqrt{A^2 + B^2} \cos(2 \beta + 2 \delta) \bigg\},
\label{deltaJy}
\end{equation}
where 
\begin{eqnarray}
A\equiv 1-\cos(\theta)^{N-2},\quad B\equiv 4 \sin(\theta/2) \cos(\theta/2)^{N-2}, \quad \delta\equiv \frac{1}{2}\arctan(B/A) .\label{ABdelta}
\end{eqnarray}
 The uncertainty along the $y$ axis is minimized exactly with respect to $\beta$ for $\beta_{\text{opt}}=\pi/2 - \delta $. Note that this depends on the squeezing angle $\theta$ through $\delta$. Following Kitagawa and Ueda \cite{Kita1993}, we can then perform a perturbative expansion in the large $N$, $|\theta| \ll 1 $ regime to prove that $\Delta J_y$ is minimal for $\theta_{\text{opt}} \approx 12^{1/6}\, 2^{2/3}\, N^{-2/3}$. Interestingly, we find a \emph{small constant factor difference} with respect to the optimal squeezing angle reported in  
\cite{Kita1993} and used in \cite{FelixPRA}, ${\theta}^{\text{KU}}_{\text{opt}} \approx 24^{1/6} 2^{2/3}\, N^{-2/3}$. Finally, we can expand the analytic formula for $\beta_{\text{opt}}$, leading to the following optimal values: 
\begin{equation}
\theta_{\text{opt}} \approx 12^{1/6}\, 2^{2/3}\, N^{-2/3}, \quad \beta_{\text{opt}} \approx \pi/2 - 3^{-1/6} N^{-1/3}- 3^{1/6} 2^{-1} \;N^{-2/3},  \quad N\gg1. 
\label{ideal}
\end{equation}
{\color{rev}
As we explicitly show in \ref{app:OATS}, $\theta_{\text{opt}}, \beta_{\text{opt}}$ give the best possible short-time scaling of the uncertainty in the presence of collective dephasing and, in fact, also even-odd dephasing.} 

Due to the fact that the OATS initial state is entangled, an exact evaluation of the expressions in Eqs.\,(\ref{eqnmomentsJy1})-(\ref{eqnmomentsJy2}) is lacking. To compute the mean values, we perform a cumulant expansion {\em over the qubit operators}, as opposed to the bath {\color{rev} (see also \ref{noncolcum} for details, where the derivations are further carried {\em without} assuming the collective limit).} Second-order truncation is justified when the qubit state is nearly Gaussian in the $N \gg1$ regime, and yields analytic results. In the short-time limit, we find
\begin{eqnarray}
\omega_c \tau_{\text{opt,coll}}^{\text{OAT}} &=3^{1/3} 2^{-1/2}\;\chi_0^{-1}\; N^{-5/6} , \nonumber \\ 
\Delta \hat{b}_{\text{opt,coll}}^{\text{OAT}} & \approx 3^{1/6} 2^{1/4}\;\chi_0^{1/2}\; ( \omega_c/ T)^{1/2} N^{-5/12}  ,
\label{felix-oat}
\end{eqnarray}
which is a sub-SQL performance. This is to be contrasted with the noiseless scenario, in which $\theta_{\text{opt}}, \beta_{\text{opt}}$ are also optimal and yield a superclassical $N^{-5/6}$ scaling \cite{Monika}. Bath-induced anti-squeezing, which manifests itself by means of a non-zero phase coefficient $\Psi(t)$, annihilates all the quantum advantage. 

\section{Non-collective spin-boson dephasing: Even-odd Ramsey metrology}
\label{sec:eo}

\subsection{Spin-boson reduced dynamics under even-odd couplings}

The first step toward analyzing Ramsey frequency estimation beyond the collective noise setting is to particularize the even-odd model introduced in the ZMGSD framework within the spin-boson context. The evolution of matrix element $\langle \vec{\alpha}| \rho(t)| \vec{\beta} \rangle=\langle \vec{\alpha}| \rho_0| \vec{\beta} \rangle 
e^{-\gamma(t)} e^{i \varphi_0(t)} $ for $N=2J$ even-odd ZMGSD noise is characterized by:
  \begin{eqnarray}
    \gamma  (t)^{\text{eo}}=\Big\{ \kappa_{\text{ee}}(t) \left(m_{\text{e}}-m'_{\text{e}}\right){}^2 &+\kappa_{\text{oo}}(t)\left(m_{\text{o}}-m'_{\text{o}}\right){}^2 
    +2 \kappa_{\text{eo}}(t) \left(m_{\text{e}}-m_{\text{o}}\right) \left(m'_{\text{e}}-m'_{\text{o}}\right)\Big\},\nonumber \\
   \varphi _0(t)^{\text{eo}}= \Big\{ \xi_{\text{ee}}(t)    (m_{\text{e}}^2      -m_{\text{e}}^{'2})&+\xi_{\text{oo}}(t) (m_{\text{o}}^{'2}       - m_{\text{o}}^{'2})   
    +2 \xi_{\text{eo}}(t) \left(m_{\text{e}} m_{\text{o}}-m'_{\text{e}} m'_{\text{o}}\right) \Big\},
     \label{EOcoefs0}
\end{eqnarray}
where 
$$2 m_{{\text{e}}}= \sum_{n=1}^J \alpha_n, \quad 
2m'_{\text{e}} =\sum_{n=1}^J \beta_n, \quad
2m_{\text{o}} =\sum_{n= J+1}^{2J} \alpha_n, \quad 
2m'_{\text{o}}= \sum_{n=J+1}^{2J} \beta_n,$$ 
are the $z$ angular momentum component of $|\vec{\alpha}\rangle, |\vec{\beta}\rangle$ for even (e) qubits (labeled $n=1,\ldots, J$) and odd (o) qubits (labeled $n=J+1,\ldots, 2J$), respectively.

From the general consideration of Sec.\,\ref{sbred}, we have the simplification $\vartheta_{\text{eo}}(t)=0$ under the assumptions we make. The dynamic coefficients corresponding to qubits in the same (s) cluster, which can be obtained from Eqs.\,(\ref{spchi})-(\ref{spxi}) with $t_{nm}=0$, are equal, and do not differ from their collective counterparts:  $\kappa_{\text{ee}}(t)= \kappa_{\text{oo}}(t) \equiv \kappa_s(t)$, $\xi_{\text{ee}}(t) = \xi_{\text{oo}}(t) \equiv \xi_s(t)$. On the other hand, the coefficients corresponding to qubits in different (d) clusters may be obtained by using again Eqs.\,(\ref{spchi})-(\ref{spxi}), now with a single non-vanishing transit time $t_{\text{eo}}= r_{\text{eo}}/v$, proportional to cluster the distance $r_{\text{eo}}$, that is, $\kappa_{\text{eo}}(t) \equiv \kappa_d(t)$, $\xi_{\text{eo}}(t) \equiv \xi_d(t)$. Let us focus on the one-dimensional case. We can then write:
\begin{eqnarray}
\kappa_s (t)&= \frac{1}{4} \int_0^{\infty} \!J(\omega) \frac{1-\cos(\omega t)}{\omega^2} \;d\omega,\quad
\kappa_d(t)= \frac{1}{4} \int_0^{\infty} \!J(\omega) \frac{1-\cos(\omega t)}{\omega^2} \cos(\omega t_{\text{eo}}) \;d\omega ,
\label{chi0eo}\\
\xi_s(t)&= \frac{1}{4} \int_0^{\infty} \!J(\omega) \frac{\omega t-\sin(\omega t)}{\omega^2} \;d\omega,\quad 
\xi_d(t)= \frac{1}{4} \int_0^{\infty} \!J(\omega) \frac{\omega t-\sin(\omega t)}{\omega^2} \cos(\omega t_{\text{eo}}) \;d\omega .
\end{eqnarray}
Accordingly, even-odd dynamics in Eq.\,(\ref{EOcoefs0}) specializes to the $D=1$ spin-boson setting of interest as follows:
\begin{eqnarray}
    \gamma (t)^{\text{eo}} &= \kappa_s (t)\Big[ \left(m_{\text{e}}-m'_{\text{e}}\right){}^2 +\left(m_{\text{o}}-m'_{\text{o}}\right){}^2 \Big] +2 \kappa_d(t)  \left(m_{\text{e}}-m_{\text{o}}\right) \left(m'_{\text{e}}-m'_{\text{o}}\right)  ,  \nonumber \\
   \varphi _0(t)^{\text{eo}} &= \xi_s(t) \Big[ (m_{\text{e}}^2 -m_{\text{e}}^{'2}) + (m_{\text{o}}^{2} - m_{\text{o}}^{'2})\Big]
    +2 \xi_d (t) \left(m_{\text{e}} m_{\text{o}}-m'_{\text{e}} m'_{\text{o}}\right).
     \label{EOcoefsSB} 
 \end{eqnarray}
From Eq.\,(\ref{rel}), the decay and phase coefficients are then $\chi_i(t)= 4\kappa_i(t)$, $\Psi_i(t)= 4 \xi_i(t)$, with $i \in \{\text{s}, \text{d}\}$. We can now evaluate the even-odd frequency uncertainty $\Delta \hat{b}(t)$ for CSS and OATS initial preparations.

\subsection{CSS scaling}

For an initial CSS state of $N$ qubits, the calculation of the uncertainty can still be carried out exactly, resulting in the following expression:
\begin{eqnarray*}
\text{$\Delta \hat{b}$} (t)^2 &=  
\frac{(N+2)\,e^{\chi _s(t)}- (N-2)\,e^{-\chi _s(t)} \cos(2 \Psi _s(t)){}^{N/2-2} \cos(2 \Psi _d(t)){}^{N/2}}{N \;T\,  t\,  \cos(\Psi _s(t)){}^{N-2}  \cos(\Psi _d(t)){}^{N} }\nonumber\\
  &+ \frac{   e^{\chi _d(t)}  \cos(\Psi _s(t)-\Psi _d(t)) {}^{N-2}-e^{-\chi _d(t)}  \cos(\Psi _s(t)+\Psi _d(t)){}^{N-2}
  }{ \;T\,  t\,  \cos(\Psi _s(t)){}^{N-2}  \cos(\Psi _d(t)){}^{N}}. 
\label{dbsqeo}
\end{eqnarray*}
In the short-time regime of metrological interest, $\omega_c t \ll 1$, the decay coefficients are quadratic in time, $\chi_s(t) \approx \chi^2_{s0}\, (\omega_c t)^2$ and $\chi_d(t) \approx \chi_{d0}^2(x)\, (\omega_c t)^2$, whereas the phase coefficients are cubic, $\Psi_s(t) \approx \Psi_{s0}^3\, (\omega_c t)^3$ and $\Psi_d(t) \approx \Psi_{d0}^3(x)\, (\omega_c t)^3$. Note that when the qubits are in different clusters, we have explicitly written the dimensionless factors as $\chi_{d0}^2(x), \Psi_{d0}^3(x)$ to highlight the dependence on the (dimensionless) transit time parameter $x \equiv \omega_c t_{\text{eo}}$. Expanding the above expression up to the sixth order with respect to time, we find: 
\begin{eqnarray*}
\Delta \hat{b}(t,x) &\approx \frac{\sqrt{N+\frac{1}{2} N^2 (\omega_c t)^2 \left(\text{$\chi $}_{s0}^2+\text{$\chi $}_{d0}^2(x)\right)}}{N \sqrt{t T}} \left[1+\frac{N^2}{128} \left( \Psi_{s0}^3 + \Psi_{d0}^3(x) \right)^2 (\omega_c t)^6 \right]. 
\label{deltab}
\end{eqnarray*}
Here, the first term in the right hand-side corresponds to the NQN uncertainty short-time expansion ($\Psi_{s0}=\Psi_{d0}\equiv 0$), whereas the remaining term, of order $\mathcal{O}(N^2\,\omega_c t^6)$, gives the leading quantum corrections, $Q^{\text{\,eo}}_{\text{CSS}}(t,x)$. In the limit $N\gg 1$, the optimal measurement time and minimum uncertainty are given by:
\begin{eqnarray}
 \omega_c \tau_{\text{opt,eo}}^{\text{CSS}}(x)& \approx 2^{1/4} \left(\text{$\chi$}_{s0}^2+\text{$\chi$}_{d0}^2(x)\right)^{-1/2} N^{-1/2},
\label{dbopteo0}\\
\Delta \hat{b}_{\text{opt,eo}}^{\text{CSS}} (x) & \approx 2^{1/4}(\omega_c/T)^{1/2}\left(\chi_{s0}^2 + \chi_{d0}^2(x)\right)^{1/4} N^{-1/4}, 
\label{dbopteo}
\end{eqnarray}
with quantum corrections vanishing at a rate $N^{-1}$ at $\tau_{\text{opt,eo}}^{\text{CSS}}$. Thus, the uncertainty scales as in the collective noise case, $N^{-1/4}$, which is worse than the SQL.  
 
  \begin{figure}[t!]
    \centering
    \includegraphics[width=14.5cm]{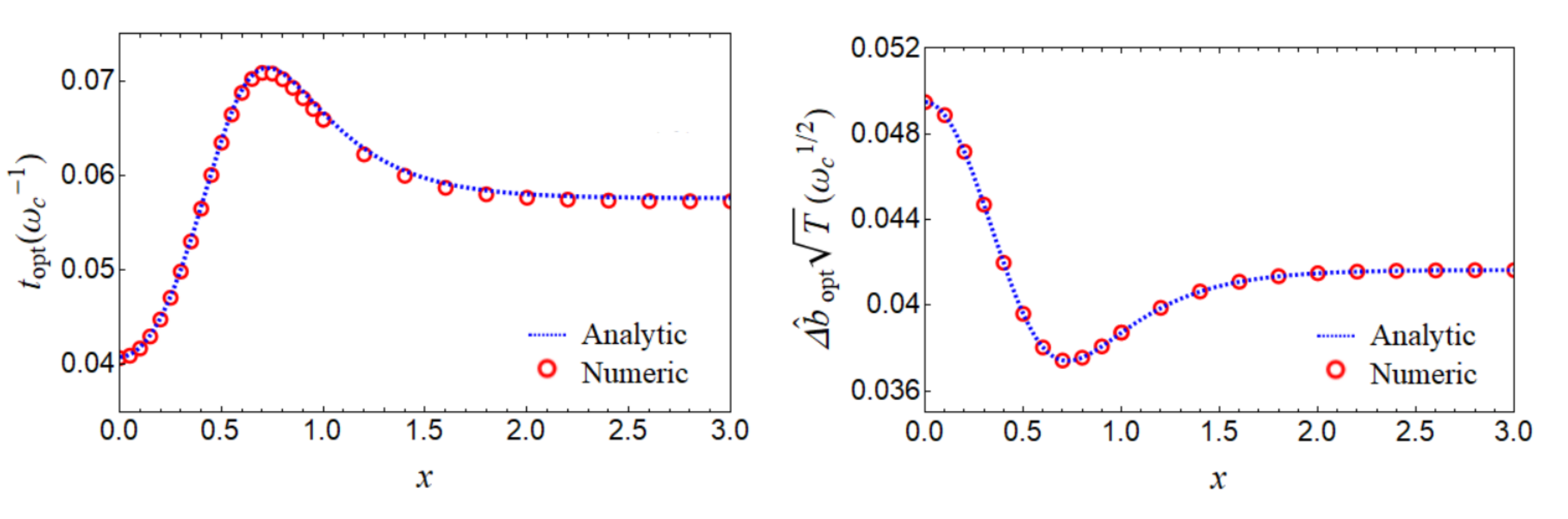}
    \vspace*{-3mm}
    \caption{{\bf Initial CSS with $\bm{N=20,000}$ qubits under even-odd bosonic dephasing.} Optimal measurement time (left) and minimum uncertainty (right) as a function of adimensional transit time $x$. Solid blue curves: Analytic expressions from the short-time expansion. Red dots: Optimal 	 time and uncertainty obtained from numerical minimization. Numerically, the minimum is found to occur at $x_{\text{opt}} \approx 0.76778$. Noise parameters are as in Fig.\,1.}
    \label{deltabx}
\end{figure}

The behavior of both $t_{\text{opt,eo}}^{\text{CSS}}$ and $\Delta \hat{b}_{\text{opt,eo}}^{\text{CSS}}$ as a function of the ``non-collectivity'' parameter $x$ is shown in Fig.\,\ref{deltabx} for fixed system size. Several conclusions may be drawn. First, one can see that the analytic expressions given by Eqs.\,(\ref{dbopteo0})-(\ref{dbopteo}) accurately fit the numerical results over the full range of values of the adimensional transit time or, equivalently, qubit separation. This is expected, as we established that all the quantum corrections are asymptotically vanishing. Secondly, the minimum uncertainty achievable under collective noise, $x=0$, is always larger than the one in the non-collective regime, $x \neq 0$. In this sense, we can regard {\em non-collectivity as a resource}, improving the estimation precision by an $N$-independent factor. Thirdly, for large enough $x$ the system approaches a second collective limit, which is also expected: the qubit clusters eventually no longer interact with each other. This manifests in vanishing even-odd decay and phase coefficients $\chi_{d}(t), \Psi_d(t)$, since the term $\cos(\omega\, t_{\text{eo}})$ in the frequency integrals is rapidly oscillating when $t_{\text{eo}}\gg 1$. In this regime, we are effectively carrying out two collective measurements over two clusters of $N/2$ qubits each; accordingly, using Eq.\,(\ref{felix-css}), we have $\Delta \hat{b}^{\text{CSS}}_{\text{opt,eo}} (x\gg1) \approx 2^{-1/4} \Delta \hat{b}^{\text{CSS}}_{\text{opt,coll}}$, in agreement with Fig.\,\ref{deltabx}. 

Finally, it is clear from both the plot and Eq.\,(\ref{dbopteo}) that maximum metrological advantage can be achieved for an intermediate, {\em finite} value of the non-collectivity parameter $x_{\text{opt}}$. Analytically, $x_{\text{opt}}$ corresponds to the value that minimizes $\chi^2_{d0}(x)$. Using the short-time expression for the decay parameter given in Eq.\,(\ref{chist}), we can in turn establish that $\chi_{d0}^2(x) \propto {\langle \{ B_{\text{e}}(0), B_{\text{o}}(0)\} \rangle_{\rm B}} / {\omega_c^2},$ for $\omega_c t \ll 1.$ Under the assumptions of Sec.\,\ref{sbred}, this \textit{spatial} equal-time correlation function can be written as an integral over frequency,
$$ \chi_{d0}^2(x)  =
 \int_0^{\infty} \!d\omega  J(\omega) \cos(\omega t_{\text{eo}})= 
 \, \alpha \Gamma(s+1) \,\frac{\cos \left((s+1) \arctan\left(x \right) \right))}{(1+x^2)^{(s+1)/2}} , $$
where we have assumed an exponential cutoff for $J(\omega)$ in the second equality to obtain an analytic expression. Importantly, depending on the value of $x$, this function can be made \emph{negative} in principle. Thus, lowering the qubits spatial correlations proves useful for achieving a sensing advantage over the collective noise scenario.

\subsection{OATS scaling}

We now turn our attention to the short-time uncertainty scaling of an OATS corresponding to the optimal squeezing and rotation angles given in Eq.\,(\ref{ideal}). As in the collective noise setting, an exact evaluation of the expressions in Eqs.\,(\ref{eqnmomentsJy1})-(\ref{eqnmomentsJy2}) is not available, due to non-trivial multiqubit entanglement. However, the desired mean values can still be computed by performing a cumulant expansion over the qubit operators, which we detail in \ref{noncolcum}. In the $N\gg1 $ limit where second-order truncation is warranted, we obtain analytic (albeit unwieldy) formulas. Expanding these in the short-time regime yields the following expression for the estimation precision:
\begin{eqnarray*}
\Delta \hat{b}(t,x) & \approx \!\frac{\sqrt{\frac{3^{2/3}}{2} N^{1/3} +\frac{1}{2} N^2 (\omega_c t)^2 \left(\text{$\chi$}_{s0}^2+\text{$\chi $}_{d0}^2(x)\right)}}{N \sqrt{t \,T}} 
\bigg[1 + \frac{N^{3/4}}{3^{5/6} 2^{17/12}} (\Psi_{s0}^3 + \Psi_{d0}^3(x)) (\omega_c t)^3 \bigg],
\label{deltabOATS}
\end{eqnarray*} 
where, as before, the second term in square parenthesis gives the quantum corrections to the NQN short-time formula ($\Psi_{s0}=\Psi_{d0}\equiv 0$). The best possible uncertainty and optimal measurement time have the same sub-SQL scaling as they do for an OATS evolving under collective noise, namely:
\begin{eqnarray}
\omega_c \tau_{\text{opt,eo}}^{\text{OATS}} (x) &= 3^{1/3}\, \left(\text{$\chi_{s0}$}^2+\text{$\chi_{d0}$}^2(x)\right)^{-1/2}N^{-5/6}, \nonumber\\
\Delta \hat{b}_{\text{opt,eo}}^{\text{OATS}}(x) &=3^{1/6}\, (\omega_c/T)^{1/2}\; (\chi_{s0}^2 + \chi_{d0}^2(x))^{1/4}N^{-5/12}.
\label{eoOATS}
\end{eqnarray}

\begin{figure}[t!]
    \centering
   \includegraphics[width=14.5cm]{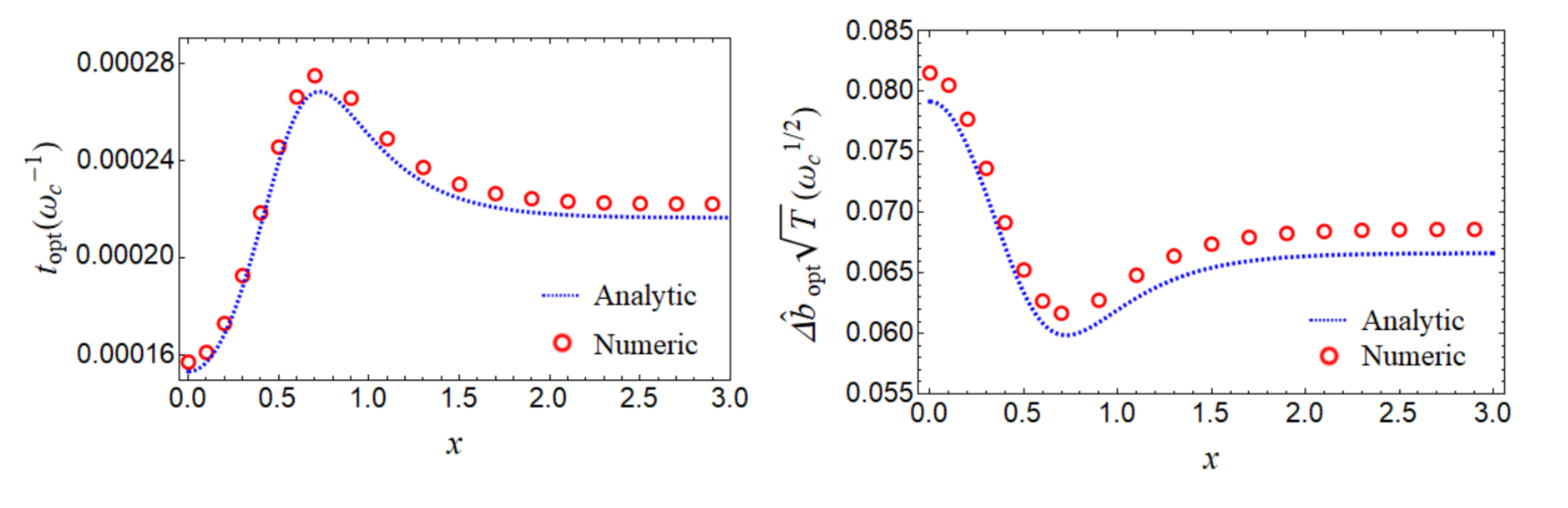}
   \vspace*{-3mm}
    \caption{{\bf Initial OATS with \bm{$N=20,000$} qubits under even-odd bosonic dephasing.}  Optimal measurement time (left) and minimum uncertainty (right) as a function of adimensional transit time $x$. Blue dashed curves: Analytic expressions from the short-	time expansion. Red dots: Optimal time and uncertainty obtained from numerical minimization. Noise parameters are as in the previous 	figures, and ideal squeezing and rotation angles $(\theta_{\text{opt}},\beta_{\text{opt}})$ given by Eq.\,(\ref{ideal}) are used.
}  
\label{sweet}
\end{figure}  

Representative results are shown in Fig.\,\ref{sweet}. Overall, the agreement between analytics and numerics is still very good, although the optimal measurement time and uncertainties predicted by expanding to second order the NQN formula have a small offset with respect to their exact values. However, the  significance of the offset disappears in the $N \gg 1$ limit: For $N=2000$ the relative difference between analytic and numeric {\em optimal} uncertainty values is already below 3$\%$ (data not shown), and for $N=20,000$ as in Fig.\,\ref{sweet}, the value is 1.33$\% $. Quantum corrections at the optimal time are, again, asymptotically negligible, $N^{3/4} (\omega_c t_{\text{opt,eo}}^{\text{OATS}})^3 \propto  N^{-7/4} \ll 1$. Importantly,  the $x$-dependence of the numerical prefactors for both the optimal measurement time and the minimum uncertainty is identical to the CSS case under even-odd noise. Thus, the ideal value $x_{\text{opt}}$ that optimizes the constant-factor advantage with respect to the collective setting is still given by the separation that {\em minimizes the spatial correlation} $\chi_{d0}^2(x)\approx \langle \{ B_{\text{e}}(0), B_{\text{o}}(0)\} \rangle_{\rm B}/\omega^2_c$. Similarly, a metrological advantage over collective noise is also retained in the limit $x\gg 1$, where $\chi_{d0}(x) \approx 0$. Comparing to Eq.\,(\ref{felix-oat}), $\Delta \hat{b}^{\text{OATS}}_{\text{opt,eo}} (x\gg1) \approx 2^{-1/4} \Delta \hat{b}^{\text{OATS}}_{\text{opt,coll}}$, again in agreement with the plot.

We can then conclude that, despite the ensuing different $N$-scaling, the benefit that non-collectivity affords for OATS and CSS is similar: in both cases, the even-odd constant-factor advantage we can gain, asymptotically, with respect to the collective case is maximized for the same finite value $x_{\text{opt}}$ that makes the spatial noise correlation $\chi_{d0}^2(x)$ as small as possible.

\section{Non-collective spin-boson dephasing: Randomized Ramsey metrology}
\label{sec:random}

The above analysis brings forward the importance of spatial noise correlations to improve sensitivity of the chosen initial states in comparison to the collective setting. However, though capable of producing a constant factor advantage, the simple departure from collectivity afforded by the even-odd setting fails to improve the uncertainty scaling with respect to $N$. This prompts to look for different ways to break the collective permutational symmetry. By making again reference to the spin-boson Hamiltonian in Eq.\,(\ref{eqnSBH}), and in the absence of prior knowledge about the noise properties, we consider a \textit{randomized coupling} (RC) setting, whereby the spatial locations of the $N$ probe qubits are independently and identically distributed according to an isotropic zero-mean Gaussian probability distribution. That is:
\begin{eqnarray}
{\mathbb{P}}(\vec{r}_n)= \frac{ e^{-{|\vec{r}_n|^2}/({2 \epsilon^2})}  }{(2 \pi \epsilon^2)^{D/2}} , \quad \forall n, D\in \{1,2,3\}, \qquad 
{\mathbb{P}}(\vec{r}_1,\ldots,\vec{r}_N) \equiv {\mathbb{P}}(\vec{r})= \prod_{n=1}^N {\mathbb{P} }(\vec{r}_n).
\label{sprob}
\end{eqnarray}
In what follows, we shall include $\vec{r}$ in the argument of the ensuing functions to signal any dependence with respect to an arbitrary subset of the qubit positions $\{\vec{r}_1,\ldots,\vec{r}_N\}$, and denote the classical average of a function $f(\vec{r},t)$ with respect to distribution (\ref{sprob}) by $\mathbb{E}\{ f(\vec{r},t) \}$. 

Before proceeding, it is worth noting that a method for achieving Heisenberg scaling that shares superficial similarities with the above was proposed in Ref.\,\cite{Mogilev}. As in our case, dephasing from a bosonic environment is considered; however, the focus is on a {\em continuum limit} where sums over qubit indexes can be approximated by spatial integrals with respect to a Gaussian qubit density. While such a procedure may be suitable to model dense atomic-cloud sensors, it is justified in a large-density limit (formally, $N/V \rightarrow \infty$) which we do not {\em a priori} require \footnote{Technically, the fact that individual qubit labels are retained in our analysis also implies additional (diagonal) contributions to the relevant decay factors, {\color{rev} which are {\em absent} in the approach of \cite{Mogilev}; see, in particular, Eq.\,(\ref{rcchi0}).}}. Importantly, our approach allows the precision scaling of {\em any} initial state to be evaluated in principle, whereas the mechanism for engineering decoherence suppression discussed in \cite{Mogilev} relies explicitly upon the two-component nature and symmetry properties of the GHZ state.

\subsection{Spatially averaged two-point noise correlation functions} 
\label{sub:randomcorr}

In the RC setting, improvements may be expected, on average, if sufficiently many configurations where noise correlations are negative are sampled during the sensing process. Some initial qualitative insight may be gained by considering the spatially averaged two-point noise correlation functions. That is, with reference to Eq.\,(\ref{eqCorr}), we consider: 
$$ \mathbb{E}\{ C_{nm}(\vec{r},t)\} \equiv  \mathbb{E} \{
\left\langle B_n(\vec{r} ,t) B_m(\vec{r},0)\, \right\rangle_{{\mathrm B}}\} , \quad 
B_n(\vec{r},t)= \sum_{\vec{k}} g_{k} (e^{i \vec{k} \cdot \vec{r}_n} e^{i \Omega_{k}t}  b_{\vec{k}}^{\dagger}+e^{-i \vec{k} \cdot \vec{r}_n} e^{-i \Omega_{k}t} b_{\vec{k}} ),$$
where the expectation is taken over all random realizations of the qubit positions. Let, as before $\vec{r}_{nm} = \vec{r}_n-\vec{r}_m$ denote the relative spatial separation between qubits $n,m$. Since, for $n \neq m$, we have  
\begin{equation}
\mathbb{E}\{ \cos(\vec{k}\cdot \vec{r}_{nm})\}  = e^{-k^2 \epsilon^2} ,
\label{spatial_cos}
\end{equation}
the desired average may be carried out as follows: 
\begin{eqnarray*}
\mathbb{E}\{\langle B_n(\vec{r},t)B_m(\vec{r},0)\rangle_{{\mathrm B}} \} &= \sum_{\vec{k}}|g_{k}|^2 e^{-k^2 \epsilon^2} [ \coth(\beta \Omega_k/2) \cos(\Omega_{k}\tau) - i \sin(\Omega_{k}\tau] ) \\
&\approx \sum_{\vec{k}}|g_{k}|^2 e^{-k^2 \epsilon^2} [ \cos(\Omega_{k} t) - i \sin(\Omega_{k} t) ]
= \!\int_0^{\infty} \!\! \!d\omega\, J(\omega) e^{-\omega^2 \epsilon^2/v^2} e^{-i\omega t}, 
\end{eqnarray*}
where, in the last equality, we have assumed for simplicity that $D=1$. By noticing that the parameter ${v}/{\epsilon}$ has units of frequency, it is useful to further introduce the dimensionless ``smallness parameter`` $\eta\equiv (v/\epsilon)/{\omega_c},$ which may be thought as the ratio $\eta \equiv \ell_{\mathrm B}/\ell_{\mathrm S}$ between the characteristic length scale $\ell_{\mathrm S}=\epsilon$ associated to the spatial randomization and a characteristic length scale $\ell_{\mathrm B} = v/\omega_c \equiv v \tau_c/2\pi$ that a bath excitation propagating at speed $v$ covers during the ``memory time'' $\tau_c = 2\pi/\omega_c \equiv 1/\nu_c$ of the bath itself.

\begin{figure}[t!]
    \centering
    \includegraphics[width=8.5cm]{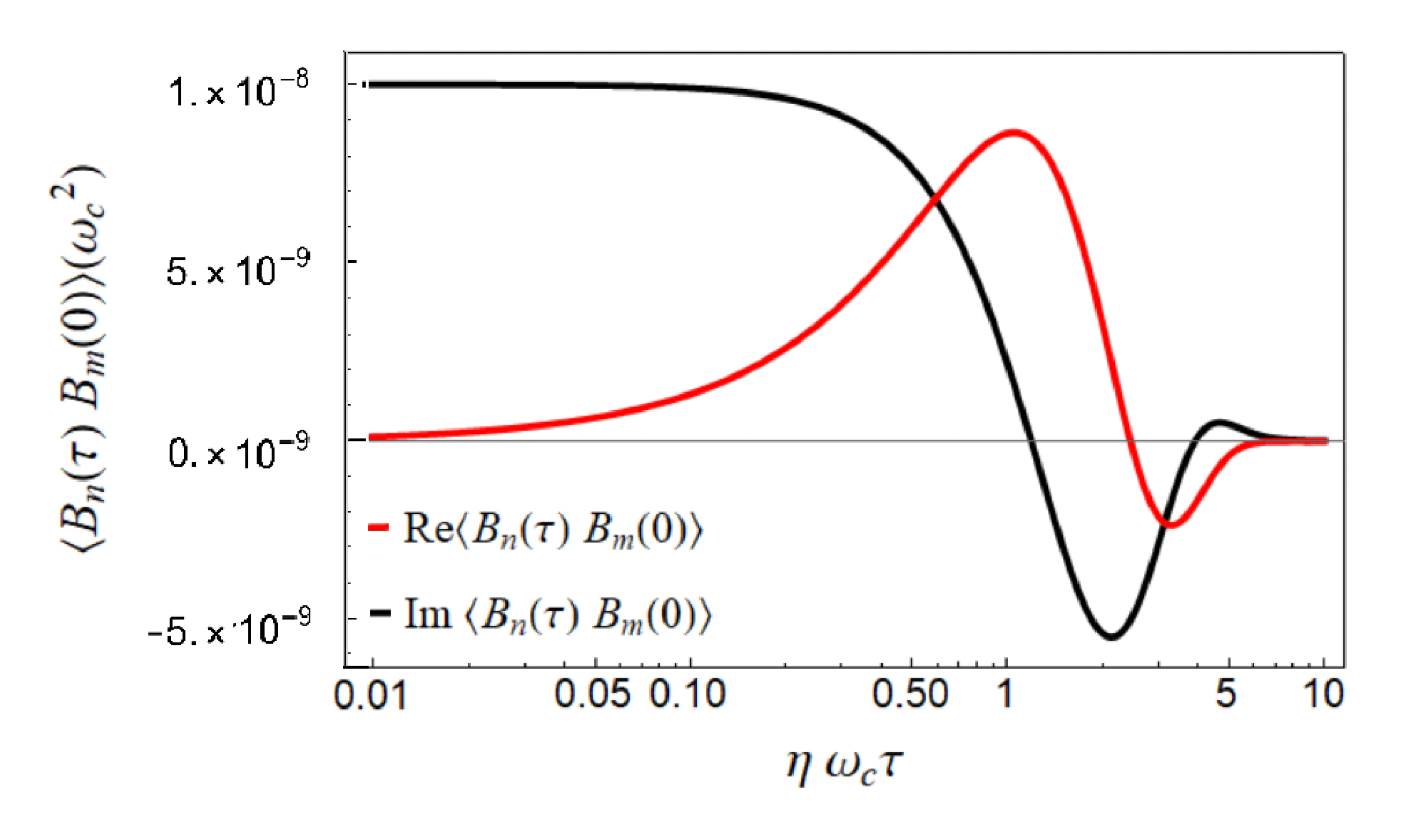}
    \vspace*{-3mm}
    \caption{{\bf Spatially-averaged two-point bath correlation function of a 1D RC bosonic model.} Real (black) and imaginary (red) 	components are shown in units of $\omega_c^2$, as a function of the dimensionless time variable $\eta \omega_c t$. Parameters are: $\eta =0.01, \epsilon=100,   \omega_c=1,s=3, v=1$. Both functions are upper bounded by an $\eta$-dependent constant which vanishes for small $\eta$, and takes the value $10^{-8}$ for this parameter choice.}
    \label{spatialtwopointcorr}
\end{figure}

Assuming a spectral density with a Gaussian cutoff, $K(\omega,\omega_c)= e^{-\omega^2/\omega_c^2}$, the above integral can be evaluated exactly, yielding 
\begin{eqnarray}
\mathbb{E}\{\langle B_n(\vec{r},t)B_m(\vec{r},0)\rangle_{{\mathrm B}} \} &=\frac{1}{2} \eta^{s+1} \omega_c^2\bigg\{  \Gamma \left(\frac{s+1}{2}\right) {}_1F_1\left(\frac{s+1}{2};\frac{1}{2};-\frac{(\eta \omega_c t)^2}{4 }\right)\nonumber\\
& \hspace{15mm}-i \left(\eta \omega_c t \right) \Gamma \left(\frac{s}{2}+1\right) {}_1F_1\left(\frac{s+2}{2};\frac{3}{2};-\frac{(\eta \omega_c t)^2}{4 }\right)\bigg\} ,
\label{sp}
\end{eqnarray}
where $_1F_1(a,b,z)$ denotes the confluent hypergeometric function of the first kind. Since the third argument in $_1F_1(a,b,z)$ is negative, the function is bounded. Thus, \textcolor{rev} {{\em the spatially-averaged two-point correlator is bounded by a constant which vanishes at a rate of $\eta^{s+1}$ in the limit $\eta \rightarrow 0$, for all times and $n\ne m$} (see also Fig.\,\ref{spatialtwopointcorr})}. 

The physical interpretation is that, when $\eta \ll 1$, averaging over the random qubit locations effectively spatially de-correlates the noise they experience on average, $\mathbb{E}\{\langle B_n(\vec{r},t)B_m(\vec{r},0)\rangle_{{\mathrm B}} \}/{\omega_c^2} \approx \delta_{nm} f(\eta, \omega_c t)$, where $f={\mathcal O}(\eta^{s+1})$. Physically, the requirement of small $\eta$ is met if the spatial dispersion $\epsilon$ of the qubits is ``large'' relative to the characteristic bath length scale, so that several system-bath coupling configurations may be sampled within $\ell_{\mathrm S}$. It is interesting to note that the idea of suppressing dephasing by ``qubit motion'' has been invoked before, by leveraging a controlled transfer of a logical qubit state across different physical qubits \cite{Averin} and, in the context of metrology, by invoking quantum teleportation to effect such a transfer and effectively reset the qubit noise environment {\color{rev} \cite{Matsuzaki2}}. Here, the strongest formal analogies and a conceptual ``duality'' stem from making contact with randomized DD schemes \cite{RandomDD}: in DD, a necessary condition for improvement is access to a ``fast-control'' regime where the characteristic timescale $\tau_{\mathrm S}$ associated to the control obeys $\omega_c \tau_{\mathrm S} \ll 2\pi$ \cite{KavehLimits}. That is, by identifying $\tau_{\mathrm B}\equiv \tau_c$, it must be $\nu_c \tau_{\mathrm S}\ll 1$ or, equivalently, the ``temporal'' smallness parameter $\tilde{\eta}\equiv \tau_{\mathrm S}/\tau_{\mathrm B} \ll 1$. Thus, just like DD is ineffective at suppressing ``temporally white'' noise for which $\tau_{\mathrm B} \rightarrow 0$, the proposed randomization has no effect on ``spatially white'' ($\vec{r}$-independent) noise for which, formally, $\ell_{\mathrm B} \rightarrow \infty$.

\subsection{Spin-boson reduced dynamics under randomized couplings}

In order to quantitatively describe how the above intuition manifests in the metrological setting of interest, we proceed to characterize the reduced qubit dynamics, which now entails taking both a quantum average corresponding to the partial trace over the bath (with respect to the initial state $\rho_{\mathrm B}$) and a classical average over the qubit position (with respect to the joint distribution in Eq.\,(\ref{sprob})). More concretely, the evolution operator acting over the total density matrix $\rho_0 \otimes \rho_B $ depends on the RC qubits positions through the bath operators $B_n(\vec{r},t)$ introduced in Eq.\,({\ref{eqnSBH}), $U(\vec{r},t)= \mathcal{T}_+ \exp \left(\sum_{n=1}^N \sigma_n^z \int_0^t \,ds\,B_n(\vec{r}, s) \right)$. We can then write:
\begin{eqnarray}
\bar{\rho}(t)& \equiv \mathbb{E}\{\rho(\vec{r},t) \}  =
\mathbb{E}\left\{ \text{Tr}_B\left[ U(\vec{r},t) \rho_0 \otimes \rho_B U^{\dagger}(\vec{r},t) \right]\right \}= 
\sum_{\vec{\alpha} \vec{\beta}} |\vec{\alpha} \rangle \langle \vec{\beta}|\;\mathbb{E}\{ \langle \vec{\alpha}| \rho(\vec{r},t)| \vec{\beta} \rangle\},\label{RCdyn}
 \end{eqnarray}
with the reduced matrix elements now carrying a spatial average: 
\begin{eqnarray}
\mathbb{E}\{ \langle \vec{\alpha}| \rho(\vec{r},t)| \vec{\beta} \rangle\} &  
=\langle \vec{\alpha}|\rho_0|\vec{\beta}\rangle \mathbb{E} \left \{ e^{-\gamma(\vec{r},t)} e^{i \varphi_0(\vec{r},t)} \right \}. 
\label{rcred}
\end{eqnarray}
Decay $\gamma(\vec{r},t)$ and phase $\varphi_0(\vec{r},t)$ are given by Eqs.\,(\ref{gamma})-(\ref{phi0}) with the spin-boson dynamic coefficients 
\begin{eqnarray}
\kappa_{nm}(\vec{r},t)&=\frac{1}{4}\sum_{\vec{k}} |g_k|^2 \frac{1- \cos \Omega_{k} t}{\Omega_{k}^2} \coth \left(\beta \Omega_{k}/2\right)   \cos (\vec{k} \cdot \vec{r}_{nm}) ,
\label{chirc} \\
\xi_{nm}(\vec{r},t)  &=\frac{1}{4} \sum_{\vec{k}} |g_{k}|^2 \frac{\Omega_{k}t- \sin \Omega_{k} t}{\Omega_{k}^2}   \cos (\vec{k} \cdot 
\vec{r}_{nm} ).
\label{xirc}
\end{eqnarray}
Note that we have kept the expressions for $\kappa_{nm}(\vec{r},t)$ and $\xi_{nm}(\vec{r},t)$ as discrete sums over the $\vec{k}$ modes,  instead of writing them in terms of an integral over frequency, as this will make it possible to leverage Eq.\,(\ref{spatial_cos}). Using Eq.\,(\ref{RCdyn}), the expectation value of any system operator in the RC setting can then be written as
\begin{eqnarray}
\langle \bar{\mathcal{O}}(t) \rangle \equiv  \mathbb{E}\{ \langle \mathcal{O}(\vec{r},t) \rangle \} = 
\text{Tr}_S\left[ \bar{\rho}(t) \mathcal{O} \right] = 
\sum_{\vec{\alpha} \vec{\beta}} \langle \vec{\beta}|\mathcal{O} |\vec{\alpha} \rangle\; 
\langle \vec{\alpha}|\rho_0|\vec{\beta}\rangle \mathbb{E} \left \{ e^{-\gamma(\vec{r},t)} e^{i \varphi_0(\vec{r},t)} \right \} .
\label{OexpRC}
\end{eqnarray}

{\color{rev}
While Eq.\,(\ref{rcred}) does not lend itself to an exact evaluation in analytical form, we can obtain the spatial average approximately, by resorting to a cumulant expansion. Assuming that truncation to the second order is appropriate (we provide {\em sufficient} conditions for that later on), we obtain:
\begin{eqnarray}
\mathbb{E}\left\{ e^{-\gamma(\vec{r},t)} e^{i \varphi_0(\vec{r},t)} \right\} \approx
e^{    - \bar{\gamma}(t)  \,+ i \bar{\varphi}_0(t) \,+\, \frac{1}{2} [  \Delta \bar{\gamma}(t)^2 - \Delta \bar{\varphi}_0(t)^2 
+ 2 i \,\text{Cov}(\varphi_0(\vec{r},t), \gamma(\vec{r},t)) ]  }\,, 
\label{cum}
\end{eqnarray}
where we have introduced $\bar{\gamma}(t)\equiv \mathbb{E}\{\gamma(\vec{r},t)\}, \bar{\varphi}_0(t)\equiv \mathbb{E}\{ \varphi_0(\vec{r},t) \}$ and $\Delta \bar{a}(t)^2 \equiv  \mathbb{E}\{ a^2(\vec{r},t)\}\, - \mathbb{E}\{a(\vec{r},t)\}^2\,$;  $\text{Cov}(a(\vec{r},t),b(\vec{r},t)) \equiv \mathbb{E}\{ a(\vec{r},t)\, b (\vec{r},t) \} \,- \mathbb{E}\{ a(\vec{r},t)\} \,\mathbb{E} \{ b(\vec{r},t) \}$ denote the variance and co-variance of position-dependent quantities. Thus, we need to calculate the cumulants up to the second order and show that the expansion yields an accurate approximation in the parameter regime of interest. Since the calculations are lengthy, we 
refer to \ref{RCapp} for derivations, and focus here on the implications of the results}. 

{\color{rev}
\subsubsection{Reduced density matrix and observable expectation values.} 
\label{reduc}

{\color{rev} 
To proceed with the analysis, we assume $D=1$, as in Sec.\,\ref{sub:randomcorr}, a spectral density with Gaussian cutoff, and work in the metrologically relevant short-time regime. The quantities $\bar{\gamma}(t)$ and $\bar{\varphi}_0(t)$ that enter Eq.\,(\ref{OexpRC}) consist of weighted sums of spatially averaged dynamic coefficients. We denote by 
$$\mathbb{E}\{\kappa_{nm}(\vec{r}, t)\}\equiv   \bar{\kappa}_1(t), \;\; \mathbb{E}\{\xi_{nm}(\vec{r}, t)\}\equiv   \bar{\xi}_1(t), \; n \neq m, \quad 
\mathbb{E}\{\kappa_{nn}(\vec{r}, t)\}\equiv   \bar{\kappa}_0(t),\; n=m,$$
the mean decay and phase coefficients corresponding to distinct qubit indexes, and the mean decay for equal indexes, respectively. Importantly, as shown in Eqs.\,(\ref{chi0st})-(\ref{xi1st}) from \ref{means}, $\bar{\kappa}_1(t)$ and $\bar{\xi}_1(t)$ are proportional to powers of the smallness parameter $\eta$ and hence vanishing in the large dispersion $\eta \ll 1$ limit, whereas $\bar{\kappa}_0(t) \approx \bar{\kappa}_0 (\omega_c t)^2$ remains immune to spatial averaging. Similarly, the quantities $\Delta \bar{\gamma}(t), \Delta \bar{\varphi}_0(t), \text{Cov}(\varphi_0(\vec{r},t), \bar{\gamma}(\vec{r},t))$ entering the second order cumulant in Eq.\,(\ref{OexpRC}) depend on the spatial variance and covariance of relevant dynamic coefficients. For example, $\Delta \bar{\gamma}(t)$ can be written as a weighted sum of $\mathbb{E}\{\kappa_{nm}(\vec{r}, t)\, \kappa_{n' m'}(\vec{r}, t)\} - \mathbb{E}\{\kappa_{nm}(\vec{r}, t)\} \,\mathbb{E}\{ \kappa_{n' m'}(\vec{r}, t)\}$ terms, with only a subset of the indexes $\{n,m,n',m'\}$ yielding a non-zero result, as shown in \ref{cov}. The resulting time-dependent expressions, Eqs.\,(\ref{FFF1})-(\ref{FFG}), are all proportional to powers of $\eta$. The upshot is that, formally, {\em all of the means and (co-)variances vanish, at different rates, in the limit where $\eta \rightarrow 0$}, with the exception of the term $\kappa_0(t)$, which is insensitive to spatial randomization\footnote{Note that this contribution is {\em zero from the outset} in the continuum limit of \cite{Mogilev}.}.  

\smallskip

Our task now is twofold. First, we wish to prove that in the short-time, large-dispersion regime, the first-order cumulants dominate, so we can safely discard higher orders; next, we need to establish under what conditions all the exponentials carrying terms proportional to powers of $\eta$ can be safely equaled to one. We have the following: 

\smallskip

{\bf Claim.} \label{claim} \em Assume that the short-time-large-dispersion limit $\eta \ll1$ is obeyed. Then:}

{\bf (i)} {\em First-order cumulants provide a good approximation to the full spatial average provided that the additional condition $\eta \left(\omega_c t\right)^2 N \ll 1$ is met.} 

{\bf (ii)} \emph{In the regime of validity of (i), all terms with exponents proportional to powers of $\eta$ can be made arbitrarily close to unity provided that the condition $\eta^{s+1} (\omega_c t)^2 N^2 \ll 1$ is met.}

\smallskip

The proof is carried in \ref{proofclaim}. Accordingly, when the conditions {\bf(i)}-{\bf(ii)} are both obeyed {\color {rev} -- we will henceforth refer to this limit as the \textit{small-$\eta$ regime} --}} the spatially-averaged reduced matrix elements can be written in terms of an {\em $\eta$-independent decay proportional to} $\bar{\kappa}_0(t)$:
\begin{equation}
\mathbb{E}\{\langle \vec{\alpha}| \rho(t)|\vec{\beta}\}  
\stackrel{\text{(i)}}{\approx} 
\langle \vec{\alpha}| \rho_0|\vec{\beta}\rangle \,
 e^{-\bar{\gamma}(t)} e^{i \bar{\varphi}_0(t)} 
\stackrel{\text{(ii)}}{\approx}  
\langle \vec{\alpha}| \rho_0|\vec{\beta}\rangle  \, e^{-4 N \sin \left(\theta_{\alpha \beta}/2\right)^2 \bar{\kappa}_0^2 \,\left(\omega_c t\right)^2 }. 
\label{param}
\end{equation}
Given these expressions, spatially-averaged expectation values of a relevant observable $\mathcal{O}$ may be computed via Eq.\,(\ref{OexpRC}). In particular, this means we can circumvent the use of effective propagators of Eqs.\,(\ref{prop1})-(\ref{prop2}) to evaluate $\langle \overline{J_y}(t)\rangle, \langle \overline{J^2_y}(t)\rangle$ in the small-$\eta$ regime. Additionally, note that when Eq.\,(\ref{param}) holds,  the only non-negligible time dependence in  $\langle \bar{\mathcal{O}} (t)\rangle$ stems from $\bar{\kappa}_0(t)\approx \bar{\kappa}_0^2 \, (\omega_c t)^2$, and all terms coming from qubit pairs with $n \neq m$ in the mean values and the second-order cumulants of $\gamma(\vec{r},t), \varphi_0(\vec{r},t)$ may be disregarded. The remaining non-vanishing term in Eq.\,(\ref{param}) arises {\em solely from the dynamic coefficients corresponding to pairs with $n=m$}. In other words, this analysis confirms that, under the validity of conditions {\bf (i)}-{\bf (ii)}, the net effect of randomizing the qubit positions is that the noise entering the reduced dynamics remains temporally correlated but becomes {\em spatially uncorrelated and classical} ($e^{i \bar{ \varphi}_0(\vec{r}, t)\} }  \approx 1$), consistent with the simplified picture for spatially-averaged noise correlations in Sec.\,\ref{sub:randomcorr}. Loosely speaking, the emergence of local reduced dynamics occurs because, when the spatial probability profile is wide enough, we are sampling as many qubit positions leading to negative bath correlations as positive ones. Let us emphasize however, that the above rather stringent conditions are merely \textit{sufficient} for obtaining asymptotically local dynamics and, in fact, {\em not} necessary. As we shall see in the coming sections, the regime where the qubits are effectively de-correlated may extend well out of the short-time limit.

\subsubsection{Randomization overhead.}  In a randomized Ramsey protocol, the uncertainty in estimating the target parameter $b$ may again be computed by means of the propagation formula, Eq.\,(\ref{eqnObs}), where now the expression for the operators' mean values involves both a trace and a spatial integral, that is, 
\begin{eqnarray}
\Delta \hat{b}(\tau)_{\text{RC}} = \frac{\Delta \overline{J_y} (\tau)}{ \sqrt{\nu} {|\partial\langle{\overline{J_y}(\tau)\rangle}/\partial b|}},\qquad \Delta \overline{J_y} (\tau) \equiv \sqrt{\langle{\overline{J^2_y}(\tau) \rangle}- \langle{\overline{J_y}(\tau)}\rangle^2}.
\label{eqnObsRC}
\end{eqnarray}
In practice, however, randomizing the sensors' positions necessarily entails an overhead in terms of required experiments: the sensors positions are first randomly chosen according to the Gaussian probability profile of Eq.\,(\ref{sprob}), and $\nu$ repetitions of the usual Ramsey protocol are effected, each of duration $\tau$; the whole procedure is then iterated, say, $K$ times, with the qubit positions being re-sampled at random at the beginning of each run. {\color{rev} Even in the limit $\nu \rightarrow \infty$ of infinite measurement statistics for a given qubit configuration $\{\vec{r}^{(i)}\}$,} this leads to an approximate expression for the spatial expected values in term of a finite sample mean {\color{rev} over the number of said configurations} :
\begin{eqnarray}
\bar{\rho}(t)=&\mathbb{E}\{\rho(\vec{r},t)\} \approx \frac{1}{K} \sum_{i=1}^K \rho(\vec{r} ^{\;(i)},t)\equiv  \overline{\rho}(t)_K ,\\
\langle\bar{O}(t)\rangle =& \mathbb{E}\{ \langle \mathcal{O}(\vec{r},t) \rangle \} \approx \frac{1}{K} \sum_{i=1}^K \text{Tr}_S\{ \rho(\vec{r} ^{\;(i)},t)\mathcal{O}\}=\frac{1}{K} \sum_{i=1}^K  \langle \mathcal{O}(\vec{r} ^{\;(i)},t) \rangle \equiv \langle\, \bar{\mathcal{O}}(t)_K \rangle .
\label{finiteK}
\end{eqnarray}
Here, $\vec{r}^{\;(i)}$ is a shorthand for the sensors positions at the $i$th run, $\vec{r}^{\;(i)}= \{\vec{r}_1^{\;(i)},\ldots,\vec{r}_N^{\;(i)}\}$, and $\overline{\rho}(t)_K$, $\langle \overline{\mathcal{O}}(t)_K \rangle$ denote the finite-sample approximations to $\bar{\rho}(t)$ and $\langle \bar{\mathcal{O}}(t) \rangle$ for an ensemble of size $K$. By assuming that different random realizations are independent, and invoking the central limit theorem, the ensemble average $\langle \overline{\mathcal{O}}(t)_K \rangle$ is a normally distributed random variable{\color{rev}\footnote{{\color{rev} Importantly, this assumes that the sample size is large enough for the central limit theorem to hold. We have verified that already for $K=20$ qubit configurations, as used in later plots, this is indeed the case around the optimal measurement time. It is then reasonable to disregard
bias effects coming from our finite-sample approximation to the spatial average. } }}, with a mean equal to the desired expected value $\langle \bar{\mathcal{O}}(t) \rangle$ and a standard deviation $\sigma_K=\sigma/\sqrt{K}$, where $\sigma$ is the ideal standard deviation corresponding to an infinite sample, $\sigma^2 = \mathbb{E}\{ \langle \mathcal{O}(\vec{r},t) \rangle^2  \} -\mathbb{E}\{ \langle \mathcal{O}(\vec{r},t) \rangle  \}^2.$ While an expression for $\sigma$ may be impractical to compute analytically, one may estimate it from the standard deviation of the set $\{\langle \mathcal{O}(\vec{r}^{\;(1)},t) \rangle ,\ldots, \langle \mathcal{O}(\vec{r}^{\;(K)},t) \rangle \}$. To ensure that, with probability $1-\varepsilon$, the estimation error in $\langle \overline{\mathcal{O}}(t)_K \rangle$ is less than $\delta_e >0$, the sample size must exceed a minimum value given by 
\begin{eqnarray}
K_{\text{min}} = (z_{1-\varepsilon/2} \sigma/\delta_e)^2  \approx \sigma^2/\delta_e^2, 
\label{samplesize}
\end{eqnarray}
where $z_p$ is the quantile of the standard normal distribution, that is, the value such that a normal random variable has a probability $2(1-p)$ to exceed its mean by more than $\pm z_p\sigma$ \cite{Kay}.  {\color{rev} The finite-sample approximation to the RC uncertainty, $\overline{\Delta \hat{b}(t)}_K \approx \Delta \hat{b}(t)_{\text{RC}}$ is obtained by replacing the spatial averages in Eq.\,(\ref{eqnObsRC}) by their finite-size counterparts, $\langle \overline{J_y}(t)_K\rangle$ and $\langle \overline{J^2_y}(t)_K\rangle$. As these sample averages are random variables themselves, $\overline{\Delta \hat{b}(t)}_K $ carries an intrinsic dispersion}. We quantify this dispersion by computing the standard deviation of  {\color {rev} $\overline{\Delta \hat{b}(t)}_K$, say, $\sigma_{\Delta \hat{b}}(t)$}, which is obtained through the propagation of error formula, by accounting for the variability of the sample averages in the numerator and denominator. Let us identify $x_1 \equiv \partial \langle  J_y(\vec{r}, t) \rangle/ \partial b$ and $x_2 \equiv \langle J^2_y(\vec{r}, t) \rangle -\langle J_y(\vec{r}, t)\rangle^2 $ . We can then write: 
\begin{eqnarray}
\sigma_{\Delta \hat{b}}^2 (t) =\bigg[ \bigg(\frac{\partial \,\Delta \hat{b}(t)_{\text{RC}}}{\partial \,\bar{x}_1 } \bigg|_{(\overline{x}_1)_K} \bigg)^2 \frac{\sigma_{\bar{x}_1}^2(t)}{K} + \bigg(\frac{\partial \, \Delta \hat{b}(t)_{\text{RC}}}{\partial \bar{x}_2 }\bigg|_{(\overline{x}_2)_K} \bigg)^2 \frac{\sigma_{\bar{x}_2}^2(t)}{K} \bigg], 
\label{sigmaDb}
\end{eqnarray}
where $\sigma_{\bar{x}_1}(t)$ and $\sigma_{\bar{x}_2}(t)$ correspond to the infinite sample standard deviations of $x_1$ and $x_2$, 
respectively\footnote{An alternative way to assess the RC protocol performance would be to compute the estimation precision $K$ times, each time before updating the qubit positions, resulting in the set $\{ \Delta \hat{b}(\vec{r}^{\;(1)},t),\ldots,\Delta \hat{b}(\vec{r}^{\;(K)},t)  \}$. The average performance and its deviation would then be obtained through the sample mean and variance of the set. We have verified that this procedure yields similar results for $\Delta \hat{b}(t)_{\text{RC}}$ and $\sigma_{\Delta \hat{b}}(t)$ as the one we considered, in the limit where fluctuations around the sample mean value are small enough, a condition which is satisfied in the small-$\eta $ limit we are interested in. Note, however, that the notion of a spatially averaged density matrix $\bar{\rho}(t)$, or operator $\langle \bar{\mathcal{O}}(t) \rangle$, are no longer useful in this approach.}. In what follows, we make our analysis more concrete by focusing on both GHZ and OAT initial states.  

\subsection{GHZ scaling}

{\color{rev} Thanks to its high degree of symmetry, and shown in \ref{NIS},} the GHZ state is insensitive to whether the dephasing noise is non-classical, hence $\varphi_0(t) = 0 = \bar{\varphi}_0(t)$.  Also, for spatially correlated non-Markovian dephasing, the QFI of a GHZ state may be easily inferred \cite{Dorner2012} to take the form {\color{rev} $F_Q(\tau)=N^2 \tau T e^{-\gamma(\vec{r},t)}$,} with $\gamma(\vec{r},\tau)= \sum_{n,m=1}^N \kappa_{nm}(\vec{r},\tau)/2$. By the quantum CRB of Eq.\,(\ref{qcr}), this leads to the lower bound $\Delta \hat{b}(\tau)^2 \geq e^{ \gamma(\vec{r},\tau)}/( N^2 T \tau)$. Similarly to the noiseless case, this bound can be saturated by a parity measurement \cite{SmerziRMP}. For the RC noise model, the QFI also involves an average over the spatial degrees of freedom, leading to 
\begin{equation}
\Delta\hat{b}(\tau)_{\text{RC}} = \frac{ \sqrt{ \mathbb{E} \{ e^{\gamma(\vec{r},\tau)} \}}}{N \sqrt{T \tau}}.
\label{fulldbGHZ}
\end{equation}
The above corresponds to the quantum CRB as obtained from the spatially averaged reduced density operator $\bar{\rho}(\tau)$ being a GHZ state. Let $y \equiv y(\vec{r},t)= e^{ \gamma(\vec{r},t)}$, with $\sigma_{\bar{y}}^2(t)= \mathbb{E}\{ y^2(\vec{r},t) \} - \mathbb{E}\{ y(\vec{r},t)\}^2 $. {\color{rev} In a randomized Ramsey protocol of finite sample size $K$, we have $\Delta\hat{b}(\tau)_{\text{RC}} \approx \overline{\Delta \hat{b}(t)}_K$. The approximate RC uncertainty and its standard deviation} $\sigma_{\Delta \hat{b}}(\tau)$ are given by
{\color{rev}
\begin{eqnarray}
 \overline{\Delta \hat{b}(t)}_K =  \frac{ \sqrt{\overline{y(\tau)}_K}}{N \sqrt{T \tau}}, \qquad 
\sigma_{\Delta \hat{b}}(\tau) =\sqrt{ \bigg(\frac{\partial \Delta\hat{b}(\tau)_{ \text{RC} }}{\partial \bar{y} }\bigg|_{\overline{y}_K}\bigg)^2 \;\frac{\sigma_{\bar{y}}^2(\tau)}{K}} .
\label{sigmaDbGHZ}
\end{eqnarray} }
{\color {rev}In the small-$\eta$ regime the dynamics become asymptotically spatially local, and we can further approximate 
\begin{eqnarray}
\overline{\Delta \hat{b}(t)}_K \approx \Delta\hat{b}(\tau)_{\eta= 0} =  \frac{ \sqrt{ e^{ N \bar{\kappa}_0^2/4\; (\omega_c \tau)^2}}}{N \sqrt{T \tau}} 
\label{dbGHZ}.
\end{eqnarray}}
Minimizing with respect to $\tau$ yields the following optimal measurement time and precision scaling,
\begin{eqnarray}
\omega_c \tau_{\text{opt,RC}}^{\text{GHZ}} =(2 \bar{\kappa}_0)^{-1}\;N^{-1/2}, \qquad 
{\color{rev}
\Delta \hat{b}_{\text{opt,RC}}^{\text{GHZ}} \geq (\omega_c/T)^{1/2}\;\bar{\kappa}_0^{1/2}\;e^{-1/4}\;N^{-3/4}.}
\label{GHZan}
\end{eqnarray}
This is super-classical Zeno scaling, which is the optimal asymptotic scaling for spatially uncorrelated non-Markovian noise \cite{Huelga,Macies2015}. {\color{rev} In Fig.\,\ref{dBcoll}, the finite-$K$ approximation to uncertainty $\Delta \hat{b}(\tau)_{\text{RC}}$  is shown for two RC protocols of sample size $K=20$ and differing smallness parameter $\eta=0.1, \eta=0.05$. Their performance is contrasted with the local noise limit function of Eq.\,(\ref{dbGHZ}), leading to Zeno like scaling. In both cases, the $\overline{\Delta \hat{b}(t)}_K$ behavior is close to the spatially uncorrelated uncertainty in the short-time limit and, importantly, at the optimal measurement time $\tau_{\text{opt,RC}}^{\text{GHZ}}$.} However, the curves begin to separate at longer times.  This deviation from asymptotic behavior occurs at smaller values of $\omega_c t$ when $\eta$ is bigger, as expected. For fixed sample size $K$, the uncertainty dispersion $\sigma_{\Delta \hat{b}}(t)$ grows with $\eta$ as well.  However,  given the possibility to perform a large enough number of iterations $K$, Eq.\,(\ref{dbGHZ}) implies that $\sigma_{\Delta \hat{b}}(t) $ can be reduced to arbitrarily low values in principle. 

\begin{figure}[t!]
    \centering
    \includegraphics[width=12cm]{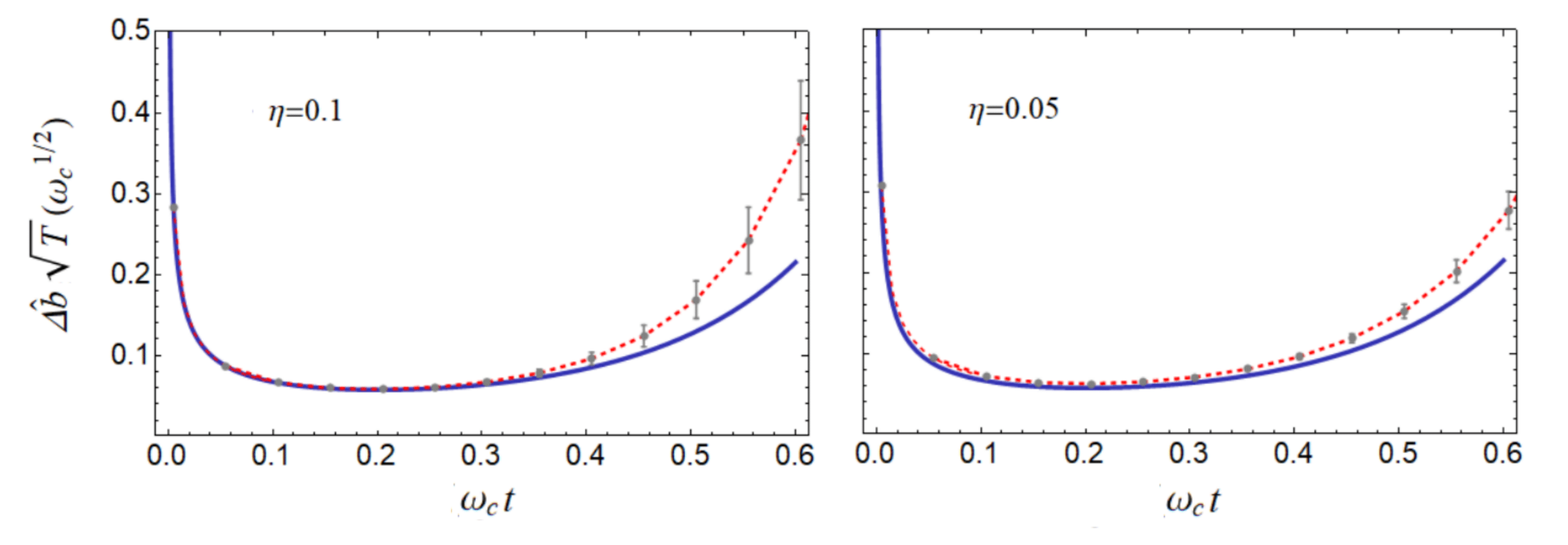}
    \vspace*{-3mm}
    \caption{{\bf Initial GHZ state with $\bm{N=50}$ qubits under 1D RC bosonic dephasing.}
    Short-time behavior of the estimation precision with $\eta=0.1$ (left) and $\eta=0.05$ (right). Blue (solid): spatially uncorrelated limit $\eta=0$. Red (dashed): RC numerical simulation for $K=20$ runs. {\color{rev}The grey bars indicate the dispersion $\sigma_{\Delta \hat{b}} (t)$, computed from Eq.\,(\ref{sigmaDbGHZ}), with $\sigma^2_{\bar{y}}(\tau)$ estimated from the mean square error}.
    A Gaussian cutoff 
    is assumed for the noise spectral density, and $s=3,\alpha=1$.}
    \label{dBcoll}
\end{figure}

\subsection{OATS scaling}   

Consider now an initial OATS as Eq.\,(\ref{oat}), with the squeezing and rotation angles given by Eq.\,(\ref{ideal}) set to minimize the noiseless uncertainty. We now show that, in the RC setting we discussed, the influence of non-collective noise is less detrimental than the collective case first analyzed \cite{FelixPRA} and does not preclude superclassical precision scaling. In the $\eta = 0$ limit of the RC model, the fully spatially local expectation values $\mathbb{E}\{\langle J_y(t)\rangle \}, \mathbb{E}\{ \langle J_y^2(t)\rangle \}$ can be computed analytically. We get:
\begin{eqnarray*}
\mathbb{E}\{\langle J_y(\vec{r},t)\rangle\}_{\eta=0} &= \frac{N}{2} \sin(b\,t)\; e^{-\bar{\kappa}_0(t)/2} \cos\left(\theta/2\right)^{N-1} ,\\
\mathbb{E}\{\langle J^2_y(\vec{r},t)\rangle\}_{\eta=0} &= \bigg\{ \frac{N}{4} + e^{-\bar{\kappa}_0(t)} \frac{N}{8}(N-1) 
\bigg[ \bigg(\frac{1}{2} \sin^2 (\beta) (1+ \cos\left(\theta\right)^{N-2}) + \cos(\beta)^2 \\
&- \sin\left(\theta/2\right) \cos\left(\theta/2\right)^{N-2} \sin(2\beta)   \bigg) 
- \bigg( \frac{1}{2} \sin(\beta)^2 
 + \cos\left(\theta\right)^{N-2} (\cos(\beta)^2 \\
& + \frac{1}{2} \sin(\beta)^2) +  \sin\left(\theta/2\right) \cos\left(\theta/2\right)^{N-2} \sin(2\beta) \bigg)  \cos (2 b\,t)  \bigg] \bigg\}.
\end{eqnarray*}
To minimize the uncertainty, we take $b\,t = n \pi$ \cite{FelixPRA}. It follows that $\mathbb{E}\{\langle J_y(t)\rangle\}_{\eta=0}=0$, so the ensemble-averaged variance simply becomes $\mathbb{E}\{ \Delta J^2_y(\vec{r},t)  \}_{\eta=0} = \mathbb{E}\{\langle J^2_y(t, \vec{r})\rangle\}_{\eta=0} $. We can then cast $\mathbb{E} \{\Delta J_y(\vec{r},t)^2\}$ into a form reminiscent of the noiseless OATS variance \cite{Kita1993}:
\begin{eqnarray}
\mathbb{E}\{ \Delta J^2_y(\vec{r},t)  \}_{\eta=0}&= \frac{N}{4} \bigg\{ 1 + e^{- \kappa_0(t)} \frac{1}{4} (N-1) \left[A+ \sqrt{A^2+B^2} \cos(2 \beta + 2 \delta) \right] \bigg\} \label{JysqOATS},
\end{eqnarray}
where $A , B$ and $\delta$ are given by Eq.\,(\ref{ABdelta}). Eq.\,(\ref{JysqOATS}) would be equal to the noiseless expression for $\Delta J_y(t)$ were it not for the $e^{-\bar{\kappa}_0(t)}$ decay factor. It can then be shown that the squeezing and rotation angles minimizing the variance in Eq.(\ref{JysqOATS}) are the same as the optimal ones in the absence of noise, Eq.\,(\ref{ideal}). This leads to 
\begin{eqnarray}
\Delta \hat{b}(\tau)^2_{\text{RC}} & \approx \Delta \hat{b}(\tau)^2_{\eta=0} = \frac{1}{N^2 T \tau } \bigg[ N \left(e^{\bar{\kappa}_0^2 (\omega_c \tau)^2}-1 \right) +  
\frac{3^{2/3}}{2}  N^{1/3} \bigg ] .
\label{dbOATS}
\end{eqnarray} 
In the limit $N \gg 1$ and $\eta \rightarrow 0$, we can show that the optimal measurement time and the corresponding uncertainty obey the following limiting scaling:
\begin{eqnarray}
\omega_c \tau_{\text{opt,RC}}^{\text{OATS}} =3^{1/3}2^{1/6}\,\bar{\kappa}_0^{-1}\; N^{-1/3}, \qquad
\Delta \hat{b}_{\text{opt,RC}}^{\text{OATS}} = 3^{1/6}2^{3/12}\,(\omega_c/T)^{1/2}\, \bar{\kappa}_0^{1/2}\;N^{-2/3},
\label{OatsRC}
\end{eqnarray}
which means we can surpass the limit set by the SQL by a potentially significant $N^{-1/6}$ factor in principle. By contrast, the loss of scaling precision in the RC protocol when compared to the noiseless scenario, for which $\Delta \hat{b}_{\text{opt}}^{\text{OATS}} \propto N^{-5/6}$, is of $N^{-1/6}$ as well. As a by-product of this analysis, we thus obtain the OATS uncertainty scaling under local non-Markovian noise, which has not been previously reported in the literature to the best of our knowledge.

The asymptotic scaling results of Eq.\,(\ref{OatsRC}) are now contrasted with an  RC simulation at finite sample size $K$ and smallness parameter $\eta$. For the $i$th iteration of the protocol, let the qubits be located at positions $\vec{r}^{\,(i)}$. Since the reduced state $\rho(\vec{r}^{\,(i)},t)$ is entangled, we resort to a cumulant expansion over the systems' degrees of freedom to evaluate the mean values $\langle J_y(\vec{r}^{\,(i)},t) \rangle$ and $\langle J^2_y(\vec{r}^{\,(i)},t) \rangle$ that enter the sample means in Eq.\,(\ref{finiteK}), as detailed in \ref{noncolcum}. Importantly, for an RC protocol with small enough $\eta$, correlations between different qubits are vanishingly small, which makes the approximate expressions for $\langle J_y( \vec{r}^{\,(i)},t) \rangle$ and $\langle J_y^2(\vec{r}^{\,(i)},t) \rangle$ remarkably accurate {\em at all times}. This is to be compared to the collective case, where the presence of strong spatial correlations induced more involved behavior, which the cumulant expansion failed to capture beyond the short-time limit \cite{FelixPRA}. {\color{rev}Finally, the finite-sample approximation to the RC uncertainty, $\Delta \hat{b}(t)_{\text{RC}} \approx \overline{\Delta \hat{b}}(t)_K$, is obtained by replacing the sample means $\langle \overline{J_y}(t)_K \rangle $, $\langle \Delta \overline{J_y}(t)_K \rangle$  in Eq.\,(\ref{eqnObsRC})}. In Fig.\,\ref{dBOATScoll} we show {\color{rev}$\overline{\Delta \hat{b}}(t)_K$} against the spatially uncorrelated $\eta=0$ curve, which obeys Eq.\,(\ref{OatsRC}) exactly. Surprisingly, unlike the GHZ case, agreement with the asymptotic behavior at longer times is excellent already for $\eta=0.1$, just an order of magnitude below unity. Further decreasing $\eta$ only decreases the uncertainty dispersion $\sigma_{\Delta \hat{b}}(t)$ at fixed sample size. {\color{rev} Subject to implementation constraints, we can then achieve a similar level of precision in estimating $\overline{\Delta \hat{b}}(t)_K$ with bigger values of $\eta$ by increasing $K$.} 

\begin{figure}[t!]
    \centering
    \includegraphics[width=12cm]{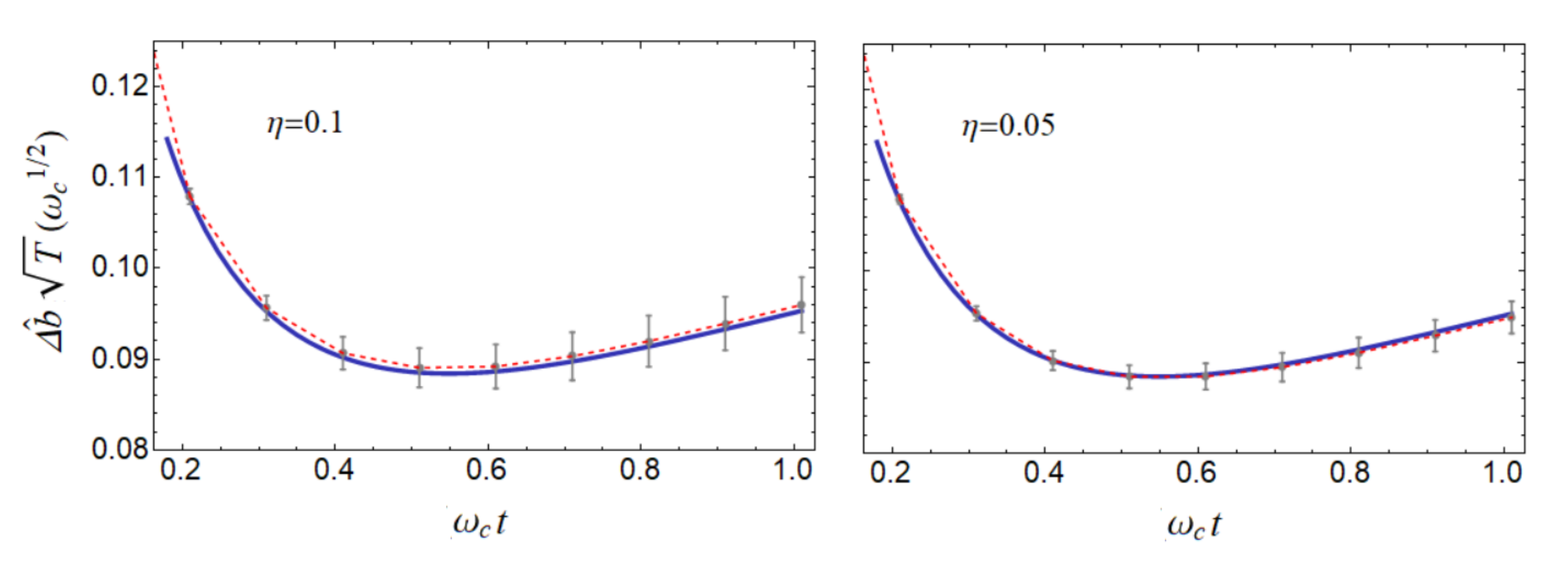}
    \vspace*{-3mm}
    \caption{{\bf Initial OAT with $\bm{N=50}$ qubits under 1D RS bosonic dephasing.}
    Short-time behavior of the estimation precision 
    with $\eta=0.1$ (left) and $\eta=0.05$ (right). Blue (solid): spatially uncorrelated limit $\eta=0$. Red (dashed): RC numerical simulation for 	$K=20$ runs. The grey bars indicate the dispersion $\sigma_{\Delta \hat{b}}^2 (t)$, computed from Eq.\,(\ref{sigmaDb}).  Noise parameters as in Fig.\,\ref{dBcoll}.}
    \label{dBOATScoll}
\end{figure}

\section{Conclusion}
\label{conclusion}

In conclusion, we have shown how relaxing the collective noise assumption can result in improved asymptotic precision scaling for Ramsey protocols subject to spatiotemporally correlated dephasing from a Gaussian quantum bath during the interrogation period. Placing the qubits in two separate clusters yields a constant factor advantage with respect to the collective case, whereas randomizing their positions can effectively spatially de-correlate the qubits on average -- provided that the characteristic length scale of the noise is small relative to the ``width'' of the probes' spatial randomization profile. Superclassical precision scaling then emerges for both OAT and GHZ states in the appropriate parameter regime, with the latter reaching the optimal Zeno scaling expected for local non-Markovian noise. For both these two non-collective scenarios, the basic insight is that breaking permutational symmetry, by allowing the probe qubits to be placed in different positions, can {\em lower the noise spatial correlations} compared to the collective case, resulting in enhanced precision. This reinforces the conclusion that the spatial structure of the noise plays a key role in metrological settings, beyond the Markovian regime considered in \cite{Jeske2014}.

While our emphasis in this work has been on a general, implementation-independent analysis, additional investigation is needed to determine whether the randomized metrology approach we introduced may have practical relevance to metrological platforms of interest, where spatial and temporal correlations may be significant. On the one hand, signatures of both spatial and temporal correlations in intrinsic noise have been reported in trapped-ion experiments \cite{BlattNoise,Mike}. On the other end, both site-selective loading and ion transport have been experimentally demonstrated in segmented traps, with reconfiguration of ion strings being also possible thanks to advanced micro-fabricated design \cite{Traps}. In the light of that, it would seem especially compelling to explore the potential of randomized strategies in the context of trapped-ion frequency estimation (or magnetometry), thereby extending the work of \cite{ruster}.

At a fundamental level, it remains an outstanding question to determine whether Heisenberg scaling may be attainable in the non-Markovian regime in the presence of spatially correlated non-classical noise and, if so, by what kind of initial states. Setups involving a pre-measurement ``countertwisting'' operation \cite{Monika,SchulteEchoes} or alternating sequences of squeezing and rotation pulses \cite{carrasco} have been recently proposed in the closely related context of phase estimation, in which case the effect of noise occurs primarily during the OAT state preparation. Notably, Heisenberg scaling has been experimentally reported using such a ``time-reversal-based'' metrology \cite{VladanSatin}. It may be interesting to determine whether similar generalized protocols remain useful for frequency estimation in our context -- with the dominant noise contribution appearing, as we described, during the encoding period instead. Finally, it is natural to ask whether there are may be more efficient ways to leverage spatial correlations as a resource, potentially yielding greater metrological gain. These are some of the avenues we are currently exploring, and on which we plan on reporting elsewhere.  

\section*{Acknowledgments}

It is a pleasure to thank Vladan Vuletic for insightful discussions and Vincent Flynn and Maryam Mudassar for a critical reading of the manuscript.  L.V. also acknowledges valuable exchange with John Gough on issues related to classicality of open quantum dynamics. Work at Dartmouth was partially supported by the US National Science Foundation through Grants No. PHY-1620541 and No. PHY-2013974. 

\appendix


\section{Additional technical details for general ZMGSD dynamics}
\label{ZMGSDAp}

\subsection{Exact evaluation of the reduced dynamics}
\label{app:exact}

Let ${H}(t)=\sum_{i}O_i \otimes B_i(t) $ and $\tilde{H}(t)=\sum_j\tilde{O}_j \otimes \tilde{B}_j(t)$ be Hamiltonians on $\mathcal{H}_S\otimes\mathcal{H}_B$, where $\{B_i(t)\}$ and $\{\tilde{B}_j(t)\}$ are bath operators and $\{O_i\}$ and $\{\tilde{O}_j\}$ are system operators. Crucially, let the system operators be \emph{mutually commuting}, that is, $[O_i,O_{i'}]=[\tilde{O}_j,\tilde{O}_{j'}]=[O_i,\tilde{O}_j]=0$, for all $i,i',j,j'$,  as required for a dephasing model. Let $U(t)$ and $\tilde{U}(t)$ be the (joint) unitary operators generated by $H(t)$ and $\tilde{H}(t)$, respectively.  

\medskip 

{\bf Claim:} Under the above assumptions, we have
\begin{eqnarray}
\label{eq::claim}
&\expect{\tilde{U}(t)^\dag U(t) }=\\&\exp\bigg\{\!\sum_{v,\tilde{v}=0}^{\infty'}\!i^{\tilde{v}}(-i)^v \!\!\int_0^t\!\!\!ds_1\ldots\!\int_0^{s_{\tilde{v}-1}}\!\!\!ds_{\tilde{v}}\int_0^t\!\!\!ds_1'\ldots\!\int_0^{s_{v-1}'}\!\!\!ds_v'
\cumu{\tilde{H}(s_{\tilde{v}})\ldots\tilde{H}(s_1)H(s_1')\ldots H(s_v')}\bigg\},\nonumber
\end{eqnarray}
where $\expect{\cdot}$ and $\cumu{\cdot}$ are moment and cumulant averages acting on the bath operators  \cite{Kubo}, and the prime over the summation indicates that either $v\neq 0$ or $\tilde{v}\neq 0$.

\smallskip

{\bf Proof.} First, we expand the unitary operators using a Dyson series,
\begin{eqnarray*}
U(t)=\sum_{m=0}^\infty\frac{(-i)^m}{m!}\int_0^t \!d\vec{s}_m^{\;'}\,\mathcal{T}_+H(s_1')\ldots H(s_m'),\quad \;
\tilde{U}(t)^\dag=\sum_{n=0}^\infty\frac{i^n}{n!}\int_0^t \!d\vec{s}_n\,\mathcal{T}_-\tilde{H}(s_1)\ldots \tilde{H}(s_n),
\end{eqnarray*}
where we have used the shorthand $\int_0^td\vec{s}_m^{\;'}\equiv \int_0^t\!ds_1'\ldots\int_0^{t}\!ds_m'$ and $\int_0^td\vec{s}_n\equiv \int_0^t\!ds_1\ldots\int_0^{t}\!ds_n$.
Substituting these expressions into $\expect{\tilde{U}(t)^\dag U(t) }$ yields,
\begin{eqnarray}
\expect{\tilde{U}(t)^\dag U(t) }\!
=\!\!\sum_{m,n=0}^\infty\!\!\frac{(-i)^m}{m!}\frac{i^n}{n!}\int_0^t \!\!d\vec{s}_n\!\!\int_0^t \!\!d\vec{s}_m^{\;'}
\expect{\mathcal{T}_-\tilde{H}(s_1)\ldots \tilde{H}(s_n)\mathcal{T}_+\,H(s_1')\ldots H(s_m')}.
\label{eq::UdagU}
\end{eqnarray}
We next write the moments above in terms of cumulants. Let $\{P_1,\ldots,P_q\}$ be a partition of the set 
$\{\tilde{H}(s_1),\dots,\tilde{H}(s_n), H(s_1'),\ldots,H(s_m')\}$ into $q$  subsets and let $\mathbb{P}_q(m+n)$ be the set of all possible $\{P_1,\ldots,P_q\}$. By the moment-cumulant relation,
it follows that 
\begin{eqnarray}\nonumber
&\int_0^td\vec{s}_n\int_0^td\vec{s}_m^{\;'}
\expect{\mathcal{T}_-\tilde{H}(s_1)\ldots \tilde{H}(s_n)\mathcal{T}_+\,H(s_1')\ldots H(s_m')}=\\
&\int_0^td\vec{s}_n\int_0^td\vec{s}_m^{\;'}\sum_{q=1}^{m+n}\sum_{\{P_1,\ldots,P_q\}\in\mathbb{P}_q(m+n)}\cumu{\mathcal{T}_-\mathcal{T}_+\prod_{H_1\in P_1}H_1}\ldots
\cumu{\mathcal{T}_-\mathcal{T}_+\prod_{H_q\in P_q}H_q}.\label{eq::momentcumulant}
\end{eqnarray}
Since $P_1,\ldots,P_q\subseteq\{\tilde{H}(s_1),\dots,\tilde{H}(s_n), H(s_1'),\ldots,H(s_m')\}$, the cumulants above contain subsets of the $\tilde{H}(s_i)$'s and $H(s_i')$'s. Each cumulant can be expressed in terms of moments of the $\tilde{H}(s_i)$'s and $H(s_i')$'s. Within each moment, the ordering symbols in the expression above indicate that the $\tilde{H}(s_i)$'s are ordered by $\mathcal{T}_-$ and are always to the left of the $H(s_i')$'s, which are ordered by $\mathcal{T}_+$. For example, a third-order cumulant containing $\tilde{H}(s_i)$, $\tilde{H}(s_j)$ and  $H(s_k')$ with $s_i>s_j$ takes the form
\begin{eqnarray*}
\cumu{\tilde{H}(s_j)\tilde{H}(s_i)H(s_k')}=&\,\expect{\tilde{H}(s_j)\tilde{H}(s_i)H(s_k')}-\expect{\tilde{H}(s_j)\tilde{H}(s_i)}\expect{H(s_k')}
\\&-\expect{\tilde{H}(s_j)H(s_k')}\expect{\tilde{H}(s_i)}-\expect{\tilde{H}(s_i)H(s_k')}\expect{\tilde{H}(s_j)}
\\&+2\expect{\tilde{H}(s_j)}\expect{\tilde{H}(s_i)}\expect{H(s_k')}.
\end{eqnarray*}

Because the time-integrals all have the same bounds and the integrands are time-ordered, Eq.\,(\ref{eq::momentcumulant}) is invariant under permutations of the $\{s_1,\ldots,s_n\}$ and  $\{s_1',\ldots,s_m'\}$. This means that there are many duplicate terms in Eq.\,(\ref{eq::momentcumulant}. In particular, two terms corresponding to partitions $\{P_1,\ldots,P_q\}$ and  $\{Q_1,\ldots,Q_q\}$ are equivalent if $|P_1|=|Q_1|,\ldots,|P_q|=|Q_q|$, that is, the subsets contain the same number of elements, and both $P_j$ and $Q_j$ contain $\tilde{v}_j$ 
$\tilde{H}(s_i')$'s and $v_j$ $H(s_i')$'s with $|P_j|=v_j+\tilde{v}_j$. The number of ways to assign the $n$ $\tilde{H}(s_i)$'s into subsets of sizes $\{\tilde{v}_1,\ldots,\tilde{v}_q\}$ is given by the multinomial coefficient 
\begin{eqnarray*}
{n\choose \tilde{v}_1,\tilde{v}_2,\ldots,\tilde{v}_q} =\frac{n!}{\tilde{v}_1!\tilde{v}_2!\ldots\tilde{v}_q!}.
\end{eqnarray*}
Similarly, the number of ways to assign the $m$ $\tilde{H}(s_i')$'s into subsets of sizes $\{{v}_1,\ldots,{v}_q\}$ is given by
\begin{eqnarray*}
{m\choose {v}_1,{v}_2,\ldots,{v}_q}=\frac{n!}{{v}_1!{v}_2!\ldots {v}_q!}.
\end{eqnarray*}
This means that, for every partition $\{P_1,\ldots,P_q\}$ with  $|P_1|=v_1+\tilde{v}_1,\ldots,|P_q|=v_q+\tilde{v}_q$, there are ${n \choose \tilde{v}_1,\tilde{v}_2,\ldots,\tilde{v}_q}{m \choose {v}_1,{v}_2,\ldots,{v}_q}$ equivalent terms. By substituting Eq.\,(\ref{eq::momentcumulant}) into Eq.\,(\ref{eq::UdagU}) and combining the equivalent terms, we can then write
\begin{eqnarray}
&\expect{\tilde{U}(t)^\dag U(t) }\nonumber
=1+\!\!\sum_{m,n=0}^{\infty'} \!\frac{(-i)^m}{m!}\frac{i^n}{n!}\int_0^t \!\!d\vec{s}_n\int_0^t \!\!d\vec{s}_m^{\;'}\sum_{q=1}^{m+n}\sum_{\{|P_1|,\dots,|P_q|\}\vdash (n+m)}
\sum_{\tilde{v}_1,v_1\atop \tilde{v}_1+v_1=|P_1|} \!\!\!\ldots \!\!\!\sum_{\tilde{v}_q,v_q \atop \tilde{v}_q+v_q=|P_q|}\\
&{n\choose \tilde{v}_1,\tilde{v}_2,\ldots,\tilde{v}_q}{m \choose {v}_1,{v}_2,\ldots,{v}_q}
\big\langle\!\big\langle\mathcal{T}_+\tilde{H}(s_{\tilde{v}_1})\ldots\tilde{H}(s_{1})\mathcal{T}_-H(s_1')\ldots H(s_{v_1}')\big\rangle\!\big\rangle_B\ldots\nonumber\\
&\cumu{\mathcal{T}_+\tilde{H}(s_{\tilde{V}+\tilde{v}_q})\ldots\tilde{H}(s_{\tilde{V}+1})\mathcal{T}_-H(s_{V+1}')\ldots H(s_{V+v_q}')},
\label{eq::UdagU2}
\end{eqnarray}
where the symbol $\vdash$ in the second sum under the time integrals means it is taken over all integer partitions $\{|P_1|,\dots,|P_q|\}$ of $m+n$ into $q$ parts, and $V\equiv\sum_{r=1}^{q-1}v_r$ and $\tilde{V}\equiv\sum_{r=1}^{q-1}\tilde{v}_r$. Note that $\{|P_1|,\dots,|P_q|\}=\{|P_{\sigma(1)}|,\dots,|P_{\sigma(q)}|\}$, where $\sigma$ is a permutation of $\{1,\ldots, q\}$. By collecting all $\tilde{v}_1,\ldots,\tilde{v}_q$ that sum to $n$ and all  ${v}_1,\ldots,{v}_q$ that sum to $m$, we find
\begin{eqnarray*}
\expect{\tilde{U}(t)^\dag U(t) }
=&1+\!\!\sum_{m,n=0}^{\infty'}\frac{(-i)^m}{m!}\frac{(i)^n}{n!}\int_0^t \!\!d\vec{s}_n\int_0^t \!\!d\vec{s}_m^{\;'}\sum_{q=1}^{m+n}\frac{1}{q!} \!\!\sum_{\tilde{v}_1,\ldots,\tilde{v}_q \atop \tilde{v}_1+\ldots +\tilde{v}_q=n}\sum_{{v}_1,\ldots,{v}_q \atop {v}_1+\ldots +{v}_q=m}\\
&\!\!\!{n \choose \tilde{v}_1,\tilde{v}_2,\ldots,\tilde{v}_q} \!{m\choose {v}_1,{v}_2,\ldots,{v}_q}\!
\big\langle\!\big\langle\mathcal{T}_+\tilde{H}(s_{\tilde{v}_1})\ldots\tilde{H}(s_{1})\mathcal{T}_-H(s_1')\ldots H(s_{v_1}')\big\rangle\!\big\rangle_B\ldots\\
&\!\!\!\cumu{\mathcal{T}_+\tilde{H}(s_{\tilde{V}+\tilde{v}_q})\ldots\tilde{H}(s_{\tilde{V}+1})\mathcal{T}_-H(s_{V+1}')\ldots H(s_{V+v_q}')}.
\end{eqnarray*}

This expression picks up a factor of $1/q!$ since the second and third sums count all $q!$ configurations of $\tilde{v}_1,\ldots,\tilde{v}_q$ and ${v}_1,\ldots,{v}_q$ such that 
$\tilde{v}_1+{v}_1=|P_{\sigma(1)}|,\ldots,\tilde{v}_q+{v}_q=|P_{\sigma(q)}|$ as distinct terms, unlike Eq.\,(\ref{eq::UdagU2}). Next, we change the variables of integration from $\{s_1,\ldots,s_n\}$ to $\{s_{1,1},\ldots s_{1,\tilde{v}_1},\ldots, s_{q,1},\ldots s_{q,\tilde{v}_q} \}$ and from  $\{s_1',\ldots,s_m'\}$ to
 $\{s_{1,1}',\ldots s_{1,{v}_1}',\ldots, s_{q,1}',\ldots s_{q,{v}_q}' \}$. Letting $\int_0^td\vec{s}_{j,j}\equiv\int_0^td{s}_{j,1}\ldots\int_0^td{s}_{j,\tilde{v}_j}$ and
$\int_0^td\vec{s}_{j,j}^{\; '}\equiv\int_0^td{s}_{j,1}'\ldots\int_0^td{s}_{j,{v}_j}'$, we expand the multinomial coefficients to produce
\begin{eqnarray*}
\expect{\tilde{U}(t)^\dag U(t) }
=&1+\sum_{m,n=0}^{\infty'}\sum_{q=1}^{m+n}
\frac{1}{q!}\sum_{\tilde{v}_1,\ldots,\tilde{v}_q \atop\tilde{v}_1+\ldots +\tilde{v}_q=n}\;\sum_{{v}_1,\ldots,{v}_q \atop {v}_1+\ldots +{v}_q=m}\\
&\!\!\!\int_0^t \!\!d\vec{s}_{1,1}\!\!\int_0^td\vec{s}_{1,1}^{\; '}\,\frac{i^{\tilde{v}_1}}{\tilde{v}_1!}\frac{(-i)^{{v}_1}}{{v}_1!}
\big\langle\!\big\langle\mathcal{T}_+\tilde{H}(s_{1,\tilde{v}_1})\ldots\tilde{H}(s_{1,1})\mathcal{T}_-H(s_{1,1}')\ldots H(s_{1,v_1}')\big\rangle\!\big\rangle_B \ldots\\
&\!\!\!\int_0^t \!\!d\vec{s}_{q,q}\!\!\int_0^td\vec{s}_{q,q}^{\; '}\,\frac{i^{\tilde{v}_q}}{\tilde{v}_q!}\frac{(-i)^{{v}_q}}{{v}_q!}
\big\langle\!\big\langle\mathcal{T}_+\tilde{H}(s_{q,\tilde{v}_q})\ldots\tilde{H}(s_{q,1})\mathcal{T}_-H(s_{q,1}')\ldots H(s_{q,v_q}')\big\rangle\!\big\rangle_B
\end{eqnarray*}

Collecting all terms with the same $q$ yields
\begin{eqnarray*}
&\big\langle\tilde{U}(t)^\dag U(t) \big\rangle_B
=1+ \\ 
& \sum_{q=1}^{\infty}\frac{1}{q!}\sum_{v_1,\tilde{v}_1=0}^{\infty\,'}
\int_0^t\!\!\!d\vec{s}_{1,1}\!\!\int_0^t\!\!\!d\vec{s}_{1,1}^{\; '}\,\frac{i^{\tilde{v}_1}}{\tilde{v}_1!}\frac{(-i)^{{v}_1}}{{v}_1!}
\big\langle\!\big\langle\mathcal{T}_+\tilde{H}(s_{1,\tilde{v}_1})\ldots\tilde{H}(s_{1,1})\mathcal{T}_-H(s_{1,1}')\ldots H(s_{1,v_1}')\big\rangle\!\big\rangle_B\\
&\qquad \!\!
\ldots\sum_{v_q,\tilde{v}_q=0}^{\infty '}\int_0^td\vec{s}_{q,q}\!\!\int_0^td\vec{s}_{q,q}^{\; '}\,\frac{i^{\tilde{v}_q}}{\tilde{v}_q!}\frac{(-i)^{{v}_q}}{{v}_q!}
\big\langle\!\big\langle\mathcal{T}_+\tilde{H}(s_{q,\tilde{v}_q})\ldots\tilde{H}(s_{q,1})\mathcal{T}_-H(s_{q,1}')\ldots H(s_{q,v_q}')\big\rangle\!\big\rangle_B\\
=&\sum_{q=0}^{\infty}\frac{1}{q!} \bigg(\sum_{v,\tilde{v}=0}^{\infty\,'}\int_0^td\vec{s}_{v}\!\!\int_0^td\vec{s}_{\tilde{v}}^{\; '}\,\frac{i^{\tilde{v}}}{\tilde{v}!}\frac{(-i)^{{v}}}{{v}!}
\big\langle\!\big\langle\mathcal{T}_+\tilde{H}(s_{\tilde{v}})\ldots\tilde{H}(s_{1})\mathcal{T}_-H(s_{1}')\ldots H(s_{v}')\big\rangle\!\big\rangle_B \bigg)^q\\
=&\sum_{q=0}^{\infty}\frac{1}{q!} \bigg(\sum_{v,\tilde{v}=0}^{\infty\,'}i^{\tilde{v}} (-i)^{{v}}\!\int_0^t \!\!ds_1 \ldots\!\int_0^{s_{\tilde{v}-1}}\!\!\!\!\!\!\!ds_{\tilde{v}} \\
&\hspace*{30mm}\int_0^t\!\!\!ds_1'\ldots\!\int_0^{s_{v-1}'}\!\!\!\!\!ds_v'
\big\langle\!\big\langle\mathcal{T}_+\tilde{H}(s_{\tilde{v}})\ldots\tilde{H}(s_{1})\mathcal{T}_-H(s_{1}')\ldots H(s_{v}')\big\rangle\!\big\rangle_B\bigg)^q,
\end{eqnarray*}
which is the Taylor series expansion of the right-hand side of Eq.\,(\ref{eq::claim}).\hfill$\blacksquare$

\medskip

Eq.\,(\ref{Leigh2}) in the main text follows as a particular case of Eq.\,(\ref{eq::claim}) when the noise operators $\{B_i(t)= \tilde{B}_i(t)\}$ are ZMGSD and we identify $U(t)$ and $\tilde{U}(t)$ with $U_{\vec{\alpha}}(t)$ and $U_{\vec{\beta}}(t)$, respectively. In that case, Gaussianity allows us to truncate the exponent at second order, greatly simplifying the calculations.

{\color{rev}\subsection{On quantum noise-insensitive subspaces}
\label{NIS}

Let us consider the evolution of the paradigmatic GHZ state in the ZMGSD noise framework. Initially, the reduced density matrix is given by
\begin{eqnarray}
\rho_{\text{GHZ}}(0)\!= \frac{1}{2} \left( |\uparrow \rangle \langle \uparrow|^{\otimes N} + |\uparrow \rangle \langle \downarrow|^{\otimes N} + |\downarrow \rangle \langle \uparrow|^{\otimes N} + |\downarrow \rangle \langle \downarrow|^{\otimes N} \right). \nonumber
\end{eqnarray}
Following Eqs.\,(\ref{gamma})-(\ref{phi1}), we see that the diagonal elements show no time evolution, while the anti-diagonal elements have vanishing phases $\varphi_0(t), \varphi_1(t)$ and a non-zero decay rate $\gamma_{\text{GHZ}}(t)= 4\sum_{n,m=1}^N \kappa_{nm}(t)$, that is, 
\begin{eqnarray}
\rho_{\text{GHZ}}(t)\!={\color{rev} \frac{1}{2} ( |\uparrow \rangle \langle \uparrow|^{\otimes N} \!+ e^{-i N b  t}\,e^{-\gamma_{\text{GHZ}}(t)} |\uparrow \rangle \langle \downarrow|^{\otimes N}\!+ e^{i N b  t}\,e^{-\gamma_{\text{GHZ}}(t)} |\downarrow \rangle \langle \uparrow|^{\otimes N} \!+ |\downarrow \rangle \langle \downarrow|^{\otimes N} )}. \nonumber
\end{eqnarray}
Thus, the GHZ state is {\em immune to the non-classical contribution of temporally correlated noise}, encoded by the phases $\varphi_0(t)$ and $\varphi_1(t)$. More generally, we can ask which matrix elements $\langle \vec{\alpha}| \rho(t)|\vec{\beta} \rangle$ in the $z$ basis are ``quantum-noise-insensitive'' (QNI). The answer depends upon the symmetries of the noise. For general non-collective dephasing, there are no further simplifications to Eqs.\,(\ref{gamma})-(\ref{phi1}). If we want both $\varphi_0(t)$ and $\varphi_1(t)$ to vanish, it is easy to see that either $\{\alpha_n=\beta_n, \,\forall\, n\}$ or $\{\alpha_n=-\beta_n, \,\forall \,n\}$. Thus, only $2^{N+1}$  of the $2^{2N}$ matrix elements, the diagonal and anti-diagonal ones, are QNI. For collective and even-odd noise, the higher degree of symmetry results in a greater number of QNI states. Let us work out the math of each of them:

\medskip

\hspace*{7mm}{\bf Collective noise.}  With reference to Sec.\,\ref{sec:general}, the collective dynamic coefficients are index-independent, with the non-vanishing decay and phase being given by
\begin{eqnarray}
\gamma(t)^{\text{coll}}= \kappa(t)  (m-m')^2 
\label{gammacoll},
\quad \varphi_0(t)^{\text{coll}}= \xi(t) (m'^2-m^2)
\label{phi0coll}. 
\end{eqnarray}
Here, $2m=\sum_{n=1}^N \alpha_n, 2m'=\sum_{n=1}^N \beta_n$ and  $\kappa(t), \xi(t)$ are obtained by replacing $B_n(t)$ with $B(t)$ in Eqs.\,(\ref{chi})-(\ref{xi}) and $m, m'$ represent the $z$ angular momentum component of $|\vec{\alpha}\rangle$, $|\vec{\beta}\rangle$. Note that permutation invariance forces $\varphi_1(t)^{\text{coll}}=0$. Thanks to its higher degree of symmetry, the reduced matrix evolution can be described exclusively in terms of the permutation-invariant Dicke states $\{|J,-J\rangle,|J,-J+1\rangle,\ldots, |J,m\rangle,\ldots,|J,J\rangle\}$, where $J= N/2$ is the total angular momentum \cite{FelixPRA}. It is clear from Eq.\,(\ref{gammacoll}) that the QNI matrix elements must have $m=\pm m'$, and are thus of the form $\langle J, m|\rho_0|J, \pm m\rangle$, for $-J \leq m \leq J$. There are  $2(2J+1)$ such elements in the Dicke basis or, noting that each $|J, m\rangle$ consists of the equally weighted sum of all ${2J \choose m+J}$ kets $|\vec{\alpha}\rangle$ such that $\sum_{n=0}^N \alpha_n= 2 m$, $2 {4J \choose 2J} $ such elements in the $z$ basis.

\medskip

{\bf Even-odd noise.} As in the main text, half of the $N=2J$ qubits (the even ones) couple to the environment via operator $B_{\text{e}}(t)$, 
and the other half (the odd ones) couple to the environment via  $B_{\text{o}}(t) \neq B_{\text{e}}(t)$. Permutation invariance is then retained only among qubits belonging to the same cluster, and we can represent ZMGSD even-odd dynamics in terms the basis states $|J/2, m_{\text{e}}; J/2, m_{\text{o}} \rangle$, with  $-J/2 \leq m_{\text{e}}, m_{\text{o}} \leq J/2$, for a total of $(J+1)^2$ elements. The even-odd decay and phase $\gamma(t)^{\text{eo}}, \varphi_0(t)^{\text{eo}}$ are now given by:
 \begin{eqnarray}
    \gamma  (t)^{\text{eo}}&=\Big\{ \kappa_{\text{ee}}(t) \left(m_{\text{e}}-m'_{\text{e}}\right){}^2+\kappa_{\text{oo}}(t)\left(m_{\text{o}}-m'_{\text{o}}\right){}^2 
    +2 \kappa_{\text{eo}}(t) \left(m_{\text{e}}-m_{\text{o}}\right) \left(m'_{\text{e}}-m'_{\text{o}}\right)\Big\},\nonumber \\
   \varphi _0(t)^{\text{eo}}&= \Big\{ \xi_{\text{ee}}(t)    (m_{\text{e}}^2      -m_{\text{e}}^{'2})+\xi_{\text{oo}}(t) (m_{\text{o}}^{'2}       - m_{\text{o}}^{'2})   
    +2 \xi_{\text{eo}}(t) \left(m_{\text{e}} m_{\text{o}}-m'_{\text{e}} m'_{\text{o}}\right) \Big\}\nonumber,\\
\varphi _1(t)^{\text{eo}}&= \Big\{  
    2 \vartheta_{\text{eo}}(t) \left(m_{\text{e}} m'_{\text{o}}-m'_{\text{e}} m_{\text{o}}\right) \Big\},    
\end{eqnarray}
It follows that in order for $\varphi_0(t)^{\text{eo}}$ and $\varphi_1(t)^{\text{eo}}$ to be zero, we must have $\{m_{{\text{e}}}=m'_{{\text{e}}}, m_{{\text{o}}}=m'_{\text{o}}\}$ or $\{m_{\text{e}}=-m'_{\text{e}}, m_{\text{o}}=-m'_{\text{o}}\}$. Let us consider the former case first. If $m_{\text{e}}+J/2=k$ spins in the even cluster are up $(0 \leq k \leq J)$, we have ${J \choose k}$ combinations of spin up/down states giving $m_{\text{e}}$, and ${J \choose k}^2$ combinations making $m_{\text{e}}= m'_{\text{e}}$. For each of these combinations, we must also have that $m_{\text{o}}=m'_{\text{o}}$. If $m_{\text{o}}+J/2=q$ spins in the odd cluster are up ($0 \leq q \leq J$), this gives  ${J \choose q}^2$ possibilities. Thus, for fixed $m_{\text{e}}$ and $m_{\text{o}}$, there are ${J \choose k}^2 {J \choose q}^2$ matrix elements fulfilling $\{m_{\text{e}}=m'_{\text{e}}, m_{\text{o}}=m'_{\text{o}}\}$. Summing over all possible values of $k,q$ and multiplying by two to account for the 
$\{m_{\text{e}}=-m'_{\text{e}}, m_{\text{o}}=-m'_{\text{o}}\}$ case, we get:
$$2 \bigg(\sum_{k=0}^{J}{J \choose k}^2\bigg) \bigg(\sum_{q=0}^{J}{J \choose q}^2\bigg)= 2 {2J \choose J}^2. $$
Thus, there are $ 2 {2J \choose J}^2$ matrix elements that are QNI, and ${2J \choose J}^2$ that altogether feel no noise. We can further extend this reasoning to $L$ clusters of $J/L$ qubits, resulting in a number of $2 {4J/L \choose 2J/L}^L $ QNI matrix elements.}

{\color{rev}
\subsection{Classicality of the noisy reduced dynamics} 
\label{sub:RU}

The possibility to exactly represent, or ``simulate'' purely dephasing quantum dynamics in terms of stochastically fluctuating {\em classical} fields has been long investigated as part of an effort to obtain a deeper understanding of nonclassicality notions -- resulting in the concept of {\em random unitary} (RU) dynamics \cite{RU1}. As an application of the explicit form of the reduced density operator that the above cumulant-approach enables, we now establish a {\em sufficient} condition on the noise for the reduced dynamics to be RU. 

Suppose that the evolution of a quantum system up to time $t>0$ is described by a completely-positive, trace-preserving map, 
$$\rho_0 \mapsto \rho(t) = \mathcal{E}_t [\rho_0]= \sum_i K_i(t) \rho_0 K_i^\dagger (t),  \qquad \sum_i K_i^\dagger(t) K_i (t)= {\mathbb I}.$$
Recall that the dynamics are RU if the evolution may be expressed as a convex sum of unitary maps,  
$$\mathcal{E}_t [\rho_0]= \sum_{i \in \mathcal{I}} p_i\, U_i(t)\, \rho_0\, U_i^{\dagger}(t), 
\quad  U_i(t) U_i^{\dagger}(t)={\mathbb I}, 
\qquad \sum_{i \in \mathcal{I}} p_i=1, p_i >0, $$  
or, more generally, as an integral $\mathcal{E}_t [\rho_0]= \int_{\Lambda}\,\mathbb{P}[d \lambda]\, U_{\lambda}(t)\, \rho_0\, U^{\dagger}_{\lambda}(t)$, for an appropriate probability measure ${\mathbb P}$. Note that, by construction, RU dynamics is necessarily {\em doubly-stochastic} (unital), since the identity operator is preserved. Thus, $\sum_i K_i (t) K_i^\dagger (t)= {\mathbb I}$ as well. Unlike in the classical dynamical setting, however, not every unital quantum map $\mathcal{E}_t$ arises as a mixture of unitaries, which makes the problem of determining whether or not a given map on a system of arbitrary dimension is RU highly challenging \cite{RU2} and computationally NP-hard in general \cite{Watrous}. We prove the following result for ZMGSD:

\smallskip

{\bf Claim.} {\em The reduced dynamics resulting from a ZMGSD model with $\varphi_1(t)=0$ is random unitary}.

\smallskip

{\bf Proof.} 
The proof is constructive. We start by noting that the action of the system Hamiltonian, $H_S$, and the contribution of the phase  $\varphi_0(t)$ [Eq.\,(\ref{phi0}) in the main text] can be factored out in terms of unitaries $U_0(t)= \exp \left\{ \!- i b t  \sum_{n=1}^N \sigma_n^z \right\}$ and  $U_Q(t)= \exp \left\{\!- i 4 \sum_{n,m=1}^N \xi_{nm}(t) \sigma_n^z \sigma_m^z \right\}$, respectively. It follows that the reduced density matrix at time $t$ can be expressed as $\rho(t) = U_0(t) U_Q(t)  \mathcal{E}_t\left[ \rho_0 \right] U_Q^{\dagger}(t) U_0^{\dagger}(t),$ with $$ \mathcal{E}_t\left[ \rho_0 \right]=\sum_{\vec{\alpha}, \vec{\beta}} e^{ - \sum_{n,m=1}^N \kappa_{nm}(t) (\alpha_n-\beta_n) (\alpha_m-\beta_m) } \langle \vec{\alpha}|\rho_0|\vec{\beta}\rangle |\vec{\alpha}\rangle \langle \vec{\beta}|, $$
representing the decay of coherence. The key step in our construction consists on explicitly expressing $ \mathcal{E}_t\left[ \rho_0 \right]$ as an RU map (albeit in terms of explicitly time-dependent Hamiltonians). To that end, we define the zero-mean, Gaussian set of stochastic processes $ \vec{\lambda}(t)\equiv \{\lambda_n(t)\}, \,n=1,\ldots, N$, such that 
$$\langle \lambda_n(t) \rangle_{\text{c}} =0, \forall n,\qquad \frac{1}{2}\int_0^t \!ds\! \int_0^t \!ds'\, \langle \lambda_n(s) \lambda_m(s') \rangle_{\text{c}} = \kappa_{nm}(t),$$
with $\langle f(\vec{\lambda}) \rangle_{\text{c}}= \int_{\Lambda} d \vec{\lambda} P(\vec{\lambda}) f(\vec{\lambda})$  denoting a \textit{classical} ensemble average and $f(\vec{\lambda})$ a function of any subset of $\{\lambda_1,\ldots, \lambda_N\}$. Let now $U(\vec{\lambda},t)\equiv  \exp \left(\!- i \sum_{n} \sigma_n^z \int_0^t ds\, \lambda_n(s)  \right) $. Using Gaussianity, we can evaluate 
$$\left \langle U(\vec{\lambda},t)\, \rho_0 \,U^{\dagger}(\vec{\lambda},t) \right \rangle_{\text{c}}=\sum_{\vec{\alpha}, \vec{\beta}} e^{ - \sum_{n,m=1}^N \kappa_{nm}(t) (\alpha_n-\beta_n) (\alpha_m-\beta_m) }  \langle \vec{\alpha}|\rho_0|\vec{\beta}\rangle |\vec{\alpha}\rangle \langle \vec{\beta}| = \mathcal{E}_t[\rho_0], $$
making $ \mathcal{E}_t[\rho_0]$ an RU channel, as claimed. As the composition of $ \mathcal{E}_t[\rho_0]$ with $U_0(t)$ and $U_Q(t)$ is still manifestly RU, so is the resulting dynamics.\hfill$\blacksquare$ 

\smallskip

According to the above claim, ZMGSD noise can always be mimicked by a classical phase damping process in the collective regime, consistent with the fact that, as noted in Sec.\,\ref{sec:general}, the bath operators $\{U_{\vec{\alpha}} (t)\}$ commute with one another. One may wonder whether the condition $\varphi_1(t)=0$ is also \emph{necessary} for a non-collective dephasing model to be RU.  A rigorous answer is very nontrivial and not available to the best of our knowledge.

It is natural to think that, for ZMGS non-Markovian dephasing as we consider, the phase $\varphi_1(t)$ may play a role similar to what the {\em Hamiltonian obstruction} does in the simpler setting of Markovian dephasing models \cite{Fagnola}. To make the connection with the Markovian limit clearer, we may recast the evolution of the qubit state under ZMGSD noise in terms of an exact non-Markovian master equation, by differentiating the expression for the reduced density matrix elements given in Eq.\,(\ref{eq::MatrixEl}). We find:
\begin{eqnarray}
\frac{d }{dt}\rho (t) = -i[H_{\text{eff}}(t), \rho(t)] + \sum_{n,m=1}^N c_{nm}(t) \Big( \sigma_n^z\, \rho(t)\, \sigma_m^z - \frac{1}{2} \{\rho(t), \sigma_n^z \sigma_m^z \} \Big),
\label{QME} \\
H_{\text{eff}}(t) = \frac{b}{2} \sum_{n=1}^N \sigma_n^z + \sum_{n,m=1}^N \dot{\xi}_{nm}(t) \sigma_n^z \sigma_m^z, \quad c_{nm}(t)= 2 \Big( \dot{\kappa}_{nm}(t) + i \dot{\vartheta}_{nm}(t) \Big)  .
\end{eqnarray} 
Explicitly, the time derivatives of the dynamic coefficients are given by: 
\begin{eqnarray*}
\dot{\xi}_{nm}(t)= \frac{1}{16} \int_{-t}^t ds\,\text{sign}(s)   \,\langle [B_n(s), B_m(0)]\rangle_{\mathrm{B}}, \label{dxidt} \\
\dot{\kappa}_{nm}(t)= \frac{1}{16} \int_{-t}^t ds\, \langle \{B_n(s), B_m(0)\}\rangle_{\mathrm{B}} , 
\quad 
\dot{\vartheta}_{nm}(t)= \frac{1}{16} \int_{-t}^t ds\,  \langle [B_n(s), B_m(0)] \rangle_{\mathrm{B}} , 
\label{dthetadt}
\end{eqnarray*}
with sign$(s)$ the sign function that is $1$ for $s>0$, $0$ for $s=0$ and $-1$ for $s<0$. Thus, one sees explicitly that the bath-induced phase $\varphi_0(t)$, which depend upon $\xi_{nm}(t)$, corresponds to a bath-mediated, entangling Ising Hamiltonian, which enters the dynamics unitarily, akin to a (non-Markovian) Lamb shift. In the Markovian limit, $\partial_t c_{nm}(t) = 0$, resulting in a constant relaxation matrix $C\equiv \{ c_{nm}\}$; Eq.\,(\ref{QME}) then reduces to the well-known Gorini-Kossakowski-Sudarshan-Lindblad form. As noted in Sec.\,\ref{sec:general}, the parameters $\vartheta_{nm}(t)$ (hence $\varphi_1(t)$) can still be non-zero as long as the noise spatial correlation matrix obeys Im$\,C\ne 0$. The latter condition immediately implies a non-vanishing Hamiltonian obstruction (see Eq.\,(28) in \cite{Fagnola}). In such a setting, vanishing of the obstruction is both sufficient and necessary for an RU representation in terms of {\em smooth, diffusive} (Wiener) processes to exist; however, a Markovian dephasing model may still be RU with a non-zero obstruction, as long as that arises only from discontinuous Poisson noise.
}

\section{Additional technical details for OATS dynamics}
\label{app:OATS}

\subsection{OATS expectation values $\langle J_y(\vec{r},t)\rangle$, $\langle J_y^2(\vec{r},t)$ for arbitrary qubit positions} \label{noncolcum}

We first evaluate $\langle J_y(\vec{r},t)\rangle$ for an OATS with qubits in positions $\vec{r} \equiv \{\vec{r}_1,\ldots, \vec{r}_n\}$. Let us write the OATS as
\begin{eqnarray}
\rho_0= |S\rangle \langle S|,\qquad |S\rangle = \exp(-i \beta J_x) \exp(-i \theta J_z^2)|+\rangle^{\otimes N}, 
\label{oats}
\end{eqnarray}
with $|+\rangle$ the positive eigenstate of $\sigma_x$. By expressing the angular momentum components in terms of single qubit matrices, $J_{\alpha}= \sum_{n=1}^N \sigma_n^{\alpha}/2$, with $\alpha \in \{x,y,z\}$, we may identify a contribution involving qubit $n$ and a second contribution which does not, that is, $J_{\alpha}= \sigma_n^{\alpha}/2 + J^{\alpha}_{\neq n}$. We can then rewrite the OATS as
\begin{eqnarray}
e^{i \theta/8} \exp(-i \beta J^{x}_{\neq n}) \exp[-i \theta (J^z_{\neq n})^2/2] \exp(-i \beta \sigma_n^x/2)\exp(-i \theta  J^z_{\neq n} \sigma_n^z/4) |+\rangle_{n} |+\rangle_{\neq n} ,
\label{prop}
\end{eqnarray}
where $|+\rangle_{\neq n}= \bigotimes_{\ell \neq n}|+\rangle_{\ell}$ is a CSS excluding qubit $n$. We also write the effective propagator corresponding to qubit $n$, Eq.\,(\ref{prop1}) in the main text, as
\begin{eqnarray}
e^{i \Phi_n(t)}= \exp\{-i [\varphi(t)+ \psi_{\neq n}(t)] \sigma_n^z  \}, \quad  \psi_{\neq n}(t)=\sum_{\ell \neq n} \Psi_{n \ell}(t) \sigma_{\ell}^z.
\end{eqnarray}
We then substitute Eqs.\,(\ref{oats}) and (\ref{prop}) into Eq.\,(\ref{prop1}), and evaluate all expectation values of operators for qubit $n$ with respect to $|+\rangle_n$ exactly. Summing over all qubits, this results in
\begin{eqnarray}
\langle J_y(\vec{r},t) \rangle &= \sum_n \frac{e^{\chi_{nn}(t)/2}}{4} \Big [\frac{e^{i \varphi(t)}}{2} \sin(\beta) (\mathcal{E}^{\neq n}_{--+}-\mathcal{E}^{\neq n}_{+--}) \nonumber\\
&+ i e^{-i \varphi(t)} \sin(\beta/2)^2 \mathcal{E}^{\neq n}_{+-+}-i e^{i \varphi(t)} \cos(\beta/2)^2 \mathcal{E}^{\neq n}_{+++} +\text{c.c.} \Big], \label{Jyr}
\end{eqnarray} 
where $ \mathcal{E}^{\neq n}_{s_1 s_2 s_3}$ represents the remaining position-dependent expectation value with respect to all qubits $\ell \neq n$:
\begin{eqnarray}
\mathcal{E}^{\neq n}_{s_1 s_2 s_3}(\vec{r},t) =\langle S_{\neq n}| \exp(i s_1 \theta J^{z}_{\neq n}/2)\exp[i s_2 \tilde{\psi}_{\neq n}(\vec{r},t)] \exp(i s_3 \theta J^{z}_{\neq n}/2) |S_{\neq n}\rangle,
\end{eqnarray}
with $ s_1, s_2, s_3 \in \{+1,-1\}$, and 
\begin{eqnarray}
\tilde{\psi}_{\neq n}(\vec{r},t)= \exp(i \beta J^x_{\neq n})\psi_{\neq n}(\vec{r},t) \exp(-i \beta J^x_{\neq n}), \quad |S_{\neq n}\rangle =\exp[-i \theta(J^z_{\neq n})^2]|+\rangle_{\neq n}.
\end{eqnarray}

So far, no approximation has been made. Note that Eq.\,(\ref{Jyr}) no longer involves operators involving qubit $n$. The mean values $\mathcal{E}^{\neq n}_{s_1 s_2 s_3}(\vec{r},t)$ can be approximately evaluated by performing a generalized cumulant expansion in the sense of Kubo \cite{Kubo}\footnote{{\color{rev} Note that, unlike the cumulant expansion carried out in \ref{app:exact}, which was a way to re-express an \emph{exact} average taken \emph{over the bath degrees of freedom}, the generalized cumulant operation considered here is a tool to \emph{approximately} evaluate mean values \emph{over the system degrees of freedom} (excluding qubit $n$). }}. Truncating this expansion to the second order and expressing cumulants in terms of moments then gives
\begin{eqnarray}
&\langle J(\vec{r},t)\rangle \simeq \exp \Big \{ i(s_1+s_3) \frac{\theta}{2} \langle J^z_{\neq n} \rangle + i s_2 \langle \tilde{\psi}_{\neq n} \rangle - \frac{\theta^2}{4}(1+s_2 s_3) \left( \langle (J^z_{\neq n})^2 \rangle -\langle J^z_{\neq n} \rangle ^2  \right) \label{Jyaprox}\\
&-\frac{1}{2} [ \langle (\tilde{\psi}_{\neq n})^2 \rangle -\langle \tilde{\psi}_{\neq n} \rangle^2 ] \!-\frac{\theta}{2} \left[s_1 s_2 \langle  J^z_{\neq n} \tilde{\psi}_{\neq n}\rangle +s_2 s_3 \langle \tilde{\psi}_{\neq n}  J^z_{\neq n} \rangle-(s_1 s_2 +s_2 s_3) \langle \tilde{\psi}_{\neq n}  \rangle \langle J^z_{\neq n} \rangle \right] \Big\} , 
\nonumber
\end{eqnarray}
where all expectation values are taken with respect to $|S_{\neq n}\rangle$. Evaluating these expectations and substituting in Eq.\,(\ref{Jyaprox}) then yields an approximate expression for $\langle J_y(\vec{r},t)\rangle$ for arbitrary spatial correlations:
\begin{eqnarray}
\langle J_y(\vec{r},t)\rangle &= \sum_{n=1}^N e^{-\chi_{nn}(t)} \exp[- \langle \tilde{\psi}_{\neq n}  \rangle /2 ] \bigg\{ e^{-\theta^2 (N-1)/8} \Big[\cos\left(\frac{\beta}{2} \right)^2 e^{-\theta \langle R\rangle} +\sin\left(\frac{\beta}{2} \right)^2 e^{\theta \langle R \rangle}\Big] \nonumber \\
&+\sin(\beta) \sin \Big[ \theta \sin(\beta) \cos\left(\frac{\theta}{2}\right) \sum_{\ell \neq n}  \Psi_{n \ell}(t) \Big]  \bigg\} \sin(\varphi(t)) \label{JyOATS},
\end{eqnarray}
where 
\begin{eqnarray}
\langle \tilde{\psi}_{\neq n}  \rangle &= \sum_{\ell}' \Psi_{n \ell}(t)^2 + \sum_{\ell_1,\ell_2 \neq \ell_1} ' \Psi_{n \ell_1}(t) \Psi_{n \ell_2}(t) \bigg[ \frac{1}{2} \sin \left( \frac{\beta}{2} \right)^2 (1- \cos(\theta)^{N-3}) \nonumber\\
&+ \sin(2 \beta) \sin \left( \frac{\theta}{2} \right) \cos \left( \frac{\theta}{2} \right)^{N-3} \bigg] ,\\
\langle R \rangle &= \sum_{\ell}' \Psi_{n \ell}(t)^2 \cos(\beta)+ \sum_{\ell_1,\ell_2 \neq \ell_1 } ' \Psi_{n \ell_2}(t)  \sin(\beta) \sin \left( \frac{\theta}{2}\right) \cos \left( \frac{\theta}{2} \right)^{N-3},
\end{eqnarray}
are the position-dependent relevant mean values, and $\sum'$ stands for a summation excluding the index $n$.

Evaluation of $\langle J_y^2(\vec{r},t) \rangle$ proceeds along similar steps as above. To treat qubits $n,m$ separately from qubits $\ell \neq n,m$ we introduce a new collective spin operator
\begin{eqnarray}
J^{\alpha}_{\neq n} \equiv \sum_{\ell \neq n m} \frac{\sigma_{\ell}^{\alpha}}{2}, \quad \alpha \in \{x,y,z\}.
\end{eqnarray}
Using $J^{\alpha}= J^{\alpha}_{\neq n} +  (\sigma^{\alpha}_n + \sigma^{\alpha}_m)/2 $, we then write the initial OATS in Eq.\,(\ref{oats}) as
\begin{eqnarray}
|S\rangle &= e^{i \theta/4} \exp[-i \beta J^x_{\neq nm}]\exp[-i \theta( J^z_{\neq nm})^2/2] \exp[-i \beta (\sigma_n^x + \sigma_m^x)/2] \nonumber \\ & \qquad \quad \exp[-i \theta J^z_{\neq nm}(\sigma_n^x + \sigma_m^x)] \exp[-i \theta \sigma_n^z \sigma_m^z/4] |+\rangle_n |+\rangle_m |+\rangle_{\neq nm}, 
\label{Snm}
\end{eqnarray}
where $|+\rangle_{\neq nm}= \bigotimes_{\ell \neq nm}|+\rangle_{\ell}$. We also write the effective propagator corresponding to qubits $n$ and $m$, Eq.\,(\ref{prop2}) in the main text, as
\begin{eqnarray}
e^{-i \Phi_{nm}(t)}=\exp[-i\varphi(t)(\sigma^z_n + \sigma^z_m)-\chi_{nm}(t) \sigma_n \sigma_m - i \psi_{\neq nm}(t) \sigma_n^z - i \psi_{\neq mn}(t)\sigma_m^z], 
\label{propnm}
\end{eqnarray}
where $\psi_{\neq nm}(t) \equiv \sum_{\ell \neq n,m}\Psi_{n \ell}(t)$. We then substitute Eqs.\,(\ref{Snm}) and (\ref{propnm}) into Eq.\,(\ref{prop2}) for $\langle J^2_y (\vec{r},t) \rangle$ and evaluate all expectation values of operators for qubits $n$ and $m$ with respect to $|+\rangle_n |+\rangle_m$ exactly: 
\begin{eqnarray}
\langle J^2_y(\vec{r},t)\rangle & \simeq  \frac{N}{4} + \frac{1}{2} \tilde{\sum_{\ell_1, \ell_2 \neq \ell_1}} e^{-\chi_{nn}(t)-\chi_{mm}(t)} \bigg [\, e^{\chi_{nm}(t)+\chi_{mn}(t)}\, \langle \Theta_{+-}|\Theta_{-+}\rangle \nonumber\\
 &- \cos(2 \varphi(t))\, e^{-\chi_{nm}(t)-\chi_{mn}(t)}\, \langle \Theta_{++}|\Theta_{--}\rangle \bigg] \label{Jy2OATS}, 
\end{eqnarray}
where $\tilde{\sum}$ excludes indexes $n,m$ from the summation,  and $\langle \Theta_{+-}|\Theta_{-+}\rangle$, $\langle \Theta_{++}|\Theta_{--}\rangle $ are expectation values of operators acting on all  qubits  $ \neq n, m$, taken with respect to the state $|S\rangle_{\neq nm} \equiv \exp[-i \theta (J^z_{\neq nm})^2/2]|+\rangle_{\neq nm} $. As before, these can be approximately evaluated by performing a cumulant expansion and truncating to the second order, yielding the following expressions:
\begin{eqnarray*}
&\langle \Theta_{+-}|\Theta_{-+}\rangle  \simeq \bigg\{ \frac{1}{8} \sin \left(\beta \right)^2 \Big [\cos(2 \theta \langle I_-\rangle) + \cosh(2 \theta \langle R_-\rangle) e^{-2 \theta^2 \langle F_x^2\rangle}\Big]-\frac{1}{2} \sin( 2\beta) \nonumber \\
&\bigg[e^{-\theta \langle R_-\rangle} \sin\!\bigg(\frac{\theta}{2}(1-2\langle I_- \rangle) \bigg)+ e^{\theta \langle R_- \rangle} \sin\!\left(\frac{\theta}{2}(1+2\langle I_- \rangle)   \right) \bigg] e^{-\theta^2 \langle F_x^2  \rangle/2} + \frac{1}{4} \cos(\beta)^2 \bigg\} e^{-\langle g_-^2 \rangle/2} , \nonumber\\
&\langle \Theta_{++}|\Theta_{--}\rangle  \simeq \bigg\{ \frac{1}{4} \bigg[ \cos\! \left(\frac{\beta}{2} \right)^4 e^{-2 \langle R_+\rangle} + \sin\!\left(\frac{\beta}{2} \right)^4 e^{2 \langle R_+\rangle} \bigg] e^{-2 \theta^2 \langle F_x^2 \rangle} - \frac{1}{2} \cos \!\left( \frac{\beta}{2} \right)^2 \sin \!\left( \frac{\beta}{2} \right)^2 \nonumber \\
& \quad \cos(2 \theta \langle I_+\rangle) + \frac{1}{2} \sin(\beta) \bigg [ \cos \!\left( \frac{\beta}{2}\right)^2  \sin\! (\theta (1-2 \langle I_+\rangle)) e^{- \theta \langle R_+ \rangle} - \sin \!\left( \frac{\beta}{2} \right) \sin\! (\theta (1+2 \langle I_+\rangle)) \nonumber \\
& \quad e^{ \theta \langle R_+ \rangle} \bigg] e^{-\theta^2 \langle F_x^2 \rangle /2} + \frac{1}{4} \sin\!\left( \beta \right)^2 \bigg\} e^{- \langle g^2_+ \rangle /2}. \label{c2}
\end{eqnarray*}
Finally, introducing the notation $\Delta \Psi_{\ell}(t)^{\pm}\equiv \Psi_{n \ell}(t) \pm \Psi_{m \ell}(t)$, the position-dependent moments and covariances entering the above equation can be expressed as:
\begin{eqnarray*}
\langle R_{\pm} \rangle &=\frac{\cos(\beta)}{8} \tilde{\sum}_{\ell} \Delta \Psi_{\ell}^{\pm} +\frac{1}{8} \sin(\beta) \sin\!\left(\frac{\theta}{2} \right) \cos\!\left(\frac{\theta}{4} \right)^{\!N-4} \!\!\tilde{\sum_{\ell_1, \ell_2 \neq \ell_1}}
\Delta \Psi_{\ell_2}^{\pm}, \\
\langle I_{\pm} \rangle &= -\frac{\sin(\beta)}{8} \cos \left(\frac{\theta}{2} \right)^{\!N-3}  \tilde{\sum_{\ell}} \Delta \Psi_{\ell}^{\pm},    \\
\langle g_{\pm}^2 \rangle &=\frac{1}{16} \tilde{\sum_{\ell}} (\Delta \Psi_{\ell}^{\pm})^2 + \frac{1}{16} \tilde{\sum_{\ell_1, \ell_2 \neq \ell_1}}  \Delta \Psi_{\ell_1}^{\pm}  \Delta \Psi_{\ell_2}^{\pm} \bigg[ \frac{1}{2} \sin(\beta)^2(1- \cos(\theta)^{N-4}) + \nonumber \\ 
&\quad\; \sin(2 \beta)\sin\left(\frac{\theta}{2} \right) \cos \left( \frac{\theta}{2} \right)^{N-4} \bigg], \\
\langle F_x^2 \rangle &= (N-2)/4 .
\end{eqnarray*}
{\color{rev} Putting things together, we obtain the desired expression for $\Delta \hat{b}(t)$ by replacing Eqs. (\ref{JyOATS}) and (\ref{Jy2OATS}) into the error propagation formula of Eq.\,(\ref{eqnObs}).}

\subsection{Optimality of noiseless OATS angles}
\label{KUopt}

Given an initial OATS subject to collective or even-odd spin-boson {\color{rev} dephasing noise, we claimed in the main text that the best possible short-time uncertainty scaling is achieved by the squeezing and rotation angles $(\theta_{\text{opt}}, \beta_{\text{opt}})$ of Eq.\,(\ref{ideal}) that minimize the uncertainty along the $y$ axis in the noiseless case. To justify the claim, consider $N$ qubit probes prepared in an OATS as in Eq.\,(\ref{oats})}, subject to the Ramsey interferometry protocol. After measuring along the $y$ axis, we can approximate $\Delta \hat{b}(t)$ around $t = 0$ as follows:
\begin{equation}
\Delta \hat{b} (t) \approx \Delta \hat{b}_{\text{app}}(t)= \frac{\sqrt{a_0(\theta,\beta,N) + a_2(\theta,\beta,N) (\omega_c t)^2}}{N/2\, \sqrt{T\; t}\; h_0(\theta,\beta,N)} ,
\label{ST}
\end{equation}
where the $a_0(\theta,\beta,N), a_2(\theta,\beta,N)$ and $h_0(\theta,\beta,N)$ coefficients come from Taylor expanding $\Delta J^2_y(t)$ and $\langle J_y(t)\rangle$ with respect to time, to second and zero order, respectively. Their specific functional form depends on the nature of the noise. Remarkably, the agreement between the collective and even-odd uncertainties with their short-time approximation (\ref{ST}) is excellent for \textit{arbitrary} squeezing and rotation angles. 
The simple time dependence of $\Delta \hat{b}_{\text{app}}(t)$ allows us to  easily minimize it with respect to $t$. We find that:
\begin{equation}
\omega_c \tau_{\text{opt}}= \frac{\sqrt{a_0 (\theta,\beta,N)}}{\sqrt{a_2(\theta,\beta,N)}},\qquad\Delta \hat{b}_{\text{opt}}=\sqrt{\frac{\omega_c}{T}} \frac{\sqrt{2} \left[ a_0(\theta,\beta,N)\; a_2(\theta,\beta,N)\right]^{1/4}}{h_0(\theta,\beta,N)\,N/2 } .
 \label{STopt}
 \end{equation}
Let us now tackle each case separately.

\smallskip

{\bf Collective noise.} We find, using {\em Mathematica}:
\begin{eqnarray*}
a_0(\theta,\beta,N)&= N^2 \cos (\beta ) e^{-N \theta ^2/2 } \left[2+\cos (\beta ) \left(e^{N \theta ^2/2 }-1\right)-4 \sin (\beta ) \sin \left(\theta /2\right) e^{N \frac{3}{8} \theta ^2 }\right] ,\\
a_2(\theta,\beta,N)&=\frac{1}{16} N^2  \text{$\chi_0$}^2 e^{- N\theta ^2/2} \left( 3+ \cos (2 \beta )+2 \sin(\beta )^2 e^{N\theta ^2/2}+4 \sin (2 \beta ) \sin \left(\theta/2\right) e^{N \frac{3}{8} \theta ^2}\right) ,\\
h_0(\theta,\beta,N)&=e^{-N \frac{1}{8} \theta ^2} ,
\end{eqnarray*}
where we allow for the squeezing and rotation angles to depend on $N$, $\{\theta, \beta\} \rightarrow \{\theta (N), \beta(N) \}$. 
 Note that, unless $\theta$ scales like $\theta(N) \propto N^{-\alpha}$ with $\alpha> 1/2$, then $e^{-N \theta^2/2} \ll 1$. Having $e^{-N \theta^2/2} \ll 1$ leads, in turn, to the following limiting expressions:
\begin{eqnarray*}
a_0(\theta,\beta,N) &= N \Big(1+ \frac{N}{2}\, \cos (\beta)^2 \Big),\;\;\;a_2(\theta,\beta,N)= \frac{1}{8} N^2 \chi_0^2 \sin(\beta)^2,
\;\;\;h_0=e^{-\frac{1}{8}N \theta^2},
\end{eqnarray*}
and replacing in Eq.\,(\ref{STopt}) the optimal uncertainty becomes 
$$ \Delta \hat{b}_{\text{opt}}=\sqrt{\frac{\omega_c}{T}}\;e^{\frac{1}{8} N \theta^2}\; \frac{ \left[ \chi_0^2  N^3/8 \left(1+N/2\; \cos (\beta)^2 \right) \sin(\beta)^2\right]^{1/4}}{N/2}, $$
which grows exponentially with $N$. Thus, we require $\theta(N)$ to scale like $N^{-\alpha}$ with $\alpha>1/2$. 
We then expand $a_0(\theta,\beta,N), a_2(\theta,\beta,N)$ in this limit, keeping the angles $N$ dependence implicit for now. We would like to see whether there is some set of angles $\{\theta(N), \beta(N) \}$ that reduces the scaling order of either of those quantities. Replacing $e^{\pm \theta N^2/4}\approx 1$ and $\sin\left(\frac{\theta}{2}\right) \propto \frac{1}{2} N^{-\alpha} \approx 0$ in the above expression for $a_2(\theta,\beta,N)$, we get
$$a_2(\theta,\beta,N) \approx\frac{1}{16} N^2  \text{$\chi_0$}^2  \left[ 3+ \cos (2 \beta )+2 \sin(\beta )^2 \right]=\frac{1}{4} N^2 \chi_0^2 ,$$
which is clearly always positive, making $a_2(\theta,\beta,N) \propto N^2$ to leading order, regardless of  our choice of squeezing and rotation angles. However, noting that $a_0(\theta,\beta,N)$ is the value of $\Delta J_y^2(t) $ at $t=0$, we know that for the optimal squeezing and rotation angles, $a_0(\theta_{\text{opt}},\beta_{\text{opt}},N) \propto N^{1/3}$, and this scaling is best possible\footnote{Strictly speaking, $a_0(\theta,\beta,N)$ is the approximation to variance $\Delta J_y^2(0)$ of Eq.\,(\ref{deltaJy}) coming from a cumulant expansion with respect to the qubit operators, see \ref{noncolcum}. It can be checked, however, that it matches the exact formula if $N \gg 1$.}. Replacing $a_0(\theta_{\text{opt}}, \beta_{\text{opt}},N) \propto N^{1/3}, a_2(\beta_{\text{opt}}, \theta_{\text{opt}},N) \propto N^2$ in Eq.\,(\ref{STopt}) leads to the $N^{-5/12}$ scaling of the uncertainty. As the short-time expansion in Eq.\,(\ref{ST}) reproduces the uncertainty very accurately, we can then conclude it is also best possible for $\Delta \hat{b}(t)$. 

\medskip

{\bf Even-odd noise.} Proving optimality of $\{\theta_{\text{opt}},\beta_{\text{opt}}\}$ in the even-odd setting  can be shown analogously. Performing the short-time expansion in this case, we have the same values for $a_0(\theta, \beta,N)$ and $h_0(\theta,\beta,N)$ as in the collective case analyzed above, whereas $a_2(\theta,\beta,N)$ in the $N \gg 1$ limit is given by
\begin{eqnarray*}
a_2(\theta, \beta, N,x)&=\frac{N }{16} \Big\{-N \chi_0^2\left(\cos (2 \beta ) \left(e^{\theta ^2 N/2}-1\right)-4 \sin (2 \beta ) \sin \left(\theta/2\right) e^{\frac{3}{4} \theta ^2 N/2}\right)\nonumber \\
&-\chi_0^2 \left(e^{\theta ^2 N/2}- 2N +3\right)+ N \chi_1^2(x) \left(e^{\theta ^2 N/2}+1\right)\Big\}e^{\theta ^2 (-N/2)}.
\end{eqnarray*}
Again, we must require $\theta(N) \propto N^{-\alpha}$ with $\alpha >1/2$ for the uncertainty not to grow exponentially in the asymptotic $N\gg1$ regime. In this limit $e^{\pm \theta N^2}\approx 1$ and $\sin\left(\frac{\theta}{2}\right) \propto \frac{1}{2} N^{-\alpha} \approx 0$, leading to an $a_2(\theta, \beta,N,x)$ that scales like $N^2$ regardless of the $\{ \theta, \beta \}$ values:
$$ a_2(\theta, \beta, N,x)\approx  \frac{N^2}{2} (\chi_0^2 + \chi_1^2(x))\geq 0, \;\;\;\forall x.$$
This implies that the optimal choice for rotation and squeezing angles must once again minimize $a_0(\theta,\beta, N)$, and the result follows. For both of these noise regimes, a numerical optimization of $\Delta \hat{b}(t)$ over $\theta, \beta$ for fixed $N$ confirmed the analytic results we just derived.  

{\color{rev}

\section{Additional technical details for randomized coupling dynamics} 
\label{RCapp}

As stated in the main text, we carry out the calculations for $D=1$, assuming a spectral density with a Gaussian cutoff, $K(\omega, \omega_c) \equiv e^{-\omega^2/\omega_c^2}$ and in the short time regime $\omega_c t \ll 1$.

\subsection{First order cumulants under spatial averaging}
\label{means} 

The means $\mathbb{E} \{\gamma(\vec{r},t)\}, \mathbb{E}\{\varphi_0(\vec{r},t)\}$ \,are easily computed. For $n \neq m$ the spatial average in Eqs.\,(\ref{chirc})-(\ref{xirc}) in the main text may be evaluated by using Eq.\,(\ref{spatial_cos}), which makes it apparent why it was useful to express the dynamic coefficients in Eqs.\,(\ref{chirc})-(\ref{xirc}) in terms of discrete sums. It becomes then possible to express the mean of the dynamic coefficients as frequency integrals, namely:
\begin{eqnarray}
\mathbb{E} \{ \kappa_{nn}(t)\} &\equiv &  \bar{\kappa}_0(t)=\frac{1}{2} \int_{0}^{\infty} J(\omega) \frac{1- \cos \omega t}{\omega^2} d\omega , 
\label{rcchi0}\\
\mathbb{E}\{\kappa_{nm}(\vec{r}, t)\}&\equiv &  \bar{\kappa}_1(t)=\frac{1}{2} \int_{0}^{\infty} J(\omega) \frac{1- \cos \omega t}{\omega^2} e^{- \omega^2 \epsilon^2/v^2 } d\omega , \quad n \neq m , 
\label{rcchi1}\\
\mathbb{E} \{\xi_{nm}(\vec{r}, t)\} &\equiv & \bar{\xi}(t)=\frac{1}{2} \int_{0}^{\infty} J(\omega) \frac{\omega t- \sin \omega t}{\omega^2} e^{- \omega^2 \epsilon^2/v^2 } 
d\omega, \quad n \neq m.
\label{rcxi}
\end{eqnarray} 
Note that  $\bar{\kappa}_0(t)$, which corresponds to the $n=m$ case, is unaffected by the spatial average. The integrals can be given an approximate analytic expression in the short time regime $1 \ll \omega_c t$,
\begin{eqnarray}
 \bar{\kappa}_0(t)& \approx \frac{1}{8} \Gamma \left(\frac{s+1}{2}\right)  \left(\omega_c t\right)^2  \equiv \bar{\kappa}_0^2 \left(\omega_c t\right)^2 ,\label{chi0st}\\
\bar{\kappa}_1(t)&\approx \eta^{s+1}\frac{1}{8}\Gamma \left(\frac{s+1}{2}\right) \left(\omega_c t\right)^2 \equiv \eta^{s+1} \bar{\kappa}_0^2 \left(\omega_c t\right)^2 ,
\label{chi1st}\\
\bar{\xi}(t)& \approx \eta^{s+2} \frac{1}{48} s\; \Gamma \left(\frac{s}{2}\right) ( \text{$\omega $}_c t)^3 \equiv \eta^{s+2} \bar{\xi}_0^{\,3} \,( \text{$\omega $}_c t)^3 \label{xi1st}.
\end{eqnarray}
Therefore, we see that if $\eta \ll 1$, both $\bar{\kappa}_1(t)$ and $\bar{\xi}(t)$ are vanishingly small, whereas the ``qubit-local'' $\bar{\kappa}_0(t)$ term is unaffected by the procedure, consistent with intuition.

Introducing the state-dependent angle $ N \cos \left(\theta_{\alpha \beta}\right) \equiv \sum_{i=1}^N \alpha_n \beta_n $, 
we can write $\mathbb{E}\{ \gamma(\vec{r}, t) \}, \mathbb{E}\{ \varphi_0(\vec{r}, t)\}$ in terms of $\theta_{\alpha \beta}$ and the $J_z$ angular momentum components of $|\vec{\alpha}\rangle$ and $|\vec{\beta}\rangle$, $m$ and $m'$:  \begin{eqnarray*}
\bar{\gamma}(t)&= 4 \Big[ N \sin \left(\theta_{\alpha \beta}/2\right)^2 \bar{\kappa}_0(t) + \Big( (m-m')^2 -  N \sin \left(\theta_{\alpha \beta}/2\right)^2\Big)\bar{\kappa}_1(t) \Big]   ,
\label{meangamma} \\
\bar{\varphi}_0(t) &= 4(m^2-m'^2) \bar{\xi}(t).
\end{eqnarray*}

\subsection{Second order cumulants under spatial averaging}
\label{cov}

The calculation of the phase and decay variances and covariance is more involved, but follows similar lines. The resulting expressions are still products of a state-dependent structure, captured by functions of $m,m'$ and the angle $\theta_{\alpha \beta}$ introduced above, and a time-dependent contribution, represented by coefficients $F_{\ell}(t), G_{\ell}(t), FG_2(t)$, as given below:
\begin{eqnarray}
\Delta \bar{\gamma}(t)^2 &=32 \Big\{ \Big[ N \sin \left(\theta_{\alpha \beta}/2\right)^2 ( N \sin \left(\theta_{\alpha \beta}/2\right)^2 -1) \Big] F_1(t)
\label{gammasp}  \\
 &+ 2 \Big[ (m-m')^2 ( N \sin \left(\theta_{\alpha \beta}/2\right)^2-2) -  N \sin \left( \theta_{\alpha \beta}/2\right)^2 ( N \sin\left( \theta_{\alpha \beta}\right)^2- 1 )  \Big] F_2(t) \Big\} , \nonumber \\
\Delta \bar{\varphi}_0(t)^2 & = 4 N  \Big\{ N  \sin\left(\theta_{\alpha \beta}\right)^2 G_1(t) 
\label{phisp} \\ 
& +  2  \Big[ ( m^2-2 m m' \cos \left(\theta_{\alpha \beta}\right) +  m'^2) 
- N \sin\left(\theta_{\alpha \beta}\right)^2 \Big]  G_2(t) \Big\} , \nonumber 
\end{eqnarray}
\begin{eqnarray} 
\text{Cov}(\gamma(\vec{r}, t),&\varphi_0(\vec{r}, t)) = 32  \Big[ (m^2-m'^2)  \left( N \sin \left( \theta_{\alpha \beta}\right)^2 - 1 \right) \Big] FG_2(t). 
\label{FG}
\end{eqnarray}
Here, $F_{\ell}(t), G_{\ell}(t), FG_2(t)$, $\ell=\{1,2\}$, are the variance and covariance of the relevant dynamic coefficients:
\begin{eqnarray*}
F_1(t)= & \hspace*{-3mm}\mathbb{E} \{ \chi_{nm}(\vec{r}, t)^2 \} - \mathbb{E} \{ \chi_{nm}(\vec{r},t)\} ^2 ,\quad 
F_2(t)=\mathbb{E} \{\chi_{nm}(\vec{r},t) \chi_{np}(\vec{r},t)\}-\mathbb{E}\{ \chi_{nm}(\vec{r},t)\}^2,\\
G_1(t)= & \hspace*{-3mm}\mathbb{E} \{ \xi_{nm}(\vec{r},t)^2 \} - \mathbb{E} \{\xi_{nm}(\vec{r},t)\} ^2,\quad 
\;G_2(t)=\mathbb{E}\{ \xi_{nm}(\vec{r},t) \xi_{np}(\vec{r},t)\}-\mathbb{E}\{ \xi_{nm}(\vec{r},t) \}^2,\\
FG_2(t)=& \mathbb{E}\{ \chi_{nm}(\vec{r},t) \xi_{np}(\vec{r},t) \}-\mathbb{E}\{ \chi_{nm}(\vec{r},t)\}\; \mathbb{E}\{ \xi_{np}(\vec{r},t) \},
\end{eqnarray*}
with $n \neq m \neq p$. For $D=1$, in particular, they can be written as double integrals in the frequency domain:
\begin{eqnarray}
F_{\ell}(t)&=\frac{1}{4} \int_{0}^{\infty}\!\!\int_{0}^{\infty} \!\!J(\omega) J(\omega') e^{-(\omega^2 + \omega'^2) \epsilon^2} f(\omega,t) f(\omega',t) h(\omega, \omega'), 
 \label{F}\\
G_{\ell}(t)&=\frac{1}{4} \int_{0}^{\infty}\!\!\int_{0}^{\infty} \!\!J(\omega) J(\omega') e^{-(\omega^2 + \omega'^2) \epsilon^2} g(\omega,t) g(\omega',t) h(\omega, \omega'), 
\label{G}\\
FG_2(t)&=\frac{1}{4} \int_0^{\infty} \!\!\int_{0}^{\infty} \!\!J(\omega) J(\omega') e^{-(\omega^2 + \omega'^2) \epsilon^2} f(\omega,t) g(\omega',t) [\cosh(\omega \omega' \epsilon^2/v^2)-1] \,d\omega d\omega'  ,
\label{FGbis}
\end{eqnarray}
where $\ell \in \{1,2\}$ and the functions $f(\omega,t), g(\omega,t)$, and $h(\omega,\omega')$ are given by 
$$f(\omega,t)= {[1- \cos(\omega t)]}/{\omega^2}, \quad  g(\omega, t)= [{\omega t- \sin (\omega t)]}/{\omega^2}, \quad 
h(\omega, \omega') =\cosh[(3-\ell)\omega \omega' \epsilon^2/v^2]-1.$$ 

In the short-time regime, we have $f(\omega,t) \approx\frac{1}{2} t^2$, $g(\omega,t)\approx \frac{1}{6} \omega t^3$.
Introducing these approximations into Eqs.\,(\ref{F})-(\ref{FGbis}) {\color{rev}
the dynamic coefficients variances and covariance can also be evaluated, resulting in the following expressions:  
\begin{eqnarray}
F_1(t) &\approx \eta \; \pi ^{3/2} \frac{ \sec (\pi  s)}{ 2^{s+\frac{13}{2}}\Gamma \left(\frac{1}{2}-s\right) } \left( \omega_c t\right)^4,
\label{FFF1}\\
G_1(t)&\approx -\eta\; \pi ^{3/2} \frac{1}{36} \frac{ \sec (\pi  s)}{\; 2^{s+\frac{11}{2}}  \Gamma \left(-s-\frac{1}{2}\right) } \left( \omega_c t\right)^6, \\
F_2(t) &\approx \eta^{2 (s+1)}\; \frac{1}{64} \Gamma \left(\frac{s+1}{2}\right)^2  \left[\, _2F_1\left(\frac{s+1}{2},\frac{s+1}{2};\frac{1}{2};\frac{1}{4}\right)-1\right] \left( \omega_c t\right)^4 ,\\
G_2(t)&\approx \eta^{2 (s+2)}\; \frac{1}{576}  \Gamma \left(\frac{s}{2}+1\right)^2  \left[\, _2F_1\left(\frac{s+2}{2},\frac{s+2}{2};\frac{1}{2};\frac{1}{4}\right)-1\right] \left( \omega_c t\right)^6,\\
FG_2(t)&\approx\eta^{2 s+3}\;\frac{1}{192}\; \Gamma \left(\frac{s}{2}+1\right) \Gamma \left(\frac{s+1}{2}\right) \left[2^s \left(1+3^{-(s+1)}\right)-1\right] \left(\omega_c t\right)^5 
\label{FFG}, 
\end{eqnarray}
where $_2F_1(a;b;z)$ denotes the confluent hypergeometric function of the second kind.

\smallskip

We now outline the explicit calculations to obtain Eqs.\,(\ref{gammasp})-(\ref{FG}). Consider $\Delta \bar{\gamma}(t)^2= \mathbb{E}\{ \gamma(\vec{r},t)^2\}- \mathbb{E}\{ \gamma(\vec{r},t)\}^2$. The position-dependent decay rate $\gamma(\vec{r},t)$ in Eq.\,(\ref{gamma}) has dynamic coefficients $\kappa_{nm}(\vec{r}, t)$ given by Eq.\,(\ref{chirc}). The desired variance is then:
\begin{eqnarray}
\Delta \bar{\gamma}(t)^2= \!\!\!\!\sum_{n,m,
n',m'=1}^N \!\!\!\!(\alpha_n - \beta_n) (\alpha_m - \beta_m)(\alpha_{n'} - \beta_{n'})\, \text{Cov}\left( \kappa_{nm}(\vec{r}, t)\,\kappa_{n'm',}(\vec{r}, t)\right)
\label{vargamma}
\end{eqnarray}
where we recall that the covariance of arbitrary functions $f(\vec{r},t)$ and $g(\vec{r},t)$ is given by $\text{Cov} \left(  f(\vec{r},t)\,g(\vec{r},t) \right) = \mathbb{E} \{ f(\vec{r},t)\; g(\vec{r},t)\} - \mathbb{E}\{ f(\vec{r},t)\} \mathbb{E}\{g(\vec{r},t)\}$. Using Eq.\,(\ref{chirc}) can then write:
\begin{eqnarray}
& \text{Cov}\left( \kappa_{nm}(\vec{r}, t)\,\kappa_{n'm'}(\vec{r}, t)\right) = \frac{1}{16}\sum_{\vec{k},\vec{k'}} |g_k|^2 |g_{k'}|^2 \frac{1- \cos \Omega_{k} t}{\Omega_{k}^2} \coth \left(\beta \,\Omega_{k}/2\right) \nonumber\\
& \qquad\frac{1- \cos \Omega_{k'} t}{\Omega_{k'}^2} \coth \left(\beta\, \Omega_{k'}/2\right)\,  \text{Cov}\left( \cos (\vec{k} \cdot \left( \vec{r}_n - \vec{r}_m ) \right) \cos (\vec{k} \cdot \left( \vec{r}_{n'} - \vec{r}_{m'})\right) \right) 
\label{eqc2}
\end{eqnarray}
The spatial dependence is then fully captured by $\text{Cov}\left( \cos (\vec{k} \cdot \left( \vec{r}_n - \vec{r}_m ) \right) \cos (\vec{k} \cdot \left( \vec{r}_{n'} - \vec{r}_{m'})\right) \right) $. It is straightforward to see that this quantity vanishes unless the indexes obey:
\begin{enumerate}
\item $\{ n=n', m=m' \}$ or $\{ n=m', m=n'\}$, with $ n \neq m$. Let us call this subset of the indexes $\mathcal{S}_1$. We then have for the first case:
$$ \text{Cov} \left( \cos (\vec{k} \cdot \left( \vec{r}_n - \vec{r}_m ) \right) \cos (\vec{k} \cdot \left( \vec{r}_{n} - \vec{r}_{m})\right) \right) = e^{-\epsilon^2 (k^2 + k'^2)} \prod_{i=1}^D \cosh(2 \epsilon^2 k_i k_i')-1], $$
with the $\{ n=m', m=n'\}$ case yielding the same result by symmetry.
\item  Three different indexes out of the four ones, with the matching pair being either $\{n=n'\}$, $\{n=m'\}$, $\{m=n'\}$ or $\{m=m'\}$.  We call this subset of the indexes $\mathcal{S}_2$. Let us consider the case where $\{n=n'\}$:
$$ \text{Cov} \left( \cos (\vec{k} \cdot \left( \vec{r}_n - \vec{r}_m ) \right) \cos (\vec{k} \cdot \left( \vec{r}_{n} - \vec{r}_{m'})\right)\right) = e^{-\epsilon^2 (k^2 + k'^2)} \prod_{i=1}^D \cosh( \epsilon^2 k_i k_i')-1], $$
with the other three cases yielding the same result by symmetry.
\end{enumerate}

Replacing the spatial average in Eq.\,(\ref{eqc2}), we obtain an expression for the second order time-dependent coefficients,
\begin{eqnarray}
F_{\ell}(t)=  &\frac{1}{16}\sum_{\vec{k},\vec{k'}} |g_k|^2 |g_{k'}|^2 \frac{1- \cos \Omega_{k} t}{\Omega_{k}^2} \coth \left(\beta \Omega_{k}/2\right) \frac{1- \cos \Omega_{k'} t}{\Omega_{k'}^2}  \coth \left(\beta \Omega_{k'}/2\right)  \nonumber\\
&  \qquad e^{-\epsilon^2 (k^2 + k'^2)} \prod_{i=1}^D \cosh( \epsilon^2 (3-\ell) k_i k_i')-1] , \quad \ell \in \{1,2\}, 
\end{eqnarray}
with $\ell=1$ when the qubit pairs have indexes in $\mathcal{S}_1$, and $\ell=2$ when they belong to $\mathcal{S}_2$.
Note that if we restrict ourselves to the isotropic $D=1$ case, as we do here, we can exploit the linear dispersion relationship $\Omega_k= k v$ to write $F_{\ell}(t)$ in terms of a double frequency integral, Eq.\,(\ref{F}).
Substituting in Eq.\,(\ref{vargamma}), this leads to  the following expression for the decay variance:
\begin{eqnarray*}
\Delta \bar{\gamma}(t)^2&= F_1(t) \Big( \sum_{\mathcal{S}_1} \mathcal{Q}(\alpha, \beta) \Big)+ F_2(t) \Big( \sum_{\mathcal{S}_2} \mathcal{Q}(\alpha, \beta)  \Big),
\end{eqnarray*}
where $\mathcal{Q}(\alpha, \beta)= (\alpha_n - \beta_n) (\alpha_m - \beta_m)(\alpha_{n'} - \beta_{n'}) (\alpha_{m'} - \beta_{m'})$ gives the state-dependent numerical factor. One may show that the sums over the restricted index subsets $\mathcal{S}_1, \mathcal{S}_2$ can be written in terms of $m, m'$ and $\theta_{\alpha \beta}$:
\begin{eqnarray*}
\sum_{\mathcal{S}_1} \mathcal{Q}(\alpha, \beta) = 32 \left[ N \sin \left(\theta_{\alpha \beta}/2 \right)^2  (N 
\sin \left(\theta_{\alpha \beta}/2)^2 -1 \right) \right] , \\
\sum_{\mathcal{S}_2} \mathcal{Q}(\alpha, \beta) = 64 \left[ (m-m')^2 ( N \sin \left( \theta_{\alpha \beta}/2 \right)^2-2) - N \sin \left( \theta_{\alpha \beta}/2\right)^2 (N \sin \left( \theta_{\alpha \beta}/2\right)^2-1 )  \right] ,
\end{eqnarray*}
leading directly to Eq.\,(\ref{gammasp}).

The phase variance $\Delta \bar{\varphi}_0(t)^2$ and covariance $\text{Cov}(\gamma(\vec{r},t), \varphi_0(\vec{r},t))$ can be evaluated analogously  by setting up the equivalent of Eq.\,(\ref{vargamma}). The phase $\varphi_0(\vec{r},t)$ is given by Eq.\,(\ref{phi0}), with the dynamic coefficient $\xi_{nm}(\vec{r},t)$ of Eq.\,(\ref{xirc}). It is then clear that the spatial dependence of the ensuing covariance $\text{Cov} \left(\xi_{nm}(\vec{r}, t), \, \xi_{n'm'}(\vec{r}, t)) \right)$ and $\text{Cov} \left( \kappa_{nm}(\vec{r}, t),\,   \xi_{n'm'}(\vec{r}, t)\right)$ is also governed by $\text{Cov}\!\left( \cos (\vec{k} \cdot \left( \vec{r}_n - \vec{r}_m ) \right) \cos (\vec{k} \cdot \left( \vec{r}_{n'} - \vec{r}_{m'}) \right) \right)$. Thus, the average is non-vanishing for the same restricted index subsets $\mathcal{S}_1, \mathcal{S}_2$ we found for $\Delta \bar{\gamma}(t)^2$. The second-order time-dependent coefficients entering $\Delta \bar{\varphi}_0(t)^2$ and $\text{Cov}(\gamma(\vec{r},t), \varphi_0(\vec{r},t))$ are then: 
\begin{eqnarray*}
G_{\ell}(t)=&\hspace*{-3mm}\frac{1}{16}\sum_{\vec{k},\vec{k'}} |g_k|^2 |g_{k'}|^2 \,\frac{\Omega_k t- \sin \Omega_{k} t}{\Omega_{k}^2} \,  \frac{\Omega_{k'}t- \sin \Omega_{k'} t}{\Omega_{k'}^2}\, 
e^{-\epsilon^2 (k^2 + k'^2)} \!\prod_{i=1}^{ D} \!\cosh( \epsilon^2 (3-\ell) k_i k_i')-1] , \\
FG_{\ell}(t)=& \frac{1}{16}\sum_{\vec{k},\vec{k'}} |g_k|^2 |g_{k'}|^2 \frac{1- \cos \Omega_{k} t}{\Omega_{k}^2} \coth \left(\beta \Omega_{k}/2\right)  \nonumber\\
&  \qquad \frac{\Omega_{k'}t- \sin \Omega_{k'} t}{\Omega_{k'}^2} e^{-\epsilon^2 (k^2 + k'^2)} \prod_{i=1}^D \!\cosh( \epsilon^2 (3-\ell) k_i k_i')-1] ,
\end{eqnarray*}
for $\ell = 1,2$. When $D=1$ these expressions are isotropic and can be cast into Eqs.\,(\ref{G})-(\ref{FGbis}). All that is left to do is to write the resulting state-dependent structures in the summations of $\Delta \bar{\varphi}_0(t)^2$ and Cov$(\gamma(\vec{r},t), \varphi_0(\vec{r},t))$ in terms of $m,m'$ and $\theta_{\alpha \beta}$. Let us define $\mathcal{G}(\alpha, \beta)\equiv (\alpha_{n} \alpha_m - \beta_n \beta_m)(\alpha_{n'} \alpha_{m'} - \beta_{n'} \beta_{m'})$ and $\mathcal{J}(\alpha,\beta)\equiv  (\alpha_n-\beta_n) (\alpha_m-\beta_m)(\alpha_{n'} \alpha_{m'} - \beta_{n'} \beta_{m'})$. Direct calculation shows that 
\begin{eqnarray*}
\sum_{\mathcal{S}_1} \mathcal{G}(\alpha, \beta) &= 4 N^2 \sin\left(\theta_{\alpha \beta}\right)^2, \qquad \sum_{\mathcal{S}_1} \mathcal{J}(\alpha, \beta) = 0,\\
\sum_{\mathcal{S}_2} \mathcal{G}(\alpha, \beta) &= 8 \left [ 2N \left( m^2 - 2m m' \cos \left(\theta_{\alpha \beta} \right)  + m'^2 \right) -  N^2\sin\left(\theta_{\alpha \beta} \right)^2 \right],\\
\sum_{\mathcal{S}_2} \mathcal{J}(\alpha, \beta) &= 32  \left[ (m^2-m'^2)\left( N \sin \left( \theta_{\alpha \beta}/2 \right)-1 \right) \right],
\end{eqnarray*}
which leads to the expressions for $\Delta \bar{\varphi}_0(t)^2$ and Cov$(\gamma(\vec{r},t), \varphi_0(\vec{r},t))$ given by Eqs.\,(\ref{phisp})-(\ref{FG}).
}

\subsection{Proof of claim \ref{claim} } \label{proofclaim}

{\bf Claim.} \em Assume that the short-time-large-dispersion limit $\eta \ll1$ is obeyed. Then:}

{\bf (i)} {\em First-order cumulants provide a good approximation to the full spatial average provided that the additional condition $\eta \left(\omega_c t\right)^2 N \ll 1$ is met.} 

{\bf (ii)} \emph{In the regime of validity of (i), all terms with exponents proportional to powers of $\eta$ can be made arbitrarily close to unity provided that the condition $\eta^{s+1} (\omega_c t)^2 N^2 \ll 1$ is met.}

\smallskip

{\bf Proof.} Let us start from the second-order cumulant expansion of a reduced matrix element,
$$ \langle \vec{\alpha}|\rho(t)|\vec{\beta}\rangle \approx \langle \vec{\alpha}|\rho_0|\vec{\beta}\rangle\, e^{-\bar{\gamma}(t) +\frac{1}{2}\left[\Delta  \bar{\gamma}^2(t) -\Delta \bar{\varphi}_0^2(t)\right]}\,e^{i \bar{\varphi}_0( t) + i \text{Cov}(\gamma(\vec{r}, t), \varphi_0(\vec{r}, t))}. $$
To establish part {\bf (i)}, we need to prove that for all $|\vec{\alpha}\rangle, |\vec{\beta}\rangle$ with non trivial phase and decay, 
\begin{eqnarray}
\vert \bar{\varphi}_0(t) \vert \gg \vert \text{Cov}(\gamma(\vec{r},t), \varphi_0(\vec{r}, t)) \vert, \quad 
\vert \bar{\gamma}(t) \vert \gg  \frac{1}{2}\vert \Delta \bar{\gamma}^2(t) -\Delta \bar{\varphi}_0^2(t) \vert.
\label{ineq}
\end{eqnarray}
To see where the first inequality is satisfied, we write down the expressions for $\bar{\varphi}_0(\vec{r}, t)$, $\text{Cov}(\gamma(\vec{r}, t), \varphi_0(\vec{r}, t))$:
\begin{eqnarray*}
\left|(m^2-m'^2) \xi(t) \right| &\gg 8 \left| (m^2-m'^2)  \left( N \sin \left( \theta_{\alpha \beta}\right)^2 - 1 \right) FG_2(t) \right| \Leftrightarrow
\nonumber \\
\mathcal{O}\Big(\eta^{s+2}\;  ( \text{$\omega $}_c t)^3  \Big) &\gg  8 \left|\,  N \sin (\theta_{\alpha \beta}/2)^2 -1 \,\right|\,\mathcal{O}\Big(\eta^{2 s+3}\; (\omega_c t)^5 \Big) .
\end{eqnarray*}
In order for this to hold, it suffices that $\eta^{s+1} \;(\omega_c t)^2 N \ll 1 $. Proving the second inequality in Eq.\,(\ref{ineq}) proceeds along similar lines. If $\theta_{\alpha \beta} \neq 0$, the $\eta$-independent $\bar{\kappa}_0(t)$ term in $\bar{ \gamma}(t)$, proportional to $\sin (\theta_{\alpha \beta}/2)$, dominates as long as $\eta \left(\omega_c t\right)^2 N\ll 1$ is satisfied. For the $\theta_{\alpha \beta}=0$ case, the second term in $\bar{\gamma}(t) $ dominates whenever $\eta^{s+1} (\omega_c t)^2 [1+\eta^2 (\omega_c t)^2 N ] \ll 1$, which is already fulfilled provided the condition for the $\theta_{\alpha \beta} \neq 0$ case is met. 
Note how this is also true for the first inequality, and hence {\bf{i}} is satisfied when $\eta \left(\omega_c t\right)^2 N\ll 1$.

To establish part {\bf (ii)}, note that when the exponents containing $\eta$ factors are small enough, it is valid to take a first-order Taylor expansion. We can then write $e^{i \bar{\varphi}_0( t)} \approx 1-4(m^2-m'^2) \eta^{s+2} \bar{\xi}_0^{\,3} (\omega_c t)^3$, and similarly $e^{- \eta^{s+1}\bar{\kappa}_0^2 (\omega_c t)^2 (m-m')^2} \approx 1- \eta^{s+1}\bar{\kappa}_0^2 (\omega_c t)^2 (m-m')^2$. Upper-bounding $(m^2-m'^2)$ by $N^2$ and $(m-m')$ by $N$ then yields the desired result. \hfill$\blacksquare$

\section{Non-collective Gaussian dephasing models with $\bm{\varphi_1(t) \neq 0$}}
\label{app:phi1}

As we established in Sec.\,\ref{sec:NQSB}, the spin-boson model we focused on (which fully accounted for the angular dependence in the inner product $\vec{k}\cdot \vec{r}_{nm}$) has vanishing coefficients $\vartheta_{nm}(t)=0$. As a consequence, the phase $\varphi_1(t)$ reflecting non-commutativity of the set $\{ U_{\vec{\alpha}}(t)\}$ is identically zero. However, within the broader class of ZMGSD noise models, there may be relevant instances in which $\varphi_1(t)$ plays a non-trivial role. It is then reasonable to ask how the presence of this second phase would affect the conclusions we have drawn for the even-odd and RC settings.

\smallskip

{\bf Even-odd noise.} When qubits $n$ and $m$ belong to the same (s) cluster, they interact with the bath through a single noise operator, leading to a vanishing dynamic coefficient $\vartheta_s(t)$; hence, $\Psi_s(t)$ remains unaltered. However, the phase coefficient corresponding to qubits $n$ and $m$ on different (d) clusters picks up an extra contribution: If the first qubit of the pair is even, and the second one  is odd, then we have $\Psi_{\text{eo}}(t)=4 (\xi_{d}(t)+ \vartheta_d(t))$, with $\vartheta_d(t)$ given by Eq.\,(\ref{vartheta}) with $B_n(s) \mapsto B_{\text{e}}(s)$ and $B_{m}(s') \mapsto B_{\text{o}}(s')$. If the opposite is true, we have $\Psi_{\text{oe}}(t)=4 (\xi_d(t)- \vartheta_d(t))$ instead.  Invoking Eq.\,(\ref{thetast}), we see that due to the presence of $\vartheta_d(t)$, the phase coefficient short-time behavior is now (to leading order) \textit{quadratic} rather than cubic in time: $\Psi_{\text{oe}}(t) \approx -\Psi_{\text{oe}}(t) \approx 4\vartheta_d^2(x)(\omega_c t)^4$. Expressions for deriving the scaling of both CSS and OATS states can be handled similarly as in the text. In both cases, the resulting $N$-scaling of optimal measurement time and estimation precision are not affected. However, quantum corrections $Q^{\,\text{eo}}_{\text{CSS}}(x,t)$ and $Q^{\,\text{eo}}_{\text{OATS}}(x,t)$ do differ. While for an OATS their contribution is still vanishing at $\tau_{\text{opt, eo}}^{\text{OATS}}$ in the $N \gg 1$ limit, for the CSS we have $Q^{\,\text{eo}}_{\text{CSS}}(x,t)=\frac{1}{128} N^2 \vartheta^4_d(x) (\omega_c t)^4$, which at $\tau_{\text{opt,eo}}^{\text{CSS}}$ is \textit{constant} with respect to $N$. However, as long as this constant correction is sufficiently small, $\frac{1}{128} N^2 \vartheta^4_d(x) (\omega_c \tau_{\text{opt,eo}}^{\text{CSS}})^4 \ll 1$, the $N^{-1/4}$ scaling of the uncertainty [Eq.\,(\ref{dbopteo}) in the main text] remains accurate.

\medskip

{\bf RC noise.} As the RC model (as we formulated it) hinges specifically upon having a spin-boson dephasing interaction, in order to have non-zero $\varphi_1(t)$ we must alter the hypotheses of Sec.\,\ref{sec:NQSB}. Specifically, let us approximate $\vec{k} \cdot (\vec{r}_n - \vec{r}_m) \approx \Omega_k t_{nm}$ with $\Omega_k=v k$ and  $t_{nm}= |\vec{r}_n-\vec{r}_m|/v$, disregarding angular dependence in the inner product. This leads to the following expressions for the dynamic coefficients:
\begin{eqnarray}
\chi_{nm}(t)&= \sum_{\vec{k}} |g_{k}|^2 \frac{1- \cos \Omega_{k} t}{\Omega_{k}^2} \coth \Big( \frac{\beta \Omega_{k}}{2} \Big) \cos\left( \Omega_k t_{nm} \right) ,
\label{chitt}\\
\xi_{nm}(t)&= \sum_{\vec{k}} |g_{k}|^2 \frac{\Omega_{k} t- \sin \Omega_{k} t}{\Omega_{k}^2} \cos\left( \Omega_k t_{nm} \right) ,
\label{xitt}\\
\vartheta_{nm}(t)&=\frac{1}{4}\sum_{\vec{k}} |g_k|^2 \frac{1- \cos \Omega_{k} t}{\Omega_{k}^2}   \sin \left(\Omega_k t_{nm} \right) .
\label{varthetatt} 
\end{eqnarray}
Note that, as the cosine is an even function, Eqs.\,(\ref{chitt})-(\ref{xitt}) match their Sec.\,\ref{sec:NQSB} counterparts, Eqs.\,(\ref{chirc})-(\ref{xirc}). However, in this case the argument of the sine in Eq.\,(\ref{varthetatt}) is always positive, which does not allow us to exploit the sines' odd parity to cancel the contributions corresponding to $\vec{k}$ and $-\vec{k}$. Thus, we obtain a non-vanishing $\vartheta_{nm}(t)$ and $\varphi_1(\vec{r},t) \neq 0$. As a result of the non trivial second phase, the reduced matrix elements spatial cumulant expansion, Eq.\,(\ref{cum}), is modified. Let us focus on the $D=1$ case.  It is straightforward to check that $\mathbb{E}\{ \vartheta_{nm}(t)\}$ vanishes, and so $\mathbb{E}\{ \varphi_1(\vec{r},t)\}=0$. In the same way,  $ \text{Cov}(\varphi_1(t),\gamma(t)) $ and  $\text{Cov}(\varphi_0(t),\varphi_1(t))$ are vanishing identically due to the spatial average being zero. There is, however, an additional contribution to the second-order cumulant coming from the variance of $\varphi_1(\vec{r},t)$. With that, the expression becomes
\begin{eqnarray*}
\mathbb{E} \left \{e^{-\gamma(\vec{r}, t)} e^{i \varphi_0(\vec{r}, t)} e^{i \varphi_1(\vec{r}, t)}  \right \} \approx 
e^{ -\bar{\gamma}(t) + i \bar{\varphi}_0(t)
+ \frac{1}{2} ( \Delta \bar{\gamma} (t)^2
- \Delta \bar{\varphi}_0(t)^2- \Delta \bar{\varphi}_1(t)^2 +  2 i \text{Cov}(\varphi_0(t),\gamma(t))},
\end{eqnarray*}
where we may write 
$$ \Delta \bar{\varphi}_1(t)^2=   \frac{N^2}{4} \sin(\theta_{\alpha \beta})^2  H_1(t)+ \bigg[ {N}(m^2-2 m m' \cos (\theta_{\alpha \beta}) +m'^2) -  \frac{N^2}{2} \sin(\theta_{\alpha \beta})^2 \bigg] H_2(t). $$
Here, the time-dependent quantities $H_1(t), H_2(t)$ can be evaluated in the short-time limit, leading to $H_1(t) \approx K_1\, \eta\, (\omega_c t)^4$ and $H_2(t) \approx K_2\, \eta^{2(s+1)} (\omega_c t)^4 $, with $K_1, K_2$ dimensionless constants. It is then straightforward to show that Claim \ref{claim} still holds in the presence of $\Delta \bar{\varphi}_1(t)^2$, leaving the conclusions derived in the small-$\eta$ regime -- in particular, the $N$-scaling of OATS and GHZ states -- unchanged.

\vspace{5mm}

\providecommand{\newblock}{}


\begin{thebibliography}{10}
\expandafter\ifx\csname url\endcsname\relax
  \def\url#1{{\tt #1}}\fi
\expandafter\ifx\csname urlprefix\endcsname\relax\def\urlprefix{URL }\fi
\providecommand{\eprint}[2][]{\url{#2}}

\bibitem{SethMetrology}
Giovannetti V, Lloyd S and Maccone J 2004 {\em Science\/} {\bf 306} 1330; 2011
  {\em Nat. Phot.\/} {\bf 5} 222

\bibitem{SmerziRMP}
Pezz\`e L, Smerzi A, Oberthaler M~K, Schmied R and Treutlein P 2018 {\em Rev.
  Mod. Phys.\/} {\bf 90} 035005

\bibitem{BraunRMP}
Braun D, Adesso G, Benatti F, Floreanini R, Marzolino U, Mitchell M~W and
  Pirandola S 2018 {\em Rev. Mod. Phys.\/} {\bf 90} 035006

\bibitem{Smirnereview}
Haase J~F, Smirne A, Huelga S~F, Kolodynski J and Demkowicz-Dobrzanski R 2016
  {\em Quantum Meas. Quantum Metr.\/} {\bf 5} 13

\bibitem{Huelga}
Huelga S F, Macchiavello C, Pellizzari T, Ekert A K, Plenio M B and Cirac J I
  1997 {\em Phys. Rev. Lett.} {\bf 79} 3865; Escher B M, de Matos Filho L R and
  Davidovich L 2011 { \em Nat. Phys.} {\bf 7} 406; Demkowicz-Dobrzanski R,
  Kolodynski J, and Guta M 2012 {\em Nat. Commun.} {\bf 3} 1063

\bibitem{Kita1993}
Kitagawa M and Ueda M 1993 {\em Phys. Rev. A\/} {\bf 47} 5138

\bibitem{DaveSqueezing}
Wineland D J, Bollinger J J, Itano W M, Moore F L and Heinzen D J 1992 {\em
  Phys. Rev. A} {\bf 46} R6797; Wineland D J, Bollinger J J, Itano W M, and
  Heinzen D J, 1994 {\em ibid.} {\bf 50} 67

\bibitem{Nori2}
Ma J, Wang X, Sun C~P and Nori F 2011 {\em Phys. Rep.\/} {\bf 509} 89

\bibitem{Vuletic}
Hu J, Chen W, Vendeiro Z, Urvoy A, Braverman B and Vuletic V 2017 {\em Phys.
  Rev. A} {\bf 96} 050301; Braverman B, Kawasaki A and Vuletic V 2018 {\em New
  J. Phys.} {\bf 20}, 103019

\bibitem{Ouyang}
Ouyang Y, Shettell N and Markham D 2022, {\em IEEE Trans. Inf. Theory} {\bf
  68}, 1809

\bibitem{Bollinger2}
Bollinger J J, Itano W M, Wineland D J and Heinzen D J 1996 {\em Phys. Rev. A}
  {\bf 54} R4649; Leibfried D, Barrett M D, Schaetz T, Britton J, Chiaverini J,
  Itano W M, Jost J D, Langer C, and Wineland D J 2004 {\em Science} {\bf 304}
  1476

\bibitem{Jones}
Jones, J A, Karlen S D, Fitzsimons J, Ardavan A, Benjamin S C, Briggs G A D and
  Morton J J 2009 {\em Science} \textbf{324} 1166; Sewell R J, Koschorreck M,
  Napolitano M, Dubost B, Behbood N and Mitchell M W 2012 {\em Phys. Rev.
  Lett.} \textbf{109} 253605

\bibitem{Stace}
Stace T~M 2010 {\em Phys. Rev. A\/} {\bf 82} 011611

\bibitem{Ye}
Ludlow A~D, Boyd M~M, Ye J, Peik E and Schmidt P~O 2015 {\em Rev. Mod. Phys.\/}
  {\bf 87} 637

\bibitem{Maccone}
Maccone L and Ren C 2020 {\em Phys. Rev. Lett.\/} {\bf 124} 200503

\bibitem{Abbott}
Abbott B~P and et~al 2016 {\em Phys. Rev. Lett.\/} {\bf 116} 061102

\bibitem{Catxere}
Casacio C A, Madsen L S, Terrasson A, Waleed M, Barnscheidt K, Hage B, Taylor M
  A and Bowen W P 2021 {\em Nature} {\bf 594} 201

\bibitem{Monika}
Davis E, Bentsen G and Schleier-Smith M 2016 {\em Phys. Rev. Lett.\/} {\bf 116}
  053601

\bibitem{VladanSatin}
Colombo S, Pedrozo-Penafiel, E Adiyatullin A F, Li Z, Mendez E, Shu C and
  Vuletic, V 2021, arXiv:2106.03754

\bibitem{Matsuzaki}
Matsuzaki Y, Benjamin S~C and Fitzsimons J 2011 {\em Phys. Rev. A\/} {\bf 84}
  012103

\bibitem{Chin2012}
Chin A~W, Huelga S~F and Plenio M~B 2012 {\em Phys. Rev. Lett.\/} {\bf 109}
  233601

\bibitem{Macies2015}
Macieszczak K 2015 {\em Phys. Rev. A\/} {\bf 92} 010102

\bibitem{Dorner2012}
Dorner U 2012 {\em New J. Phys.\/} {\bf 14} 043011

\bibitem{Jeske2014}
Jeske J, Cole J~H and Huelga S~F 2014 {\em New J. Phys.\/} {\bf 16} 073039

\bibitem{Layden}
Layden D and Cappellaro P 2011 {\em npj Quantum Inf.\/} {\bf 4} 1

\bibitem{Monz}
Monz T, Schindler P, Barreiro J T, Chwalla M, Nigg D, Coish W A, Harlander M,
  H\"{a}nsel W, Hennrich M and Blatt R, 2011 {\em Phys. Rev. Lett.} {\bf 106},
  130506; Postler L, Rivas A, Schindler P, Erhard A, Stricker R, Nigg D, Monz
  T, Blatt R and M\"{u}ller M 2018 {\em Quantum} {\bf 2} 90

\bibitem{Lieven2020}
Boter J~M, Xue X, Kr\"ahenmann T, Watson T~F, Premakumar V~N, Ward D~R, Savage
  D~E, Lagally M~G, Friesen M, Coppersmith S~N, Eriksson M~A, Joynt R and
  Vandersypen L~M~K 2020 {\em Phys. Rev. B\/} {\bf 101} 235133

\bibitem{qns}
Bylander J, Gustavsson S, Yan F, Yoshihara F, Harrabi K, Fitch G, Cory D,
  Nakamura Y, Tsai J S and Oliver W D 2011 {\em Nat. Phys.} {\bf 7}, 565;
  Alvarez G A and Suter D 2011 {\em Phys. Rev. Lett.} {\bf 107} 230501; Yan F,
  Gustavsson S, Bylander J, Jin X, Yoshihara F, Cory D G, Nakamura Y, Orlando T
  P and Oliver W D 2013 {\em Nature Commun.} {\bf 4} 2337; Malinowski F K,
  Martins F, Cywinski L, Rudner M S, Nissen P D, Fallahi S, Gardner G C, Manfra
  M J, Marcus C M and Kuemmeth F 2017 {\em Phys. Rev. Lett.} {\bf 118} 177702;
  Chan K W, Huang W, Yang C H, Hwang J C C, Hensen B, Tanttu T, Hudson F E,
  Itoh K M, Laucht A, Morello A and Dzurak A S 2018 {\em Phys. Rev. Appl.} {\bf
  10} 044017

\bibitem{Frey}
Frey V M, Mavadia S, Norris L M, De Ferranti W, Lucarelli D, Viola L and
  Biercuk M J 2017 {\em Nature Commun.} {\bf 8} 2189; Frey V M, Norris L M,
  Viola L and Biercuk M J {\em Phys. Rev. Appl.} {\em 14} 024021

\bibitem{Szankowski}
Sza\ifmmode~\acute{n}\else \'{n}\fi{}kowski P, Trippenbach M and
  Chwede\ifmmode~\acute{n}\else \'{n}\fi{}czuk J 2014 {\em Phys. Rev. A\/} {\bf
  90} 063619

\bibitem{Quintana2017}
Quintana C~M, Chen Y, Sank D, Petukhov A~G, White T~C, Kafri D, Chiaro B,
  Megrant A, Barends R, Campbell B, Chen Z, Dunsworth A, Fowler A~G, Graff R,
  Jeffrey E, Kelly J, Lucero E, Mutus J~Y, Neeley M, Neill C, O'Malley P~J~J,
  Roushan P, Shabani A, Smelyanskiy V~N, Vainsencher A, Wenner J, Neven H and
  Martinis J~M 2017 {\em Phys. Rev. Lett.\/} {\bf 118} 057702

\bibitem{Yan2018}
Yan F, Campbell D, Krantz P, Kjaergaard M, Kim D, Yoder J~L, Hover D, Sears A,
  Kerman A~J, Orlando T~P, Gustavsson S and Oliver W~D 2018 {\em Phys. Rev.
  Lett.\/} {\bf 120} 260504

\bibitem{Uwe}
von L\"{u}pke U, Beaudoin F, Norris L~M, Sung Y, Winik R, Qiu J~Y, Kjaergaard
  M, Kim D, Yoder J, Gustavsson S, Viola L and Oliver W~D 2020 {\em PRX
  Quantum\/} {\bf 1} 010305

\bibitem{ClerkRMP}
Clerk A~A, Devoret M~H, Girvin S~M, Marquardt F and Schoelkopf R~J 2010 {\em
  Rev. Mod. Phys.\/} {\bf 82} 1155

\bibitem{Paz2017}
Paz-Silva G~A, Norris L~M and Viola L 2017 {\em Phys. Rev. A\/} {\bf 95} 022121

\bibitem{FelixPRA}
Beaudoin F, Norris L~M and Viola L 2018 {\em Phys. Rev. A\/} {\bf 98} (R)020102

\bibitem{AncillaBased}
He W~T, Guang H~Y, Li Z~Y, Deng R~Q, Zhang N~N, Zhao J~X, Deng F~G and Ai Q
  2021 {\em Phys. Rev. A\/} {\bf 104} 062429

\bibitem{DegenRMP}
Degen C~L, Reinhard F and Cappellaro P 2017 {\em Rev. Mod. Phys.\/} {\bf 89}
  035002

\bibitem{Layden2}
Layden D, Zhou S, Cappellaro P and Jiang L 2019 {\em Phys. Rev. Lett.\/} {\bf
  122} 040502

\bibitem{multiDD}
Paz-Silva G~A, Lee S~W, Green T~J and Viola L 2016 {\em New J. Phys.\/} {\bf
  18} 073020

\bibitem{DurDD}
Sekatski P, Skotiniotis M, Ko{\l{}}ody{\'{n}}ski J and D{\"{u}}r W 2017 {\em
  Quantum\/} {\bf 1} 27

\bibitem{BraskPRX}
Brask J~B, Chaves R and Ko\l{}ody\ifmmode~\acute{n}\else \'{n}\fi{}ski J 2015
  {\em Phys. Rev. X\/} {\bf 5} 031010

\bibitem{ruster}
Ruster T, Kaufmann H, Luda M~A, Kaushal V, Schmiegelow C~T, Schmidt-Kaler F and
  Poschinger U~G 2017 {\em Phys. Rev. X\/} {\bf 7} 031050

\bibitem{Postler}
Postler L, Rivas {\'A}, Schindler P, Erhard A, Stricker R, Nigg D, Monz T,
  Blatt R and M{\"u}ller M 2018 {\em Quantum\/} {\bf 2} 90

\bibitem{QECBook}
Lidar D~A and Brun T~A 2013 {\em Quantum Error Correction\/} (Cambridge
  University Press)

\bibitem{RU1}
Buscemi F, Chiribella G and D'Ariano G M 2005 {\em Phys. Rev. Lett.} {\bf 95}
  090501; Helm, J and Strunz W T 2009 {\em Phys. Rev. A} {\bf 80} 042108; Helm
  J, Strunz W T, Rietzler S and W\"urflinger L E 2011 {\em ibid.} {\bf 83}
  042103; Crow D and Joynt R 2014 {\em Phys. Rev. A} {\bf 89} 042123

\bibitem{Kay}
Kay S~M 1993 {\em Fundamentals of Statistical Signal Processing: Estimation
  Theory\/} (Prentice Hall)

\bibitem{Helstrom}
Helstrom C~W 1976 {\em Quantum Detection and Estimation Theory\/} (Academic
  Press)

\bibitem{Rubio}
Rubio J and Dunningham J 2019 {\em New J. Phys.\/} {\bf 21} 043347


\bibitem{Chabuda} Macieszczak K, Fraas M and Demkowicz-Dobrzański R. 2014 {\em New J. Phys.\/}  {\bf 16} 113002; Chabuda K, Leroux I D and Demkowicz-Dobrzański R 2016 {\em New J. Phys.\/}  {\bf 18} 083035


\bibitem{CarlDistance}
Braunstein S L and Caves C M 1994 {\em Phys. Rev. Lett.} {\bf 72} 3439;
  Braunstein S L, Caves C M and Milburn G J 1996 {\em Ann. Phys.} {\bf 247} 135

\bibitem{Holevo}
Holevo A~S 2011 {\em Probabilistic and Statistical Aspects of Quantum Theory\/}
  (2nd ed., Edizioni della Normale, Pisa)

\bibitem{Datta}
Albarelli F, Friel J~F and Datta A 2019 {\em Phys. Rev. Lett.\/} {\bf 123}
  200503

\bibitem{LiangBound}
Zhou S, Zou C~L and Jiang L 2020 {\em Quantum Sci. Technol.\/} {\bf 5} 025005

\bibitem{GCS}
Zhang W M, Feng D H and Gilmore R 1990 {\em Rev. Mod. Phys.} {\bf 62} 867;
  Boixo S, Viola L and Ortiz G 2007 {\em EPL} {\bf 79} 40003

\bibitem{SchulteEchoes}
Schulte M, Martinez-Lahuerta V~J, Scharnagl M~S and Hammerer K 2020 {\em
  Quantum\/} {\bf 4} 268

\bibitem{Bohnet}
Bohnet J~G, Sawyer B~C, Britton J~W, Wall M~L, Rey A~M, Foss-Feig M and
  Bollinger J~J 2016 {\em Science\/} {\bf 352} 1297

\bibitem{Ober}
Muessel W, Strobel H, Linnemann D, Hume D~B and Oberthaler M~K 2014 {\em Phys.
  Rev. Lett.\/} {\bf 113} 103004

\bibitem{Lukin}
Bennett S~D, Yao N~Y, Otterbach J, Zoller P, Rabl P and Lukin M~D 2013 {\em
  Phys. Rev. Lett.\/} {\bf 110} 156402

\bibitem{Paz2019}
Paz-Silva G~A, Norris L~M, Beaudoin F and Viola L 2019 {\em Phys. Rev. A\/}
  {\bf 100} 042334

\bibitem{PazFF}
Paz-Silva G~A and Viola L 2014 {\em Phys. Rev. Lett.\/} {\bf 113} 250501

\bibitem{RU2}
Koenraad M R Audenaert and Stefan Scheel 2008 New J. Phys. {\bf 10} 023011;
  Mendl C B and Wolf M M 2009, Commun. Math. Phys. {\bf 289}, 1057; Chen H B,
  Gneiting C, Lo P Y, Chen Y N and Nori F 2018 {\em Phys. Rev. Lett.} {\bf 120}
  030403; Chen H B, Lo P Y, Gneiting C, Bae J, Chen Y N and Nori F 2019 {\em
  Nat. Commun.} {\bf 10} 1

\bibitem{Watrous}
Lee C~D~Y and Watrous J 2020 {\em Quantum\/} {\bf 4} 253

\bibitem{IschiPRB}
Ischi B, Hilke M and Dub\'e M 2005 {\em Phys. Rev. B\/} {\bf 71} 195325

\bibitem{Irene}
Hodgson T E, Viola L and Amico I 2008 {\em Phys. Rev. B} {\bf 78} 165311; 2010
  {\em Phys. Rev. A} {\bf 81} 062321; Cotlet O and Lovett B W 2014 {\em New J.
  Phys.} {\bf 16} 103016

\bibitem{Palma}
Palma G M, Suominen K A and Ekert A 1996 {\em Proc. R. Soc. Lond. A} {\bf 452}
  567; Reina J H, Quiroga L and Johnson N F 2002 {\em Phys. Rev. A} {\bf 65}
  032326

\bibitem{Mogilev}
Mogilevtsev D, Garusov E, Korolkov M~V, Shatokhin V~N and Cavalcanti S~B 2018
  {\em Phys. Rev. A\/} {\bf 98} 042116

\bibitem{RandomDD}
Viola L and Knill E 2005 {\em Phys. Rev. Lett.} {\bf 94} 060502; Santos L F and
  Viola L 2005 {\em Phys. Rev. A} {\bf 72} 062303


\bibitem{Matsuzaki2} Matsuzaki Y, Benjamin S, Nakayama S, Saito S, and Munro W J 2018 {\em Phys. Rev. Lett.\/} {\bf 120} 140501


\bibitem{Averin} Averin D V, Xu K, Zhong  Y P, Song C, Wang H and Han S 2016  {\em Phys. Rev. Lett.\/}  {\bf 116} 010501

\bibitem{KavehLimits}
Khodjasteh K, Erd\'elyi T and Viola L 2011 {\em Phys. Rev. A\/} {\bf 83} 020305

\bibitem{BlattNoise}
Monz T, Schindler P, Barreiro J T, Chwalla M, Nigg D, Coish W A, Harlander M,
  H\"{a}nsel W, Hennrich M and Blatt R 2011 {\em Phys. Rev. Lett.} {\bf 106}
  130506; Schindler P, Barreiro J T, Monz T, Nebendahl V, Nigg D, Chwalla M,
  Hennrich M and Blatt R 2011 {\em Science} {\bf 332} 1059

\bibitem{Mike}
Edmunds C~L, Hempel C, Harris R~J, Frey V, Stace T~M and Biercuk M~J 2020 {\em
  Phys. Rev. Res.\/} {\bf 2} 013156

\bibitem{Traps}
Bruzewicz C D, McConnell R, Chiaverini J and Sage J M 2016 {\em Nat. Commun.}
  {\bf 7} 13005; Decaroli C, Matt R, Oswald R, Ernzer M, Flannery J, Ragg S and
  Home, J P 2021, arXiv:2103.05978

\bibitem{carrasco}
Carrasco, S C, Goerz, M H, Li, Z and Colombo, S, Vuleti{\'c}, V and Malinovsky,
  V S 2022, arXiv:2201.01744

\bibitem{baumann1985}
Baumann K and Hegerfeldt G~C 1985 {\em Publications of the Research Institute
  for Mathematical Sciences\/} {\bf 21} 191--204

\bibitem{Fagnola}
Fagnola F, Gough J E, Nurdin H I and Viola L 2019 J. Phys. A {\bf 52} 385301

\bibitem{Kubo}
Kubo R 1962 {\em J. Phys. Soc. Japan\/} {\bf 17} 1100

\end{thebibliography}

\end{document}